\pdfoutput=1
\documentclass[11pt,twoside,a4paper,cmspaper,final,collab]{cms-tdr}
\def\svnVersion{d456329}\def\svnDate{2025/02/24}\def\cmsCernNoTag{CERN-EP-2024-179}\def\cmsCernDate{\today}\def\cmsMessage{Published in the Journal of High Energy Physics as \href{http://dx.doi.org/10.1007/JHEP02(2025)097}{\doi{10.1007/JHEP02(2025)097}.}}
\begin{document}\cmsNoteHeader{HIG-19-011}

\newlength\cmsTabSkip\setlength{\cmsTabSkip}{1ex}

\newcommand{\CP}{\ensuremath{C\hspace{-.08em}P}\xspace}
\newcommand{\SD}{\text{SD}\xspace}
\newcommand{\ttH}{\ensuremath{\ttbar\PH}\xspace}
\newcommand{\tH}{\ensuremath{\PQt\PH}\xspace}
\newcommand{\tHq}{\ensuremath{\tH\PQq}\xspace}
\newcommand{\tHW}{\ensuremath{\tH\PW}\xspace}
\newcommand{\Htobb}{\ensuremath{\PH\to\bbbar}\xspace}
\newcommand{\ttLF}{\ensuremath{\ttbar\text{LF}}\xspace}
\newcommand{\ttB}{\ensuremath{\ttbar\text{B}}\xspace}
\newcommand{\ttC}{\ensuremath{\ttbar\text{C}}\xspace}
\newcommand{\ttbb}{\ensuremath{\ttbar+\bbbar}\xspace}
\newcommand{\ttborbb}{\ensuremath{\ttbar+\PQb(\PAQb)}\xspace}
\newcommand{\tttwob}{\ensuremath{\ttbar+2\PQb}\xspace}
\newcommand{\ttjets}{\ensuremath{\ttbar+\text{jets}}\xspace}
\newcommand{\tW}{\ensuremath{\PQt\PW}\xspace}
\newcommand{\ttW}{\ensuremath{\ttbar\PW}\xspace}
\newcommand{\ttg}{\ensuremath{\ttbar\PGg}\xspace}
\newcommand{\ttV}{\ensuremath{\ttbar\PV}\xspace}
\newcommand{\Zjets}{\ensuremath{\PZ+\text{jets}}\xspace}
\newcommand{\Vjets}{\ensuremath{\PV+\text{jets}}\xspace}
\newcommand{\diboson}{\ensuremath{\PV\PV}\xspace}
\newcommand{\ttsample}{\ensuremath{\ttbar}\xspace}
\newcommand{\ttbbsample}{\ensuremath{\ttbar\bbbar}\xspace}
\newcommand{\muR}{\ensuremath{\mu_{\text{R}}}\xspace}
\newcommand{\muF}{\ensuremath{\mu_{\text{F}}}\xspace}
\newcommand{\fhSevenSR}{(7\,jets, $\geq$4\,{\PQb}\,tags)\xspace}
\newcommand{\fhEightSR}{(8\,jets, $\geq$4\,{\PQb}\,tags)\xspace}
\newcommand{\fhNineSR}{($\geq$9\,jets, $\geq$4\,{\PQb}\,tags)\xspace}
\newcommand{\slSixFour}{($\geq$6\,jets, $\geq$4\,{\PQb}\,tags)\xspace}
\newcommand{\slFiveFour}{(5\,jets, $\geq$4\,{\PQb}\,tags)\xspace}
\newcommand{\dlThreeThree}{(3\,jets, 3\,{\PQb}\,tags)\xspace}
\newcommand{\dlFourThree}{($\geq$4\,jets, $\geq$3\,{\PQb}\,tags)\xspace}
\newcommand{\mi}[1]{\ensuremath{m_{#1}}\xspace}
\newcommand{\mti}[1]{\ensuremath{m_{\text{T},#1}}\xspace}
\newcommand{\mqq}{\mi{\PQq\PQq}}
\newcommand{\mll}{\ensuremath{m_{\Pell\Pell}}\xspace}
\newcommand{\ptH}{\ensuremath{\pt^{\PH}}\xspace}
\newcommand{\tflm}{\ensuremath{\text{TF}_{\text{L}}}\xspace}
\newcommand{\OANN}[1]{\ensuremath{\text{O}(#1)}\xspace}
\newcommand{\muttH}{\ensuremath{\mu_{\ttH}}\xspace}
\newcommand{\muhat}{\ensuremath{\hat{\mu}}\xspace}
\newcommand{\mutH}{\ensuremath{\mu_{\tH}}\xspace}
\newcommand{\kappat}{\ensuremath{\kappa_{\PQt}}\xspace}
\newcommand{\kappaV}{\ensuremath{\kappa_{\PV}}\xspace}
\newcommand{\kappatcp}{\ensuremath{\widetilde{\kappa}_{\PQt}}\xspace}
\newcommand{\kappatprime}{\ensuremath{\kappa^{\prime}_{\PQt}}\xspace}
\newcommand{\fCP}{\ensuremath{f_{\CP}}\xspace}
\newcommand{\cosa}{\ensuremath{\cos\alpha}\xspace}
\newcommand{\mean}[1]{\ensuremath{\langle #1 \rangle}\xspace}

\newcommand{\lumitotal}{\ensuremath{138\fbinv}\xspace}
\newcommand{\resttHSigObs}{\ensuremath{1.3}\xspace}
\newcommand{\resttHSigExp}{\ensuremath{4.1}\xspace}
\newcommand{\resttHgof}{\ensuremath{0.88}\xspace}
\newcommand{\resttHMu}{\ensuremath{0.33}\xspace}
\newcommand{\resttHMuUpDn}{\ensuremath{0.26}\xspace}
\newcommand{\resttHMuStatUp}{\ensuremath{0.17}\xspace}
\newcommand{\resttHMuStatDn}{\ensuremath{0.16}\xspace}
\newcommand{\resttHMuSystUp}{\ensuremath{0.20}\xspace}
\newcommand{\resttHMuSystDn}{\ensuremath{0.21}\xspace}
\newcommand{\resttHMuExpStatUpDn}{\ensuremath{0.17}\xspace}
\newcommand{\resttHMuExpSystUp}{\ensuremath{0.23}\xspace}
\newcommand{\resttHMuExpSystDn}{\ensuremath{0.19}\xspace}
\newcommand{\resttHttB}{\ensuremath{1.19}\xspace}
\newcommand{\resttHttBUp}{\ensuremath{0.13}\xspace}
\newcommand{\resttHttBDn}{\ensuremath{0.12}\xspace}
\newcommand{\resttHttC}{\ensuremath{1.07}\xspace}
\newcommand{\resttHttCUp}{\ensuremath{0.20}\xspace}
\newcommand{\resttHttCDn}{\ensuremath{0.19}\xspace}
\newcommand{\pvalueInclToComb}{\ensuremath{0.28}\xspace}
\newcommand{\sigmaInclToComb}{\ensuremath{1.1}\xspace}
\newcommand{\pvalueInclToSM}{\ensuremath{0.02}\xspace}
\newcommand{\sigmaInclToSM}{\ensuremath{2.4}\xspace}
\newcommand{\pvalueLeptonicToPaper}{\ensuremath{0.41}\xspace}
\newcommand{\sigmaLeptonicToPaper}{\ensuremath{0.8}\xspace}
\newcommand{\resSTXSgof}{\ensuremath{0.89}\xspace}
\newcommand{\pvalueSTXStoIncl}{\ensuremath{0.67}\xspace}
\newcommand{\sigmaSTXStoIncl}{\ensuremath{0.4}\xspace}
\newcommand{\pvalueSTXStoSM}{\ensuremath{0.21}\xspace}
\newcommand{\sigmaSTXStoSM}{\ensuremath{1.3}\xspace}
\newcommand{\restHLimit}{\ensuremath{14.6}\xspace}
\newcommand{\restHExpLimit}{\ensuremath{19.3}\xspace}
\newcommand{\restHExpLimitUp}{\ensuremath{9.2}\xspace}
\newcommand{\restHExpLimitDn}{\ensuremath{6.0}\xspace}
\newcommand{\resTwoDttH}{\ensuremath{0.35}\xspace}
\newcommand{\resTwoDtH}{\ensuremath{-3.83}\xspace}
\newcommand{\resktkVkt}{\ensuremath{+0.59}\xspace}
\newcommand{\resktkVkV}{\ensuremath{+1.40}\xspace}
\newcommand{\reskt}{\ensuremath{+0.54}\xspace}
\newcommand{\resktNegLo}{\ensuremath{-0.55}\xspace}
\newcommand{\resktNegHi}{\ensuremath{-0.24}\xspace}
\newcommand{\resktPosLo}{\ensuremath{+0.20}\xspace}
\newcommand{\resktPosHi}{\ensuremath{+0.72}\xspace}
\newcommand{\resktcpeven}{\ensuremath{+0.53}\xspace}
\newcommand{\resktcpodd}{\ensuremath{0.00}\xspace}
\newcommand{\resktcpevenComb}{\ensuremath{0.82}\xspace}
\newcommand{\resktcpoddComb}{\ensuremath{-0.65}\xspace}
\newcommand{\resfCPLimitComb}{\ensuremath{0.85}\xspace}
\newcommand{\rescosaLimitComb}{\ensuremath{0.39}\xspace}

\newcommand{\cmsTable}[1]{\resizebox{0.85\textwidth}{!}{#1}}

\cmsNoteHeader{HIG-19-011}
\title{Measurement of the \texorpdfstring{\ttH}{ttH} and \texorpdfstring{\tH}{tH} production rates in the \texorpdfstring{\Htobb}{Htobb} decay channel using proton-proton collision data at \texorpdfstring{$\sqrt{s} = 13\TeV$}{sqrt(s) = 13 TeV}}
\author{The CMS Collaboration}
\date{\today}

\abstract{An analysis of the production of a Higgs boson (\PH) in association with a top quark-antiquark pair (\ttH) or a single top quark (\tH) is presented. The Higgs boson decay into a bottom quark-antiquark pair (\Htobb) is targeted, and three different final states of the top quark decays are considered, defined by the number of leptons (electrons or muons) in the event. The analysis utilises proton-proton collision data collected at the CERN LHC with the CMS experiment at $\sqrt{s}=13\TeV$ in 2016--2018, which correspond to an integrated luminosity of \lumitotal. The observed \ttH production rate relative to the standard model expectation is $\resttHMu\pm\resttHMuUpDn = \resttHMu\pm\resttHMuStatUp\stat\pm\resttHMuSystDn\syst$. Additionally, the \ttH production rate is determined in intervals of Higgs boson transverse momentum. An upper limit at 95\% confidence level is set on the \tH production rate of \restHLimit times the standard model prediction, with an expectation of $\restHExpLimit^{+\restHExpLimitUp}_{-\restHExpLimitDn}$. Finally, constraints are derived on the strength and structure of the coupling between the Higgs boson and the top quark from simultaneous extraction of the \ttH and \tH production rates, and the results are combined with those obtained in other Higgs boson decay channels.
}

\hypersetup{%
pdfauthor={CMS Collaboration},%
pdftitle={Measurements of the ttH and tH production rates in the H-to-bb decay channel with 138/fb of proton-proton collision data at 13 TeV},%
pdfsubject={CMS, ttH, tH, bb channel},%
pdfkeywords={CMS, ttH, tH, bb channel}}

\maketitle

\section{Introduction}
\label{sec:intro}
In the standard model (SM) of particle physics, the Higgs boson (\PH) couples to fermions with a Yukawa-type interaction, with a coupling strength proportional to the fermion mass.
The coupling of the Higgs boson to the top quark, the heaviest known fermion, is particularly relevant both in the SM and in models of new physics beyond the SM (BSM).
For example, due to its size, the top-Higgs coupling strongly affects the stability of the electroweak vacuum via virtual contributions~\cite{Alekhin:2012py} and may play a special role in the mechanism of electroweak symmetry breaking~\cite{Dobrescu:1997nm,Chivukula:1998wd}.
Probing the coupling of the Higgs boson to the top quark is therefore instrumental in testing the SM and constraining BSM models, which may predict different coupling strengths and structures.

\begin{figure}[!htb]
  \centering
    \includegraphics[width=0.3\textwidth]{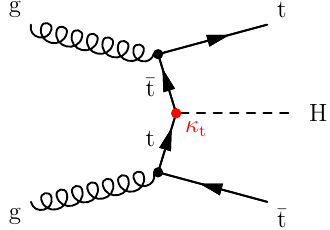} 
    \includegraphics[width=0.3\textwidth]{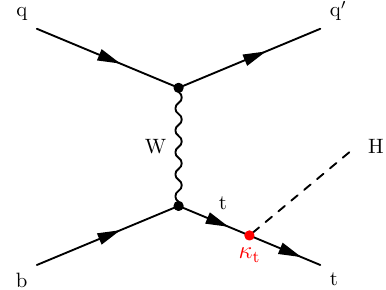} 
    \includegraphics[width=0.3\textwidth]{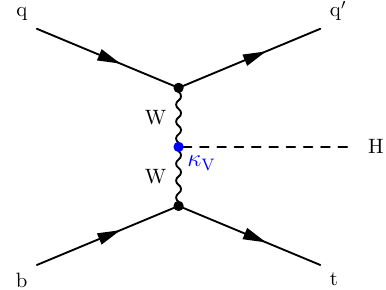} 
  \caption{
    Representative leading order Feynman diagrams for the associated production of a Higgs boson and a top quark-antiquark pair (left) and for the associated production of a single top quark and a Higgs boson in the $t$ channel, where the Higgs boson couples to the top quark (centre) or the \PW boson (right).
    The \kappat (red) and \kappaV (blue) denote the Higgs boson coupling strength to top quarks and vector bosons, respectively.
  }
  \label{fig:intro:productiondiagrams}
\end{figure}

The associated production of a Higgs boson and a top quark-antiquark pair (\ttH) provides a direct probe of the top-Higgs coupling as illustrated by the Feynman diagram in Fig.~\ref{fig:intro:productiondiagrams} (left).
The \ttH production process has been observed by the ATLAS and CMS Collaborations by combining measurements that target different decay channels of the Higgs boson using the LHC Run-1 and part of the Run-2 datasets~\cite{Aaboud:2018urx,Sirunyan:2018hoz}.
Subsequent measurements of \ttH production with the full Run-2 dataset have been performed, targeting decays of the Higgs boson into pairs of photons, \PW or \PZ bosons, tau leptons, and bottom quarks (\bbbar)~\cite{ATLAS:2022tnm,ATLAS:2020ior,CMS:2020cga,CMS:2021kom,CMS:2020mpn,ATLAS:2021qou,Sirunyan:2018mvw,Sirunyan:2018ygk}.
The associated production of a single top quark and a Higgs boson (\tH) occurs through two sets of diagrams, where the Higgs boson either couples to the top quark or to the \PW boson~\cite{Maltoni:2001hu,Farina:2012xp,Agrawal:2012ga,Demartin:2015uha}, as shown in Fig.~\ref{fig:intro:productiondiagrams} (centre and right). Consequently, \tH production offers insights into the Higgs boson's coupling to top quarks and vector bosons, as well as the interference of the two, allowing for the investigation of the relative sign of the two couplings.
Previous searches for \tH production with Run-2 data targeted Higgs boson decays to $\PGg\PGg$, $\PW\PW$, $\PZ\PZ$, $\PGt\PGt$, and \bbbar~\cite{ATLAS:2022tnm,ATLAS:2020ior,CMS:2020cga,CMS:2021kom,CMS:2020mpn,Sirunyan:2018lzm}.
In the SM, the Higgs boson has a positive eigenvalue under combined charge-parity (\CP) transformations (\CP even).
The combined measurement of \ttH and \tH production provides a probe of \CP-odd admixtures in the top-Higgs coupling through interfering contributions, which would be a clear sign of physics beyond the SM.
Previous constraints on the \CP nature of the top-Higgs coupling have been obtained from measurements of \ttH and \tH production with Run-2 data targeting Higgs boson decays to $\PGg\PGg$, $\PZ\PZ$, $\PW\PW$, $\PGt\PGt$, and \bbbar~\cite{ATLAS:2020ior,CMS:2020cga,CMS:2021nnc,CMS:2022dbt,ATLAS:2023cbt}.

In the SM, for a 125\GeV Higgs boson, the \Htobb decay has the largest branching fraction of $0.58\pm 0.02$~\cite{deFlorian:2016spz}.
Furthermore, \ttH production with \Htobb decays involves Higgs boson couplings only to fermions and thus exclusively probes the Higgs Yukawa sector.
The \Htobb decay channel is thus experimentally attractive, yet its measurement is challenging due to considerable SM backgrounds at the CERN LHC.
The previous searches for \ttH and \tH production in the \Htobb channel by the ATLAS and CMS Collaborations achieved sensitivities that correspond to observed (expected) significances of up to 1.6 (2.7) standard deviations (\SD) for \ttH production~\cite{ATLAS:2021qou,Sirunyan:2018mvw,Sirunyan:2018ygk} and observed (expected) upper limits at 95\% confidence level (\CL) on SM \tH production of approximately 90 (41) times the SM expectation~\cite{Sirunyan:2018lzm}.
At the time of submitting this paper, the ATLAS Collaboration reported a re-analysis of the Run-2 data with an observed (expected) significance for \ttH production of 4.6 (5.4)\,\SD~\cite{ATLAS:2024gth}.
Constraints on the \CP nature of the top-Higgs coupling have been obtained in this channel from the combination of information from \ttH and \tH production by the ATLAS Collaboration~\cite{ATLAS:2023cbt}.

This article describes the combined analysis of \ttH and \tH production in the \bbbar decay channel of the Higgs boson by the CMS Collaboration.
Proton-proton ($\Pp\Pp$) collision data collected during LHC Run-2 from 2016 to 2018 at a centre-of-mass energy of 13\TeV are used, corresponding to an integrated luminosity of \lumitotal.
Several signal interpretations are performed.
First, targeting solely the \ttH signal, the \ttH production rate is determined assuming the \tH production rate predicted in the SM.
The \ttH rate is determined both inclusively and in different regions of transverse momentum \pt of the Higgs boson (\ptH), following the simplified template cross section (STXS) approach~\cite{deFlorian:2016spz}.
Second, targeting the \tH signal process, an upper limit on the \tH production rate is determined assuming the \ttH production rate predicted in the SM.
Third, targeting simultaneously the \ttH and \tH production processes, the Higgs boson coupling to top quarks is analysed, and both the strength and the \CP structure of the coupling are probed.

The analysis builds upon strategies developed in previous analyses of \ttH~\cite{Sirunyan:2018mvw,Sirunyan:2018ygk} and \tH production~\cite{Sirunyan:2018lzm} using 35.9\fbinv of data collected in 2016.
The event selection aims at identifying events in which a Higgs boson is produced in association with a \ttbar pair or a single top quark, and decays to \bbbar. 
Three mutually exclusive channels are considered in the analysis, defined by the number of leptons (electrons or muons) in the \ttbar decay modes: the fully hadronic (FH) channel, in which both \PW bosons decay to a \qqbar pair, the single-lepton (SL) channel, with one \PW boson decaying to a charged lepton and a neutrino and the other \PW boson decaying to \qqbar, and the dilepton (DL) channel, with both \PW bosons decaying to a charged lepton and neutrino.
The SL channel also includes dedicated event categories targeting \tH production.
No dedicated reconstruction of hadronically decaying tau leptons is performed. 
The experimental signature of signal events is characterised by the
presence of high-\pt \PQb quark jets (\PQb jets) from the Higgs boson
and top quark decays and, depending on the channel, jets or isolated
electrons and muons, as well as missing transverse momentum arising
from the presence of undetected neutrinos from the \PW boson or tau
lepton decays.
Electrons or muons from tau lepton decays can enter the selection in the SL and DL channels, while hadronically decaying tau leptons can contribute to the FH and SL channels.
Dominant background contributions arise from SM events composed uniquely of jets produced through the strong interaction, referred to as quantum chromodynamics (QCD) multijet events, in the FH channel and from \ttjets production in all channels.
The latter includes in particular \ttbb production, where additional \PQb quarks can arise from QCD radiation or loop-induced QCD processes.
The \ttbb background remains almost irreducible with respect to \ttH, \Htobb, with both processes having four \PQb quarks in the final state, and constitutes the critical background to the analysis.

The presence of multiple \PQb jets in the final state and their
experimental resolution complicate the event reconstruction and limit
the sensitivity of the analysis using simple kinematic observables
like the Higgs boson invariant mass.
Therefore, the signal extraction is performed using a multivariate analysis based on artificial neural networks (ANNs) that simultaneously exploit the information from multiple experimental observables.
The ANNs are used to discriminate signal from background events and, in the SL and DL channels, to categorise events into signal and background control regions.
The signal rate is obtained from a combined profile likelihood fit of distributions of the ANN score, distributions of ratios of ANN scores, or of the event yield to the data, depending on the channel and category.

Several changes in the analysis strategy have been adopted with respect to the previous analyses~\cite{Sirunyan:2018mvw,Sirunyan:2018ygk,Sirunyan:2018lzm}, including usage of the state-of-the-art \ttbb event simulation~\cite{Jezo:2018yaf,Buccioni:2019plc} in the background modelling, updates in the multivariate analysis techniques, improvements in the \PQb tagging algorithm, and the exploitation of all three \ttbar decay channels.

This article is structured as follows.
The CMS detector is described in Section~\ref{sec:detector}.
In Section~\ref{sec:trigger}, the data samples and trigger selection are described, followed by a description of the simulated samples in Section~\ref{sec:mcsamples}.
The object and event reconstruction are presented in Section~\ref{sec:objects} and the event selection in Section~\ref{sec:selection}.
The modelling of the critical \ttjets background is detailed in Section~\ref{sec:bkgmodel}.
The analysis strategy in the different channels is detailed in Section~\ref{sec:strategy} and systematic uncertainties are discussed in Section~\ref{sec:systematics}. 
The statistical analysis and obtained results are presented in Section~\ref{sec:results}. 
Finally, a summary is provided in Section~\ref{sec:summary}. 
Tabulated results are provided in the HEPData record for this analysis~\cite{hepdata}.

\section{The CMS detector}
\label{sec:detector}
The central feature of the CMS apparatus is a superconducting solenoid of 6\unit{m} internal diameter, providing a magnetic field of 3.8\unit{T}.
Within the solenoid volume are a silicon pixel and strip tracker, a lead tungstate crystal electromagnetic calorimeter (ECAL), and a brass and scintillator hadron calorimeter, each composed of a barrel and two endcap sections.
Forward calorimeters extend the pseudorapidity coverage provided by the barrel and endcap detectors.
Muons are measured in gas-ionisation detectors embedded in the steel flux-return yoke outside the solenoid.
A more detailed description of the CMS detector, together with a definition of the coordinate system used and the relevant kinematic variables, can be found in Refs.~\cite{CMS:2008xjf, Phase1Pixel}.

Events of interest are selected using a two-tiered trigger system~\cite{CMS:2016ngn}.
The first level (L1), composed of custom hardware processors, uses information from the calorimeters and muon detectors to select events at a rate of around 100\unit{kHz} within a fixed latency of about 4\mus~\cite{CMS:2020cmk}.
The second level, known as the high-level trigger (HLT), consists of a farm of processors running a version of the full event reconstruction software optimised for fast processing, and reduces the event rate to a few \unit{kHz} before data storage.

\section{Data samples and trigger requirements}
\label{sec:trigger}
The analysis uses $\Pp\Pp$ collision data recorded at $\sqrt{s} = 13\TeV$ at the LHC during 2016--2018.
Only the data-taking periods during which the CMS detector was fully
operational are considered.
The corresponding total integrated luminosity amounts to \lumitotal, of
which 36.3, 41.5, and 59.7\fbinv have been recorded in 2016, 2017, and
2018,
respectively~\cite{CMS-LUM-17-003,CMS-PAS-LUM-17-004,CMS-PAS-LUM-18-002}.
In this analysis, events are selected at the HLT level by different trigger requirements defined by the
presence of several jets or one or two leptons and, in some cases, a
large scalar sum of jet \pt (\HT) in the events, depending on the analysis channel.

Novel trigger criteria have been developed at the beginning of the LHC Run-2, targeting events in the FH channel. 
They require the presence of a large number of jets, large \HT, and one, two, or three jets tagged by the online \PQb jet identification. 
To further improve the trigger efficiency, events recorded by a single-jet trigger with a high-\pt threshold or by a trigger with a high-\HT threshold are also considered.
The selection criteria are detailed in Table~\ref{tab:fhtriggers}.
In the SL channel, events were selected by triggers requiring the
presence of one muon or one electron with \pt exceeding a given threshold.
In order to compensate for the increased \pt threshold of the
single-electron trigger in 2017 and 2018, a trigger requiring both
an electron and large \HT was also used.
The selection criteria are detailed in Table~\ref{tab:sltriggers}.
In the DL channel, events were selected online by the dilepton
($\Pe\Pe$, $\PGm\PGm$, $\Pe\PGm$) triggers detailed in
Table~\ref{tab:dltriggers}, complemented by the single-lepton triggers
from Table~\ref{tab:sltriggers} to maximise the selection efficiency.
Lepton identification and isolation requirements are imposed for some
of the single-lepton and dilepton triggers, and some of the dilepton
triggers also impose an impact parameter requirement~\cite{CMS:2021yvr,CMS:2020uim}.
The trigger selection efficiencies for signal events that pass the
offline baseline selection criteria (see Section~\ref{sec:selection}) amount 
to 90\% in the FH and the SL channels, and to 98\% in the DL channel.

\begin{table}[!htbp]
  \centering
  \topcaption{
    Trigger selection criteria in the fully hadronic (FH) channel.
    Multiple criteria, each represented by one row, are used per year and combined with a logical OR.
    In the case of the four-jet trigger, the minimum jet \pt is different for each jet and separated by a slash (/).
  }
  \label{tab:fhtriggers}
  \renewcommand{\arraystretch}{1.2}
   \begin{tabular}{lcccc}
     Year & Jets & \PQb-tagged jets & Jet \pt (\GeVns) & \HT (\GeVns) \\ 
     \hline
     2016 &   $\geq$6 &   $\geq$1 &  $>$40 &  $>$450 \\
          &   $\geq$6 &   $\geq$2 &  $>$30 &  $>$400 \\
          &   $\geq$1 &       \NA & $>$450 &  \NA \\[\cmsTabSkip]
     2017 &   $\geq$6 &   $\geq$1 &  $>$40 &  $>$430 \\
          &   $\geq$6 &   $\geq$2 &  $>$32 &  $>$380 \\
          &   $\geq$4 &   $\geq$3 &  $>$75/60/45/40 & $>$300 \\
          &       \NA &       \NA & \NA & $>$1050 \\[\cmsTabSkip]
     2018 &   $\geq$6 &   $\geq$1 &  $>$36 &  $>$450 \\
          &   $\geq$6 &   $\geq$2 &  $>$32 &  $>$400 \\
          &   $\geq$4 &   $\geq$3 &  $>$75/60/45/40 & $>$330 \\
          &       \NA &       \NA &    \NA & $>$1050 \\
  \end{tabular}
  \renewcommand{\arraystretch}{1.0}
\end{table} 

\begin{table}[!htbp]
  \centering
  \topcaption{
    Trigger selection criteria in the single-lepton (SL) channel.
    Multiple criteria per lepton flavour, each represented by one row, are used per year and combined with a logical OR.
  }
  \label{tab:sltriggers}
  \renewcommand{\arraystretch}{1.2}
   \begin{tabular}{lcccc}
     Year & Flavour & Lepton \pt (\GeVns) & \HT (\GeVns) \\ 
     \hline
     2016 & $\Pe$  & $>$27 &  \NA \\
          & $\PGm$ & $>$24 &  \NA \\[\cmsTabSkip]
     2017 & $\Pe$  & $>$32 &  \NA \\
          & $\Pe$  & $>$28 &  $>$150 \\
          & $\PGm$ & $>$27 &  \NA \\[\cmsTabSkip]
     2018 & $\Pe$  & $>$32 &  \NA \\
          & $\Pe$  & $>$28 &  $>$150 \\
          & $\PGm$ & $>$24 &  \NA \\
  \end{tabular}
  \renewcommand{\arraystretch}{1.0}
\end{table} 

\begin{table}[!htbp]
  \centering
  \topcaption{
    Trigger selection criteria in the dilepton (DL) channel.
    Multiple criteria per lepton flavour, each represented by one row, are used per year and combined with a logical OR.
  }
  \label{tab:dltriggers}
  \renewcommand{\arraystretch}{1.2}
   \begin{tabular}{lccc}
     Year & Flavour & Leading/subleading lepton \pt (\GeVns) & $m_{\PGm\PGm}$ (\GeVns) \\ 
     \hline
     2016 & $\Pe\Pe$   & $>$23/12                  &  \NA \\
          & $\Pe\PGm$  & $>$23 ($\PGm$)/12 ($\Pe$) &  \NA \\
          & $\Pe\PGm$  & $>$23 ($\Pe$)/8 ($\PGm$)  &  \NA \\
          & $\PGm\PGm$ & $>$17/8                   &  \NA \\[\cmsTabSkip]       
     2017 & $\Pe\Pe$   & $>$23/12                  &  \NA \\
          & $\Pe\PGm$  & $>$23 ($\PGm$)/12 ($\Pe$) &  \NA \\
          & $\Pe\PGm$  & $>$23 ($\Pe$)/12 ($\PGm$) &  \NA \\
          & $\Pe\PGm$  & $>$23 ($\Pe$)/8 ($\PGm$)  &  \NA \\
          & $\PGm\PGm$ & $>$17/8                   & $>$3.8 \\[\cmsTabSkip]
     2018 & $\Pe\Pe$   & $>$23/12                  &  \NA \\
          & $\Pe\PGm$  & $>$23 ($\PGm$)/12 ($\Pe$) &  \NA \\
          & $\Pe\PGm$  & $>$23 ($\Pe$)/12 ($\PGm$) &  \NA \\
          & $\Pe\PGm$  & $>$23 ($\Pe$)/8 ($\PGm$)  &  \NA \\
          & $\PGm\PGm$ & $>$17/8                   & $>$3.8 \\  
  \end{tabular}
  \renewcommand{\arraystretch}{1.0}
\end{table} 

The trigger selections are also applied to the simulated events
described in Section~\ref{sec:mcsamples}, and residual differences
between the efficiencies in data and simulation are corrected.
The size of the correction factors is approximately 0.95 in the SL channel and 0.98 in the DL channel.
In the FH channel, the factors vary from approximately 0.99 in 2016 to 0.9 in 2017 and 0.95 in 2018, owing to the different trigger conditions and thresholds used in each data-taking period.
The efficiencies of the triggers in the FH channel and of the
single-electron triggers were measured in datasets collected using
single-muon triggers that present a negligible correlation with the
triggers used in this analysis.
The single-muon trigger efficiencies were measured using the
tag-and-probe technique in a control region enriched in
$\PZ\to\PGmp\PGmm$ events.
The trigger efficiency in the DL channel was measured in a dataset
collected using triggers requiring the presence of missing transverse
momentum.

During the 2016 and 2017 data-taking, a gradual shift in the timing of the inputs of the ECAL L1 trigger in the region at $\abs{\eta} > 2$, referred to as ``L1 prefiring'', caused a specific trigger inefficiency~\cite{CMS:2020cmk}.
For events containing an electron (a jet) with $\pt\gtrsim50\GeV$ ($\gtrsim100\GeV$), the efficiency loss is approximately 10--20\% in the region $2.5 < \abs{\eta} < 3$.
Correction factors were computed from data and applied to the efficiency evaluated from simulation.

\section{Simulated samples}
\label{sec:mcsamples}

Several Monte Carlo (MC) event generators, interfaced with a detailed detector simulation based on \GEANTfour~(v.\ 9.4)~\cite{Agostinelli:2002hh}, are used to model signal and background events.
Separate samples corresponding to the 2016, 2017, and 2018 data-taking conditions were produced in order to match the different LHC and detector conditions and reconstruction efficiencies.
Events are simulated at next-to-leading order (NLO) accuracy of QCD perturbation theory with the \POWHEG~(v.\,2)~\cite{Nason:2004rx,Frixione:2007vw,Alioli:2010xd,Jezo:2015aia,powheg:tth:1} or \MGvATNLO~(v.\ 2.4.2)~\cite{Alwall:2014hca} event generator, or at leading order (LO) accuracy using \PYTHIA~(v.\ 8.230)~\cite{Sjostrand:2014zea}, depending on the process.
The value of the Higgs boson mass is assumed to be 125\GeV, while the top quark mass value is set to 172.5\GeV.
In all simulated samples, the proton structure is described by the parton distribution function (PDF) set {NNPDF3.1}~\cite{Ball:2017nwa}, except for the minor background samples matching the 2016 conditions, where the PDF set {NNPDF3.0}~\cite{Ball:2014uwa} is used.
Parton showering (PS) and hadronisation are simulated with \PYTHIA.
The parameters for the underlying event description correspond to the CP5 tune~\cite{Sirunyan:2019dfx} for all signal and background processes, except the minor background processes in the 2016 samples, where the CUETP8M1~\cite{bib:CUETP8tune} tune is used.

The \ttH signal is simulated at NLO accuracy with \POWHEG and, solely for the coupling interpretation, at LO with \MGvATNLO.
The \tH signal is simulated at LO with \MGvATNLO.
It proceeds through $t$-channel production (\tHq), associated \tW production (\tHW), and $s$-channel production.
The contribution from $s$-channel \tH production is negligible and is not considered in this analysis.
Different flavour schemes are chosen to simulate the \tHq and \tHW processes.
In the five-flavour scheme (5FS), bottom quarks are considered as massless sea quarks of the proton and may appear in the initial state of $\Pp\Pp$ scattering processes.
In the four-flavour scheme (4FS), bottom quarks are considered massive and are produced by gluon splitting at the matrix-element (ME) level~\cite{Maltoni:2012pa}.
The \tHq process is simulated in the 4FS and the \tHW process in the 5FS. This selection is made to ensure that potential interference contributions from \tHW with \ttH production, which are deemed negligible within the phase space considered in this analysis~\cite{Demartin:2016axk}, are excluded from the simulation. In both the \ttH and \tH simulated samples, all relevant decay channels of the Higgs boson are considered.
In the case of the Higgs boson coupling analysis, the kinematic properties of the \tH events vary depending on the probed coupling strength, and the kinematic properties of both the \tH and the \ttH events vary depending on the probed \CP structure.
This is accounted for by event weights, following the approach described in Refs.~\cite{Gainer:2014bta,Mattelaer:2016gcx}.

Major background contributions arise from \ttjets production.
They are modelled with simulated events obtained from two different generators implemented in the \POWHEG framework and interfaced to \PYTHIA for the PS simulation.
The version and configuration of the two generators are summarised in Table~\ref{tab:ttjetssamples}.
The \POWHEG generator is used to simulate the $\Pp\Pp\to\ttbar$ process at NLO accuracy in the 5FS (referred to as \ttsample sample), while the dedicated \POWHEG-\textsc{Box-Res} program presented in Ref.~\cite{Jezo:2018yaf} together with \textsc{OpenLoops}~\cite{Buccioni:2019sur} is used to simulate the $\Pp\Pp\to\ttbb$ process at NLO accuracy in the 4FS (referred to as \ttbbsample sample).
In the \ttbbsample sample, the additional \PQb jets that do not stem from the top quark decays and that are critical for the background description are simulated at the ME level and thus expected to model more accurately the data.
The choice of the renormalisation and factorisation scales (\muR, \muF) for the \ttbbsample sample follows Refs.~\cite{Jezo:2018yaf,Cascioli:2013era,deFlorian:2016spz} and is motivated by the presence of two very different relevant scales related to the \ttbar and \bbbar systems.
Compared to Ref.~\cite{Jezo:2018yaf}, the values of the scales are reduced by a factor 0.5 in order to approximate the effect of missing higher-order corrections and attenuate theoretical uncertainties related to the ME-PS matching. This choice is motivated by the suggested scale options presented in Ref.~\cite{Buccioni:2019plc} that were derived comparing the predictions of \ttbb calculations to those of $\ttbb+1\,\text{jet}$ calculations at NLO in QCD.
The PDF sets have been chosen consistent with the flavour scheme, with values of the strong coupling constant \alpS matching the order of the ME generation in both cases. The same \PYTHIA version and underlying event tune have been used for both samples.
The combination of the two samples and the details of the \ttjets background model are described in Section~\ref{sec:bkgmodel}.
\begin{table}[!htpb]
  \topcaption{
    Generator version and configuration of the \ttsample and \ttbbsample samples.
    The parameters \mi{\PQt} and \mi{\PQb} denote the top quark and bottom quark mass, respectively, \mti{\PQt}, \mti{\PQb}, and \mti{\Pg} the transverse mass of the top quark, the bottom quark, and additional gluons, respectively, and $h_{\text{damp}}$ the parton shower matching scale.
  }
  \label{tab:ttjetssamples}
  \centering
  \renewcommand{\arraystretch}{1.4}
  \begin{tabular}{lcc}
    & \ttsample sample & \ttbbsample sample \\
    \hline
    \POWHEG version & \POWHEG~v2 & \POWHEG-\textsc{Box-Res} \\
    \PYTHIA version & 8.230 & 8.230 \\
    Flavour scheme  & 5 & 4 \\
    PDF set         & {NNPDF3.1} & {NNPDF3.1} \\
    $\mi{\PQt}$    & 172.5\GeV & 172.5\GeV \\
    $\mi{\PQb}$    & 0 & 4.75\GeV \\
    \muR            & $\sqrt{\tfrac{1}{2}\left(\mti{\PQt}^{2} + \mti{\PAQt}^{2}\right)}$ & $\tfrac{1}{2}\sqrt[4]{\mti{\PQt}\,\mti{\PAQt}\,\mti{\PQb}\,\mti{\PAQb}\vphantom{H}}$ \\
    \muF            & \muR           & $\tfrac{1}{4}\left[\mti{\PQt} + \mti{\PAQt} + \mti{\PQb} + \mti{\PAQb} + \mti{\Pg}\right]$ \\
    $h_{\text{damp}}$   & $1.379\,\mi{\PQt}$ & $1.379\,\mi{\PQt}$ \\
    Tune            & CP5 & CP5 \\
  \end{tabular}
  \renewcommand{\arraystretch}{1.0}
\end{table}

Minor backgrounds originate from single top quark production, the production of \PW and $\PZ/\PGg^{*}$ bosons with additional jets (referred to as \Vjets), \ttbar production in association with a \PW, a \PZ boson (\ttV), or a photon (\ttg), and diboson ($\PW\PW$, $\PW\PZ$, and $\PZ\PZ$ referred to as \diboson) processes.
The single top quark processes in the $t$ and \tW channels are simulated with \POWHEG~\cite{powheg:singlet:st,powheg:singlet:st2,powheg:singlet:tW}.
The $s$-channel single top quark processes, as well as the \Vjets, \ttV, and \ttg processes are simulated with \MGvATNLO.
For all of them, except for the $s$-channel single top quark sample, the matching of ME to PS jets is performed using the \textsc{FxFx}~\cite{Frederix:2012ps} prescription.
Diboson production is simulated using the \PYTHIA event generator.
Contributions from rare top quark processes, including 4\PQt, $\PQt\PW\PZ$, and 3\PQt production, are below 1\% of the total background and under 2\% of the signal in the most signal-enriched regions, and are therefore neglected.

In the FH channel, the dominant background originates from QCD multijet production.
Simulated QCD multijet events, generated with \MGvATNLO at LO accuracy using the MLM~\cite{Alwall:2007fs} prescription for the ME-PS matching, are employed solely for additional validation of observables and procedures.
For the actual analysis, its validation and optimisation, the QCD background contribution is estimated from data as described in Section~\ref{sec:strategy}.

For comparison with the observed distributions, the event yields in the simulated samples are normalised to the integrated luminosity of the corresponding data sample, according to their predicted cross sections.
The SM \ttH cross section of $507^{+35}_{-50}\unit{fb}$ is taken from calculations including QCD and electroweak corrections at NLO~\cite{deFlorian:2016spz}.
The SM cross section for \tH production is computed in the 5FS at NLO accuracy in QCD~\cite{deFlorian:2016spz}, which results in $74.3^{+5.6}_{-11.3}\unit{fb}$ for \tHq and $15.2^{+1.2}_{-1.4}\unit{fb}$ for \tHW production.
The \textrm{DR2} scheme~\cite{Demartin:2016axk} is employed in the calculation of \tHW production to remove overlapping contributions between \tHW and \ttH processes.
In this scheme, the squared double resonant term is removed from the squared amplitude of the \tHW process, while the interference term between double- and single-resonant contributions is kept.
Higgs boson branching fractions are obtained from calculations with at least NLO accuracy~\cite{deFlorian:2016spz}.
The calculated \ttH and \tH cross sections and Higgs boson branching fractions are rescaled assuming a Higgs boson mass of 125.38\GeV, corresponding to the latest measurement by the CMS Collaboration~\cite{CMS:2020xrn}.
The \ttbar cross section of 832\unit{pb} corresponds to the next-to-next-to-leading order (NNLO) calculation with resummation to next-to-next-to-leading logarithmic (NNLL) accuracy~\cite{bib:xs1,bib:xs2,bib:xs3,bib:xs4,bib:xs7,Czakon:2013goa,Czakon:2011xx}.
The cross sections of the other background processes are taken at NNLO (\Vjets), approximate NNLO (single top quark \tW-channel~\cite{bib:twchan}), and NLO (single top quark $t$- and $s$-channels~\cite{Aliev:2010zk,Kant:2014oha}, \ttV~\cite{Maltoni:2015ena}, and diboson~\cite{bib:mcfm:diboson}) accuracy.

Effects from additional $\Pp\Pp$ interactions in the same or nearby bunch crossings (pileup) are modelled by adding simulated minimum bias events to all simulated events.
The pileup multiplicity distribution in simulation is reweighted to reflect the luminosity profile of the observed $\Pp\Pp$ collisions~\cite{CMS:2020ebo}.
Further correction factors described in Section~\ref{sec:objects} are applied to the simulation where necessary to improve the description of the data.

\section{Object and event reconstruction}
\label{sec:objects}
Events are reconstructed offline based on a particle-flow (PF) technique~\cite{CMS-PRF-14-001}, which aims to reconstruct and identify each individual particle produced in $\Pp\Pp$ collisions by optimally combining information from the various elements of the CMS detector. 

The primary vertex is taken to be the vertex corresponding to the hardest scattering in the event, evaluated using tracking information alone, as described in Section 9.4.1 of Ref.~\cite{CMS-TDR-15-02}.

The reconstruction of muons relies on a combination of measurements in the tracker and in the muon detectors~\cite{CMS:2018rym}.
The muons are identified based on the quality of the combined track fit and on the number of hits in the different tracking detectors, with an efficiency of about 95\%. Electrons are reconstructed by combining the momentum measurement in the tracker with the energy measurement of the corresponding cluster in the ECAL and with the energy sum of all bremsstrahlung photons, obtained from the ECAL, spatially compatible with originating from the electron track~\cite{CMS:2020uim,CMS-DP-2020-021}.
The electrons are identified using criteria on the cluster shape in the ECAL, the track quality, and the compatibility between the tracker and ECAL measurements, corresponding to an identification efficiency of approximately 80\%, including the isolation requirements described below.
Both muons and electrons are required to lie within the acceptance of the tracker, covering the region up to $\abs{\eta}=2.4$.
Electrons reconstructed in the transition region $1.44 < \abs{\eta} < 1.56 $ between the barrel and the endcap calorimeters are discarded.
Isolation requirements are imposed on the leptons based on the scalar \pt sum of all PF objects in a cone of radius $\Delta R=0.3$ (0.4) around the track direction of the electron (muon), where $\Delta R = \sqrt{\Delta\eta^2+\Delta\phi^2}$.
The isolation is corrected by removing contributions from pileup.
Residual differences between the lepton reconstruction, identification, and isolation efficiencies in data and simulation are corrected, based on efficiency measurements in high-purity data samples of \PZ boson decays~\cite{CMS:2018rym,CMS:2020uim}.

Hadronic jets are reconstructed by clustering PF candidates using the anti-\kt algorithm~\cite{Cacciari:2008gp, Cacciari:2011ma} with a distance parameter of 0.4, omitting charged particles matched to pileup vertices.
The jet energy is corrected for the neutral-hadron contribution expected from pileup interactions~\cite{Cacciari:2008gn}.
Further energy corrections depending on the jet \pt and $\eta$ are applied, which are derived in simulation such that the average measured energy of the jets becomes identical to that of the particle-level jets.
Residual differences between the jet energy scale in data and in simulation are measured using the momentum balance in dijet, $\PGg + \text{jets}$, \Zjets, and multijet events, and appropriate corrections are made.
The jet energy resolution amounts typically to 15--20\% at 30\GeV and 10\% at 100\GeV~\cite{CMS:2016lmd}. 
Jets overlapping with an electron or muon passing the criteria described above within a cone of $\Delta R=0.4$ are discarded. 
Further selection criteria are applied to each jet to remove jets potentially dominated by instrumental effects or reconstruction failures as well as jets arising from pileup interactions~\cite{CMS:2020ebo,CMS-PAS-JME-16-003}.
Only jets within the tracker acceptance of \mbox{$\abs{\eta}<2.4$} and with \mbox{$\pt>30\GeV$} are considered in the analysis.
In addition, jets with \mbox{$\pt>40\GeV$} and \mbox{$2.4<\abs{\eta}<4.7$} (forward jets) are selected in the SL channel for the event reconstruction under the hypothesis of \tHq production, which is characterised by the presence of jets with large $\abs{\eta}$ values. 

The \PQb jets are identified using \PQb tagging techniques~\cite{BTV-16-002}.
This analysis benefits from the \textsc{DeepJet}~\cite{Bols:2020bkb,CMS-DP-2021-004} tagger, featuring an increase of the \PQb tagging efficiency by 5--10\% with respect to the tagger used in the previous analyses~\cite{Sirunyan:2018mvw,Sirunyan:2018ygk,Sirunyan:2018lzm} for the same mistag rate.
Jets are considered as \PQb tagged if they fulfil a requirement on the \textsc{DeepJet} score that corresponds to a \PQb tagging efficiency of about 75--80\% and 1.5--2\% (15--17\%) mistag rate for light-flavour (\PQc quark) jets.
Specifically for the estimation of the QCD multijet background, a second requirement on the \textsc{DeepJet} output is utilised, which corresponds to approximately 90\% \PQb\ tagging efficiency at a 10\% (43--47\%) light-flavour (\PQc quark) mistagging rate, referred to as ``loose'' \PQb\ tag.
The \textsc{DeepJet} discriminant, and thus \PQb tagging efficiencies and mistag rates, in the simulation is corrected to match the one measured in data~\cite{BTV-16-002}.
The \textsc{DeepJet} discriminant value is further used in the analysis as one of the input variables for the ANN discriminants described in Section~\ref{sec:strategy}.

The missing transverse momentum vector \ptvecmiss is computed as the negative vector sum of the transverse momenta of all the PF candidates in an event, and its magnitude is denoted as \ptmiss~\cite{Sirunyan:2019kia}.
The \ptvecmiss is recomputed to account for corrections to the energy scale of the reconstructed jets in the event.

\section{Event selection}
\label{sec:selection}
The first step of the event classification, referred to as baseline selection, defines the events assigned to the DL, SL, and FH channels, and is summarised in Table~\ref{tab:baselinecriteria} and described in the following.
\begin{table}[htbp]
  \centering
  \topcaption{
    Baseline selection criteria in the fully hadronic (FH), single-lepton (SL), and dilepton (DL) channels based on the observables defined in the text.
    Leptons and jets are ranked in \pt.
    The $\ast$ indicates that the requirement is only applied to the same-flavour DL channels.
    Where the criteria differ per year of data taking, they are quoted as three values, corresponding to 2016/2017/2018, respectively.
  }
  \label{tab:baselinecriteria}
  \renewcommand{\arraystretch}{1.1}
  \begin{tabular}{lccc}
                                           & FH channel    & SL channel             & DL channel \\
    \hline
    Number of leptons                      & 0             & 1                      & 2   \\
    Sign and flavour of leptons            & \NA           & $\Pepm$, $\PGmpm$ & $\Pep\Pem$, $\PGmpm\Pemp$, $\PGmp\PGmm$ \\
    \pt of leading electron (\GeVns)       & \NA           & $>$29/30/30            & $>$25              \\
    \pt of leading muon (\GeVns)           & \NA           & $>$26/29/26            & $>$25                 \\
    \pt of additional leptons (\GeVns)     & $<$15         & $<$15                  & $>$15            \\
    $\abs{\eta}$ of leptons                & $<$2.4        & $<$2.4                 & $<$2.4               \\
    \mll (\GeVns)                          & \NA           & \NA                    & $>$20 (and $<$76 or $>$106)$^{\ast}$    \\[10pt]
    Number of jets                         & $\geq$7       & $\geq$5                & $\geq$3                 \\
    \pt of jets (\GeVns)                   & $>$30         & $>$30                  & $>$30               \\
    \pt of $6^{\text{th}}$ jet (\GeVns)      & $>$40         & \NA                    & \NA              \\	
    $\abs{\eta}$ of jets                   & $<$2.4        & $<$2.4                 & $<$2.4              \\
    Number of \PQb-tagged jets             & $\geq$2       & $\geq$4                & $\geq$3                \\
    \mqq (\GeVns)                          & $>$30 and $<$250 & \NA                 & \NA              \\[10pt]
    \HT (\GeVns)                           & $>$500        & \NA                    & \NA               \\
    \ptmiss (\GeVns)                       & \NA           & $>$20                  & $>$40$^{\ast}$                \\
  \end{tabular}
  \renewcommand{\arraystretch}{1.0}
\end{table}

Events with exactly two opposite-sign leptons (electrons or muons) with \pt of at least 25 (15)\GeV for the (sub)leading \pt lepton and at least three \PQb-tagged jets are assigned to the DL channel.
The offline and online flavour content of the event must be consistent, meaning, \eg that $\PGm\PGm$ events are required to be recorded by single-muon or dimuon triggers.
The invariant mass of the selected lepton pair, \mll, is required to be larger than 20\GeV to suppress events from heavy-flavour resonance decays and low-mass Drell--Yan processes.
In order to reject $\PZ+\text{jets}$ events, same-flavour events with $76<\mll<106\GeV$ are discarded. 

The baseline selection for the SL channel requires the presence of exactly one electron or muon, consistent with the trigger that accepted the event.
The lepton \pt requirement is chosen based on the trigger thresholds, and therefore depends on the data-taking year.
Muons must have $\pt>26$ (29)\GeV for data recorded in 2016 and 2018 (2017), while the \pt thresholds for electrons are 29 (30)\GeV in 2016 (2017 and 2018).
Furthermore, no additional leptons with $\pt>15\GeV$ must be present in the event.
Only events with at least five jets, of which at least four are \PQb tagged, are considered. 
A minimum \ptmiss of 40 and 20\GeV is required in the DL channels with same-flavour leptons and in the SL channel, respectively, to further suppress minor background contributions from \Vjets and QCD multijet production.

All remaining events with no leptons passing the selection criteria in the DL and SL channels and with at least seven jets, of which at least two are \PQb tagged, are assigned to the FH channel.
In order to match the online trigger selection, events with $\HT<500\GeV$ or less than six jets with $\pt>40\GeV$ are discarded.
The invariant mass \mqq of the pair of non-\PQb-tagged jets for which \mqq is closest to the nominal \PW boson mass is required to lie between 30 and 250\GeV.
To estimate the dominant QCD multijet background from control samples in data in this channel, events are further split into a signal region (SR) and several control and validation regions (CRs and VRs, respectively) according to the value of \mqq, as described in Section~\ref{sec:strategy}.

The aforementioned criteria have been optimised for a high \ttH signal selection efficiency, while still retaining a sufficiently high \tH signal acceptance.
The minimum number of jets required in each channel is lower than the one expected for \ttH production at LO in order to recover events with jets falling outside of the object acceptance and to increase the efficiency for \tH signal events.

\begin{figure}[htbp]
  \centering
    \includegraphics[width=0.48\textwidth]{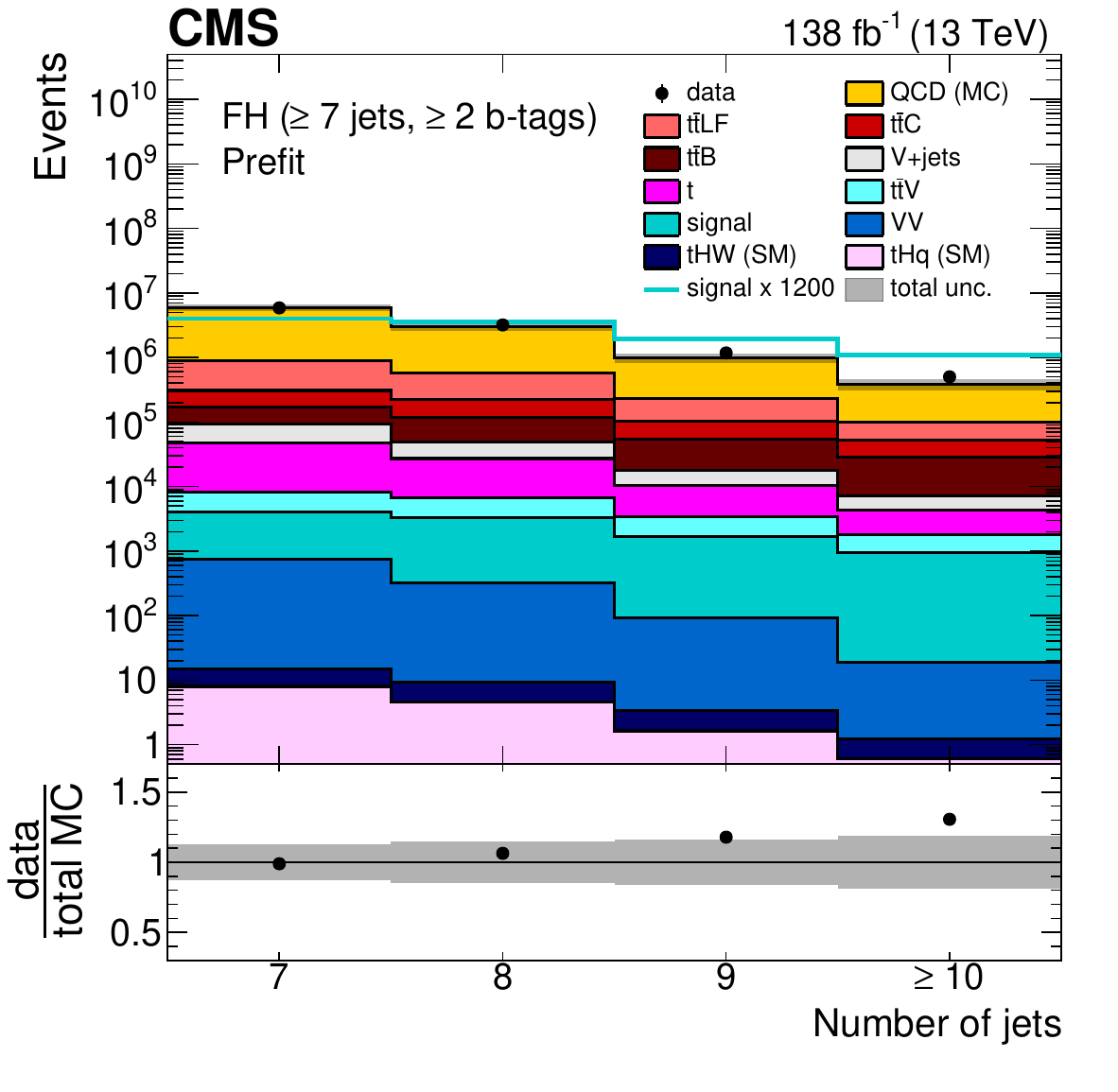} 
    \includegraphics[width=0.48\textwidth]{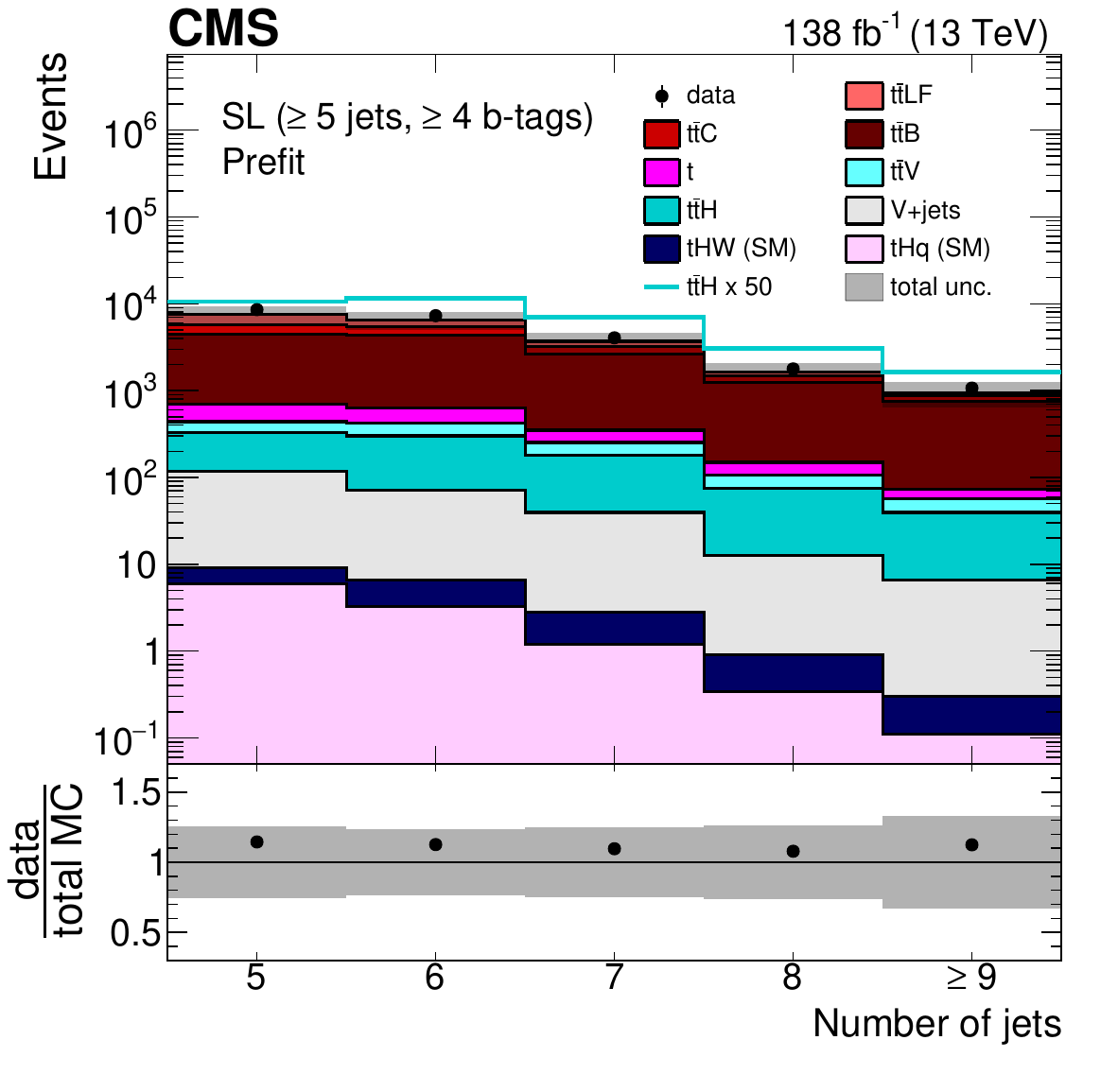} 
    \includegraphics[width=0.48\textwidth]{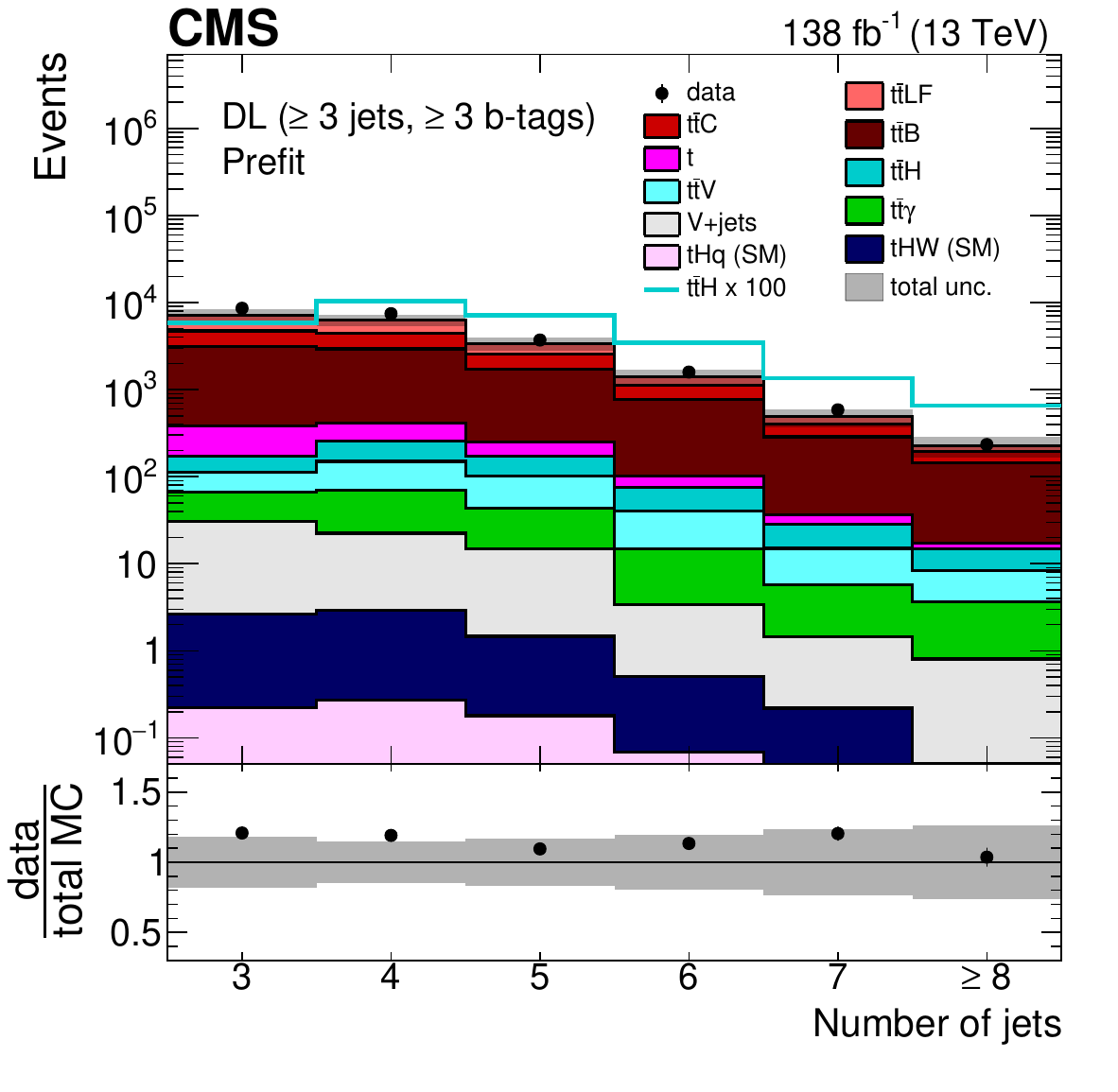}
  \caption{
    Jet multiplicity distribution in the FH (upper left), SL (upper right), and DL (lower) channels, after the baseline selection and prior to the fit to the data.  
    Here, the QCD multijet background prediction is taken from simulation. The different \ttjets background contributions (\ttLF, \ttB, and \ttC) are discussed in Section~\ref{sec:bkgmodel}.
    The expected \ttH signal contribution, scaled as indicated in the legend for better visibility, is also overlayed (line).
    The uncertainty band represents the total (statistical and systematic) uncertainty.
    The last bin in each distribution includes the overflow events.
  }
  \label{fig:selection:njets}
\end{figure}

Figure~\ref{fig:selection:njets} shows the jet multiplicity distribution for the FH, SL, and DL channels after the baseline selection and prior to the final fit to data. The expected signal and background contributions are estimated using the simulations for the various processes introduced in Section~\ref{sec:mcsamples}. Good agreement between data and simulation is observed in the SL and DL channels within the uncertainties. In the FH channel, where the dominant background contribution stems from QCD multijet production, a significant trend in data is observed towards higher jet multiplicity; the poor modelling of the QCD multijet background by the simulation is expected and the reason it is estimated directly from data.

To maximise the analysis sensitivity, two likelihood ratios are computed in each event and provided as input to the final ANN discriminants described in Section~\ref{sec:strategy}.
First, a \PQb tagging likelihood ratio (BLR) is computed that is designed to discriminate between events compatible with the presence of four and two \PQb jets.
The likelihoods are evaluated under either the hypothesis that four jets or that two jets in the event originate from \PQb quarks, based on the expected \PQb tagging discriminant probability densities from simulation, summing over all possible permutations of quark-jet associations.
The BLR is defined as the ratio of the four-\PQb-jets likelihood to the sum of the four- and the two-\PQb-jets likelihoods.
Second, a likelihood ratio comparing the \ttH, \Htobb signal and the \ttbb background hypotheses is computed based on an ME method (MEM)~\cite{Kondo:1988yd} that utilises the full kinematic properties of the event, as documented in Refs.~\cite{Sirunyan:2018ygk,CMS:2015enw}.

Furthermore, in the SL channel boosted decision trees (BDTs), referred to as reconstruction BDTs, are used to assign the jets in an event to partons under the hypothesis of the event being either a \ttH, \tHq, \tHW, or \ttbar event.
The BDTs take as input observables such as the invariant mass and \pt
of the reconstructed particles, the \PQb tagging discriminant value of jets, as well as angular distances between reconstructed particles.
They are trained separately for each hypothesis, such that the most probable jet assignment under the given hypothesis yields the largest BDT output value.
In case of the \tHq hypothesis, the forward jets are also taken into account in the event reconstruction as candidates for the light-flavour quarks, which typically have large $\abs{\eta}$ values in \tHq events.
The procedure achieves reconstruction efficiencies of approximately 43\%, 60\%, 52\%, and 66\% for \ttH, \tHq, \tHW, or \ttbar events, respectively.
The lower efficiency for \ttH events compared to \ttbar events is caused by the greater combinatorial challenge due to the presence of two additional \PQb jets for \ttH events.
The output values of the reconstruction BDTs as well as several observables that are obtained based on the chosen jet assignments are among the input variables to the ANNs used for the signal extraction in this channel.

The further event categorisation within each channel and the strategy adopted to extract the signal are described in Section~\ref{sec:strategy}.

\section{The \texorpdfstring{\ttjets background model}{tt+jets background model}}
\label{sec:bkgmodel}
After the baseline selection, contributions from \ttjets production constitute more than 95\% of the total expected background events in the leptonic channels.
Particularly critical among these are \ttbar events with additional \PQb jet production, since their final state is identical to that of the \ttH signal events, the cross section is approximately 10--20 times larger than that of the signal in the phase space considered, and the events are difficult to model theoretically due to the multiparton final state with two very different mass scales.

Several dedicated measurements~\cite{CMS:2023xjh,Sirunyan:2020kga,Sirunyan:2019jud}, as well as the observations in previous versions of this analysis~\cite{Sirunyan:2018mvw} and the observations in the control regions studied in this analysis (see Section~\ref{sec:strategy}), show that the $\ttbar+\PQb$ jets contribution present in data is larger than predicted by the simulation, by a factor of typically 1.2--1.4, depending on the MC event generator.
In Ref.~\cite{CMS:2023xjh}, also normalised differential \ttbb cross-section measurements in the one-lepton channel are presented for various observables.
For events with $\geq$6 jets and $\geq$4 \PQb-tagged jets, the shapes of the distributions observed in data in Ref.~\cite{CMS:2023xjh} mostly agree to those predicted with the \ttbbsample sample described in Section~\ref{sec:mcsamples} within the uncertainties.
Larger discrepancies, in particular in the jet and \PQb-tagged jets multiplicity distributions, are observed in Ref.~\cite{CMS:2023xjh} for events with 3 \PQb-tagged jets, which are not considered in the analysis presented in this paper.
The results in Ref.~\cite{CMS:2023xjh} were, however, obtained in a slightly different phase space, selecting for example jets with $\pt>25\GeV$, and unfolded to the particle level, such that a direct comparison with the distributions presented here is not possible.

The \ttjets background is modelled using a combination of the \ttsample and \ttbbsample samples described in Section~\ref{sec:mcsamples}.
The simulated \ttjets events are separated into three mutually exclusive processes, based on the flavour of the additional jets at the particle level that do not originate from the top quark decays. 
The jet flavour is defined using the ghost-matching procedure described in Ref.~\cite{Collaboration:2267573}, and the particle level jets need to fulfil the acceptance requirements of $\pt>20\GeV$ and $\abs{\eta}<2.4$.
The three processes are defined as:
\ttB, comprising events with at least one additional jet generated within the acceptance that contains one or more \PQb hadrons;
\ttC, for which events have at least one additional jet containing \PQc hadrons within the acceptance and no additional jets containing \PQb hadrons; or else
$\ttbar+\text{light-flavour jets}$ (\ttLF), which corresponds to events that do not belong to any of the above processes.
The \ttB events are further separated into two mutually exclusive processes.
The subset of events with exactly two additional \PQb hadrons that are close enough in direction to be inside a  single jet is denoted as \tttwob.
This separation is important because the processes are subject  to different systematic uncertainties arising from the modelling of collinear gluon splitting.
The remaining subset of events is denoted as \ttborbb.

The \ttB background events, which represent the most critical background component, are taken from the \ttbbsample sample.
Here, the additional \PQb jets are modelled by the ME calculation and are subject, in particular, to the renormalisation and factorisation scale uncertainties.
The \ttC and \ttLF events, which do not include additional \PQb jets within the acceptance but can still pass the event selection (\eg due to mistagging), are taken from the \ttsample sample.
Prior to the fit to data, the fractional contributions of the \ttB, \ttC, and \ttLF events are chosen according to the \ttsample sample in the inclusive phase space and the total yield of the \ttjets events is scaled to the NNLO+NNLL \ttbar cross section of 832\unit{pb}.
The final yields of the \ttB and \ttC contributions are free parameters in the fit to data, providing flexibility to adjust the predicted \ttB yield to the one observed in data through dedicated CRs.
The impact of potential mismodelling of the \ttB background on the extracted signal was studied using pseudo-experiments and found to be well covered by the systematic uncertainties, as described in Section~\ref{sec:systematics}.

\begin{figure}[!htpb]
  \centering
    \includegraphics[width=0.4\textwidth]{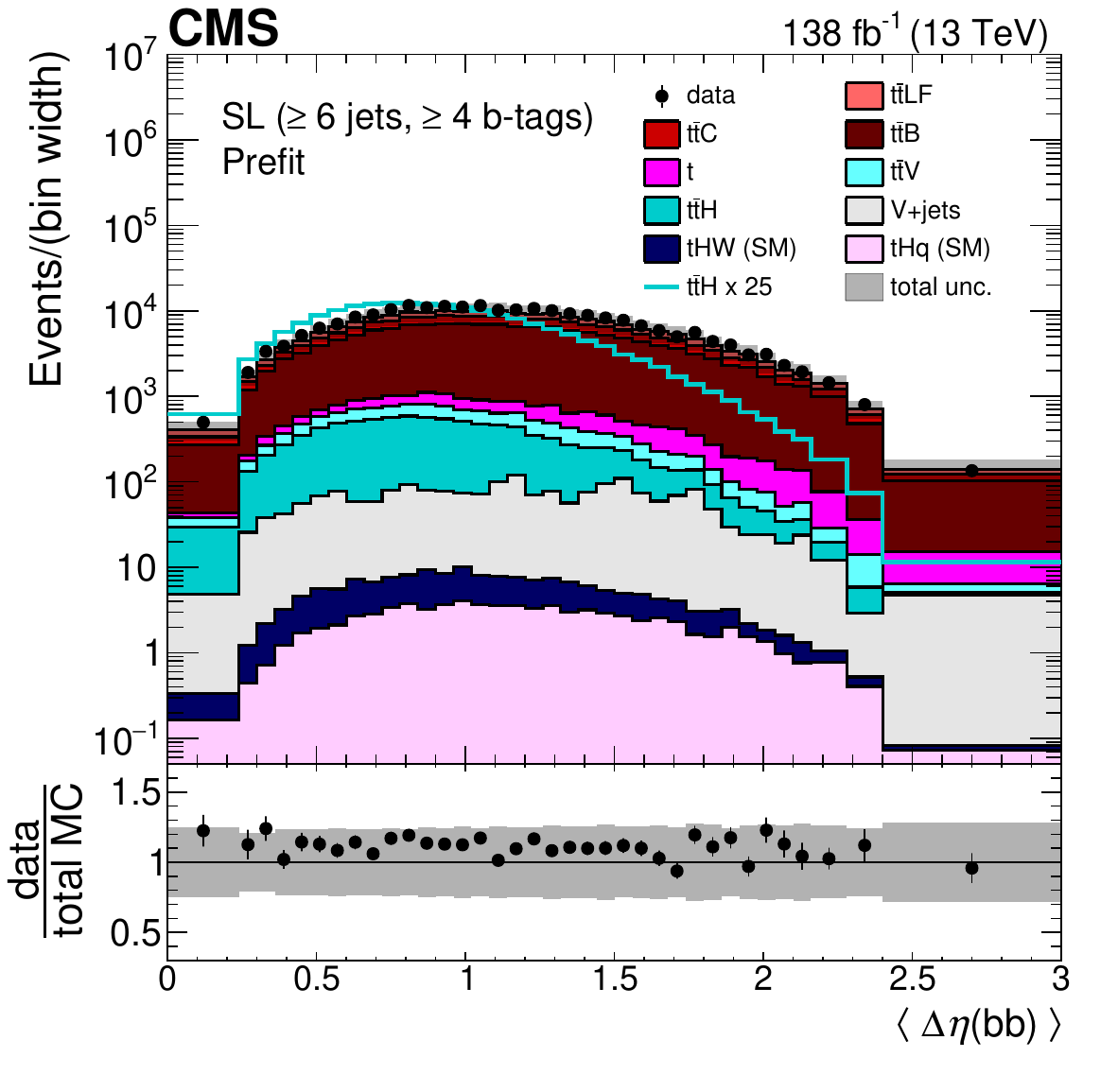}
    \includegraphics[width=0.4\textwidth]{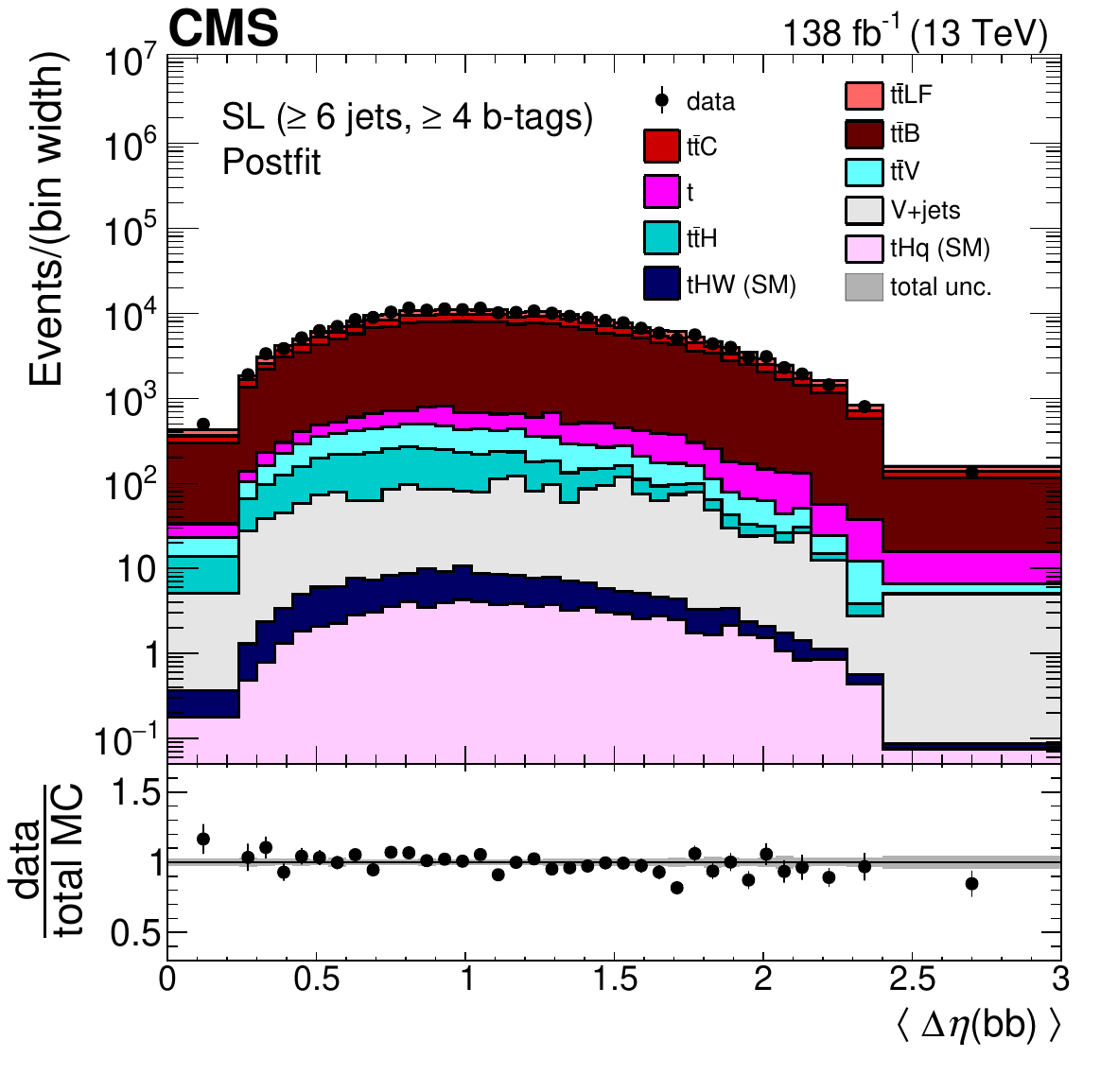} 
    \includegraphics[width=0.4\textwidth]{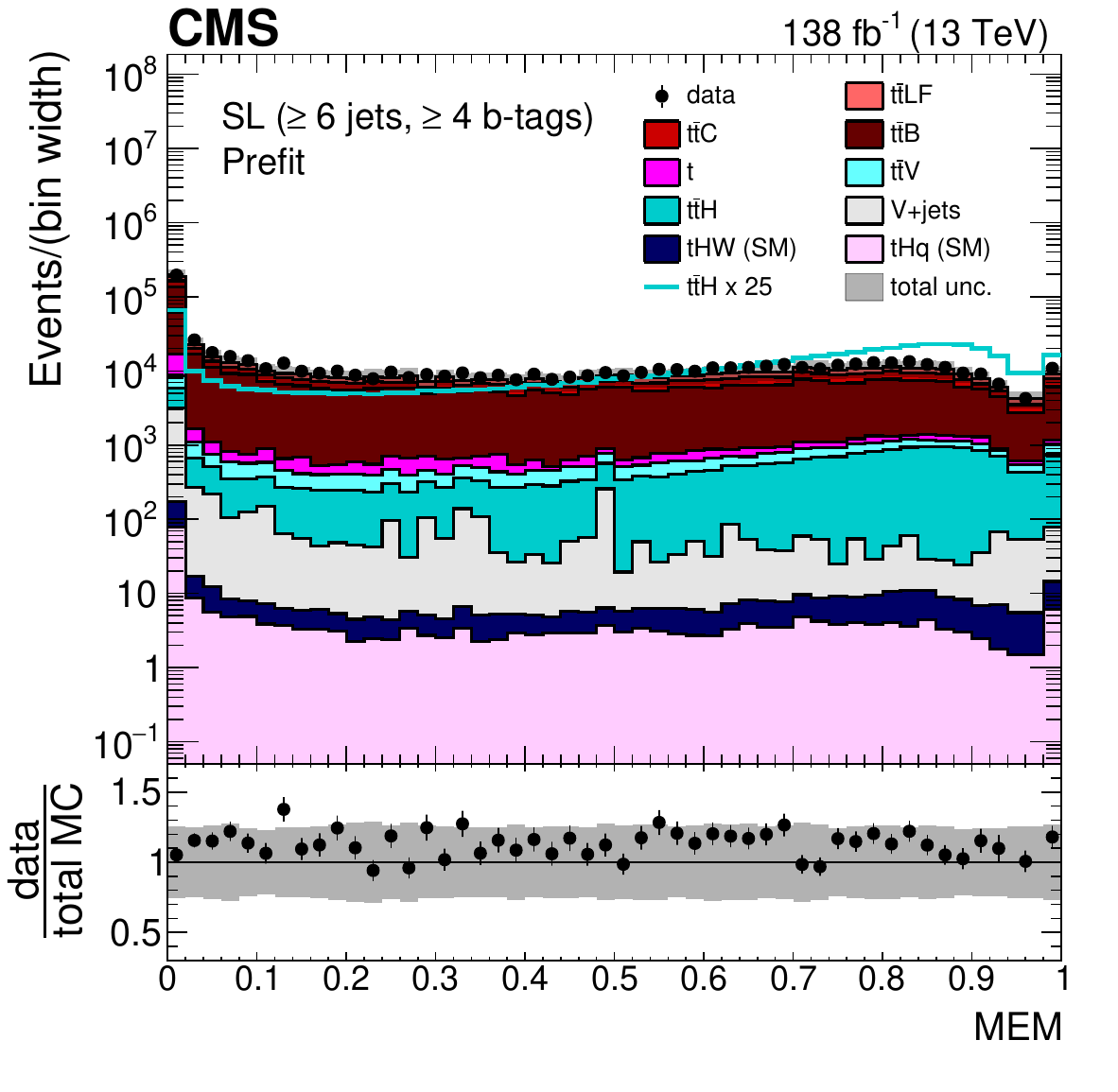} 
    \includegraphics[width=0.4\textwidth]{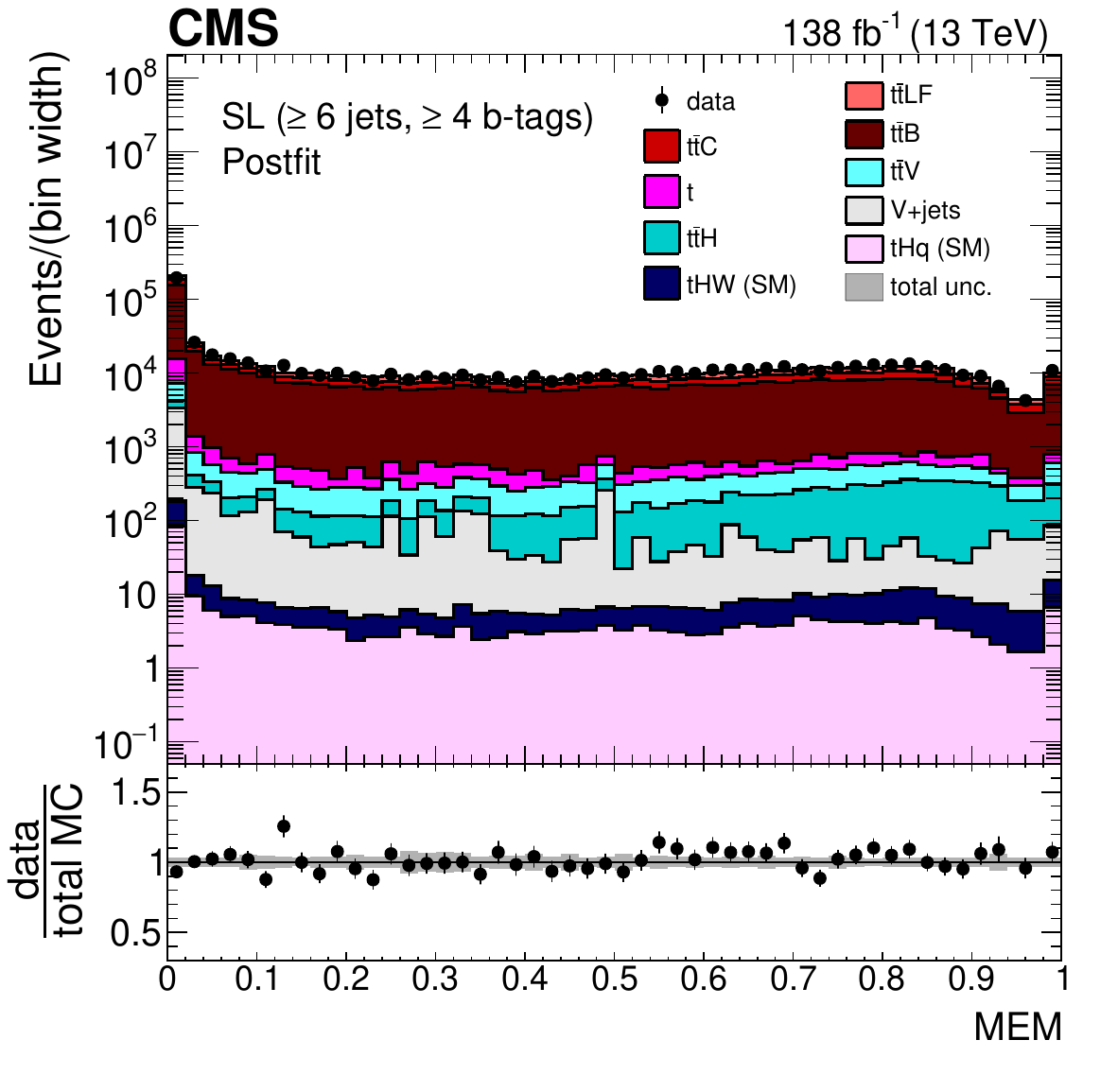} 
    \includegraphics[width=0.4\textwidth]{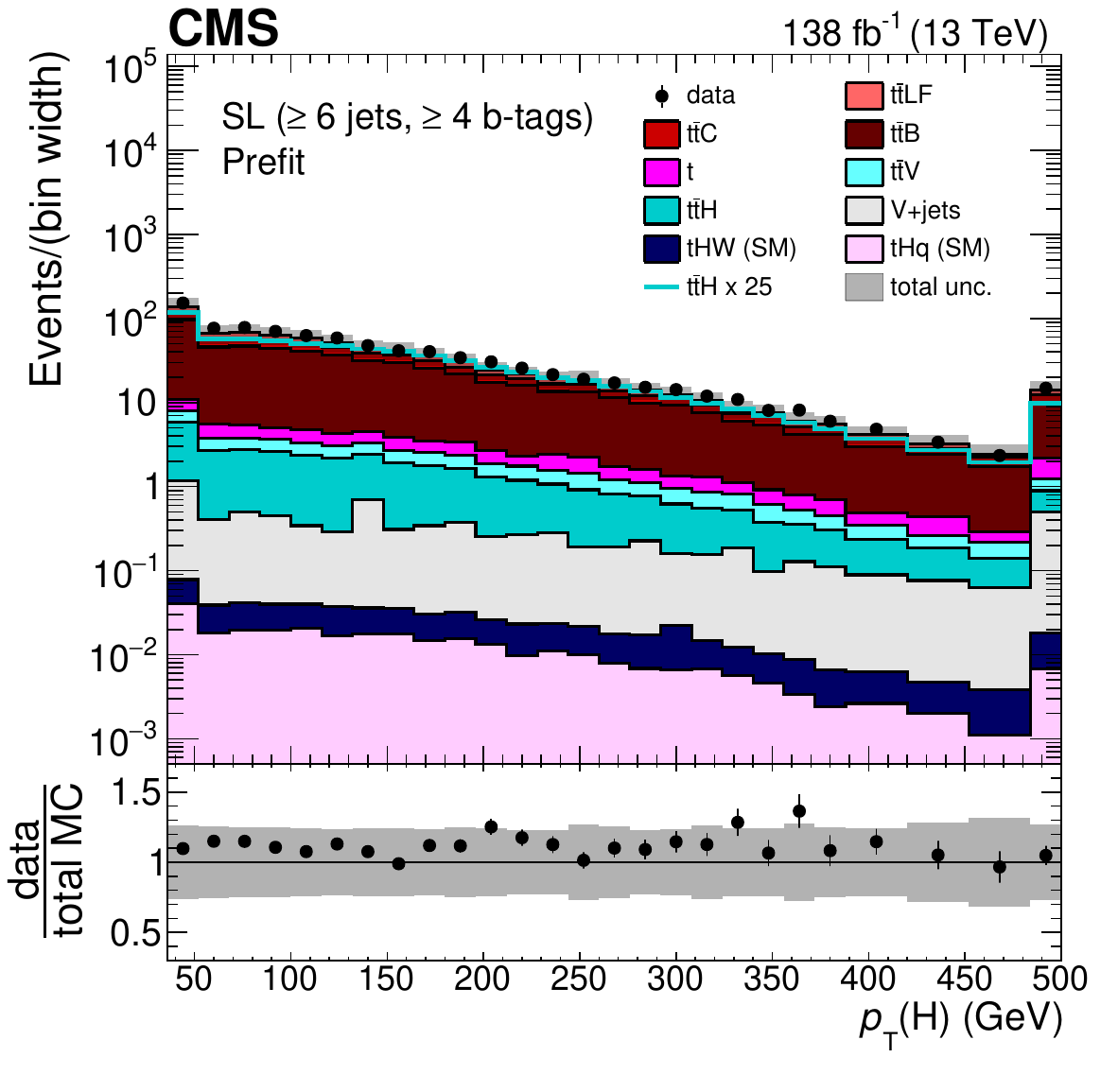} 
    \includegraphics[width=0.4\textwidth]{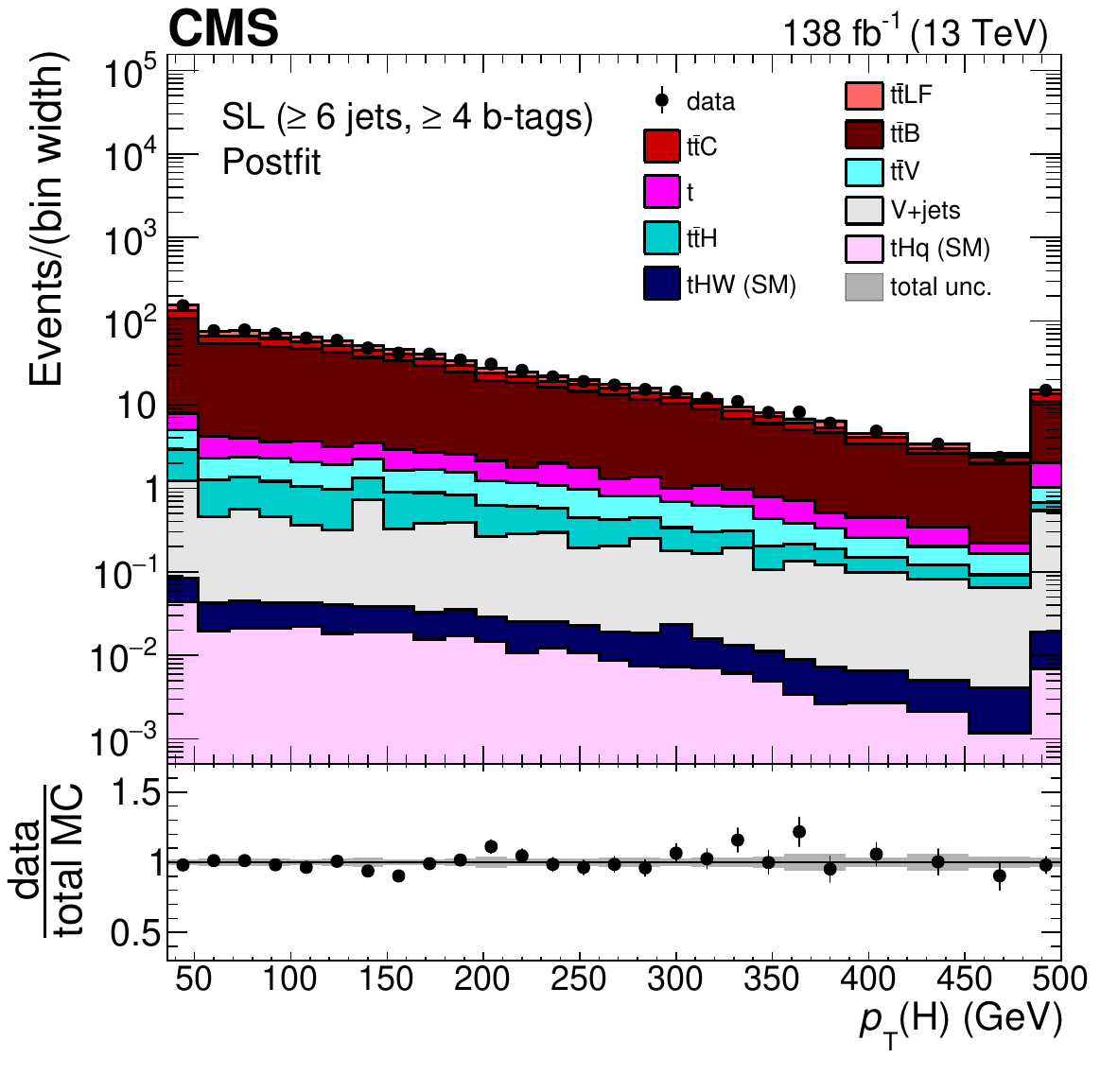} 
  \caption{
    Average $\Delta\eta$ between any two \PQb-tagged jets (upper), MEM discriminant output (middle), and \pt of the Higgs boson candidate identified with the reconstruction BDT for the \ttH hypothesis (lower) for events passing the baseline selection requirements and additionally $\geq$6 jets in the SL channel prefit (left) and with the postfit background model (right) obtained from the fit to data described in Section~\ref{sec:results}.
    In the prefit case, the \ttH signal contribution, scaled by a factor 25 for better visibility, is also overlayed (line).
    The uncertainty band represents the total (statistical and systematic) uncertainty.
    Where applicable, the last bin in each distribution includes the overflow events.
  }
  \label{fig:background:controlvariables:sl}
\end{figure}

\begin{figure}[!htpb]
  \centering
    \includegraphics[width=0.4\textwidth]{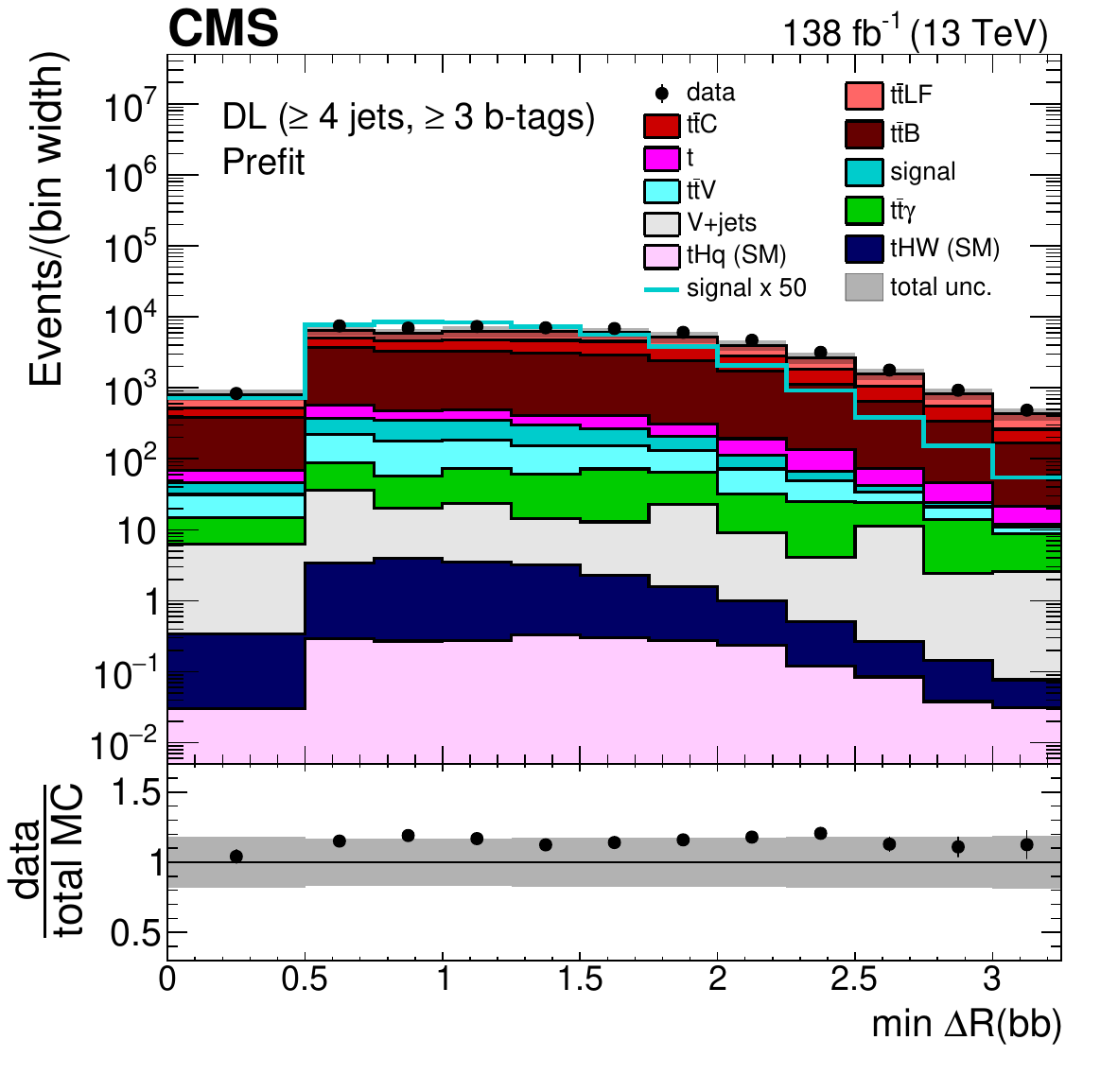}
    \includegraphics[width=0.4\textwidth]{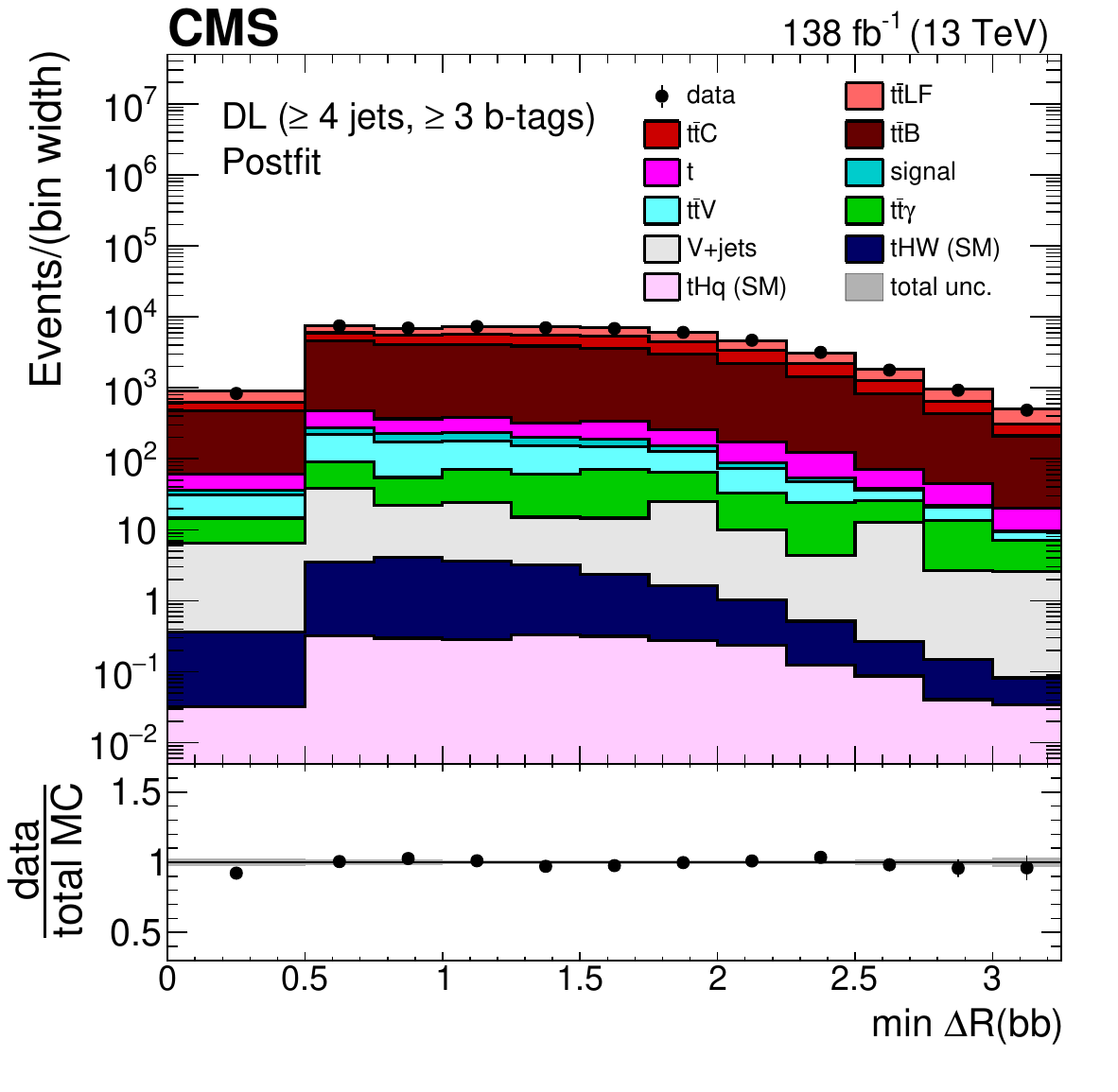}     
    \includegraphics[width=0.4\textwidth]{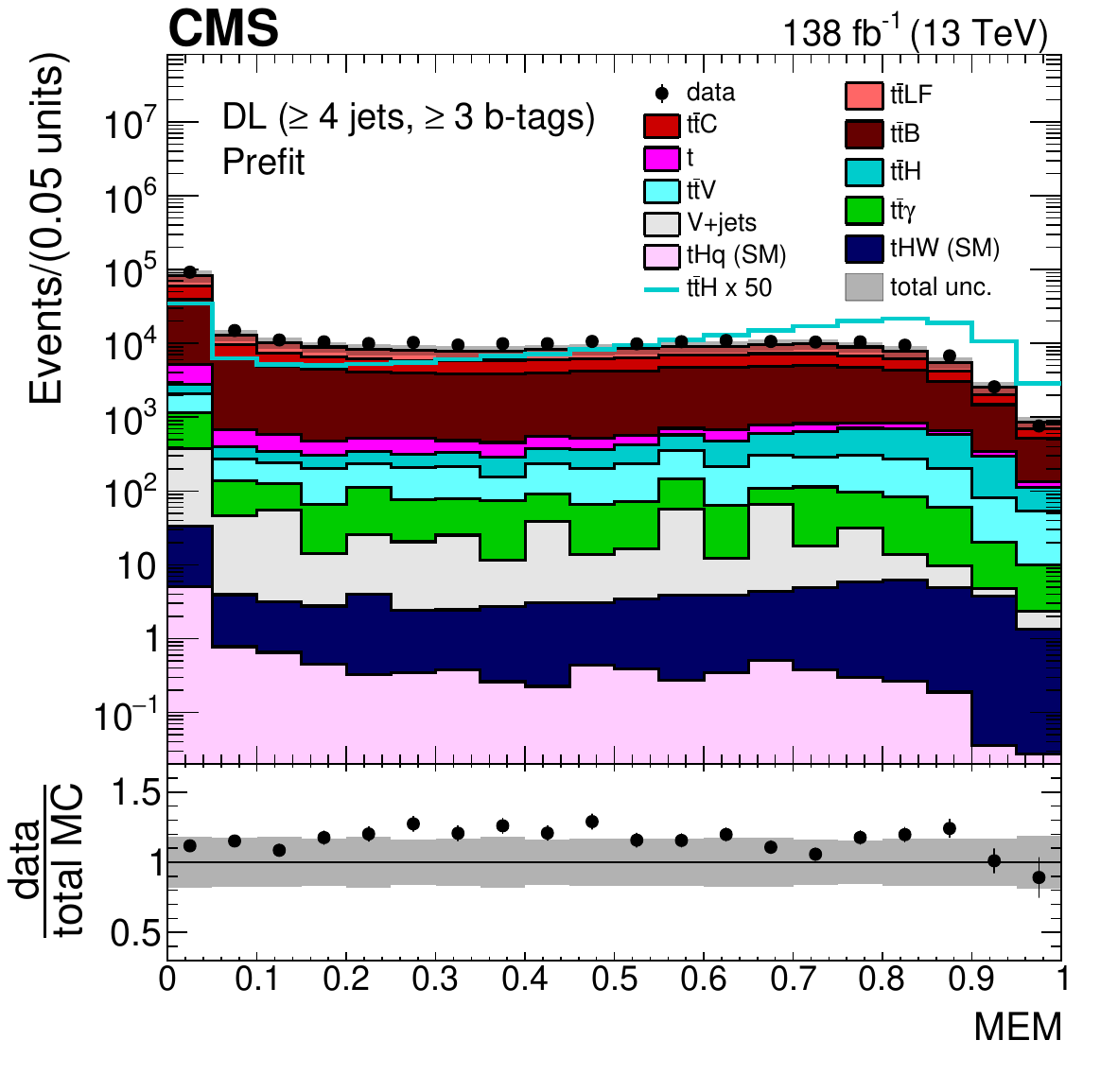}  
    \includegraphics[width=0.4\textwidth]{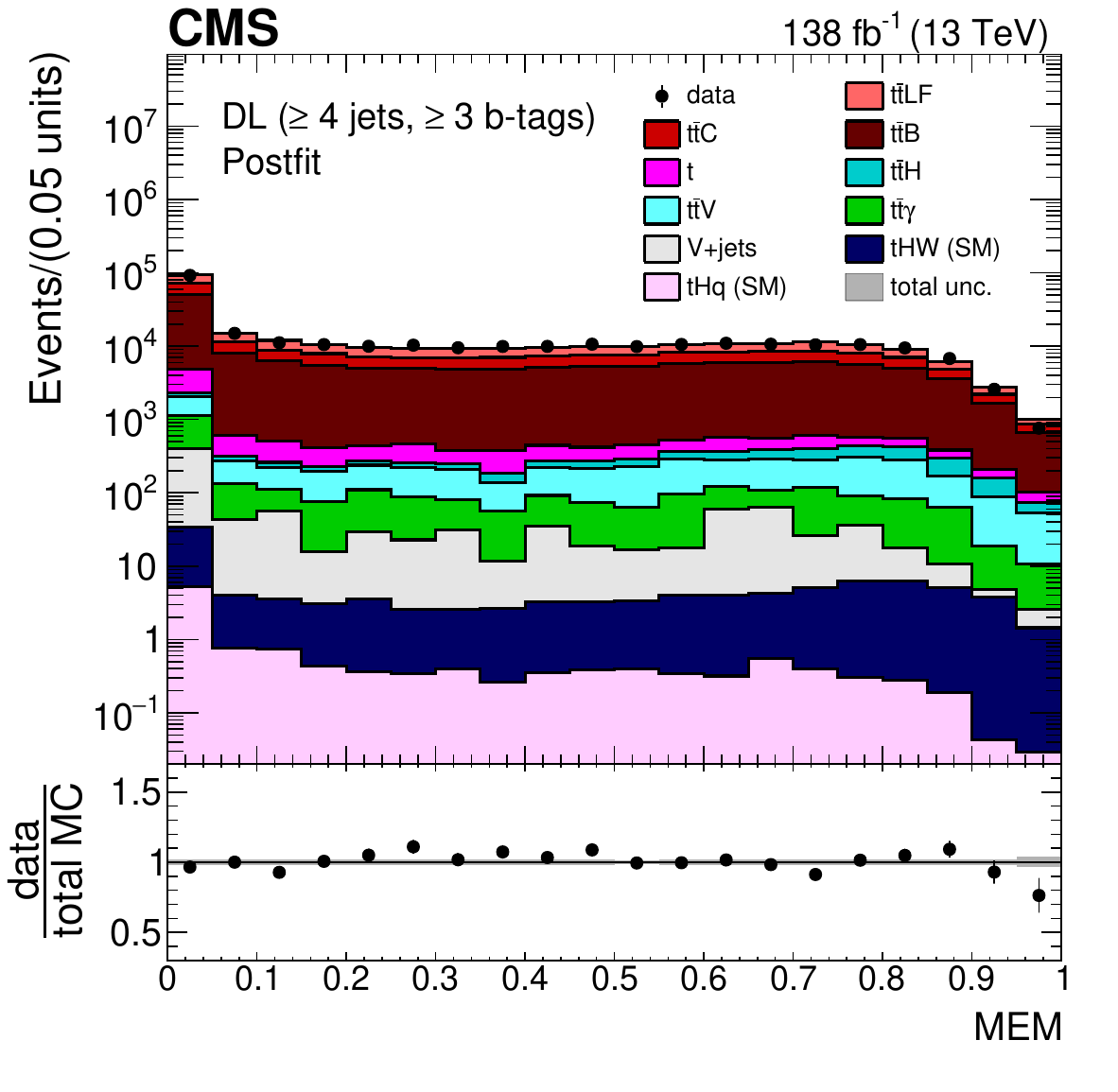} 
    \includegraphics[width=0.4\textwidth]{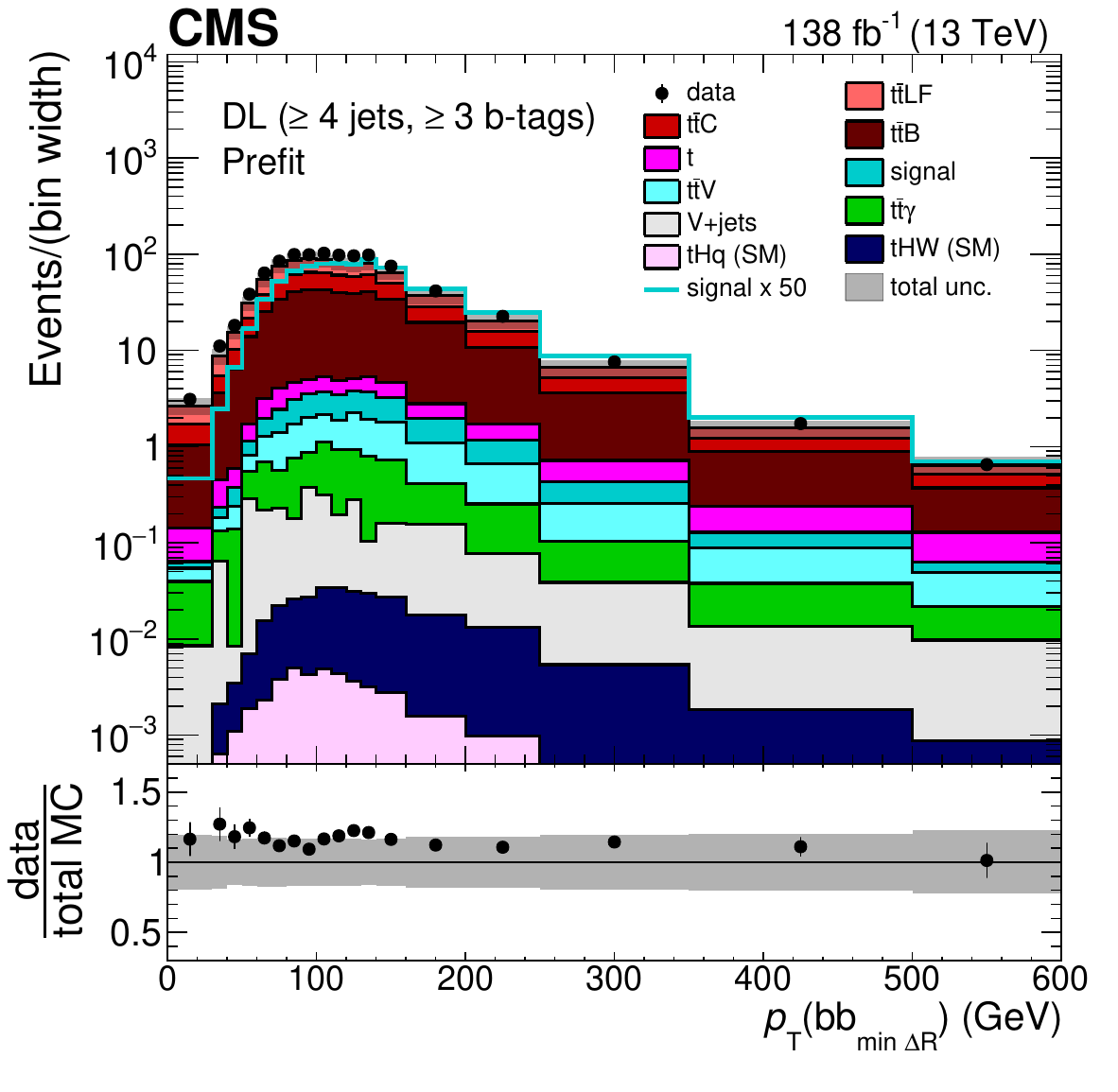}  
    \includegraphics[width=0.4\textwidth]{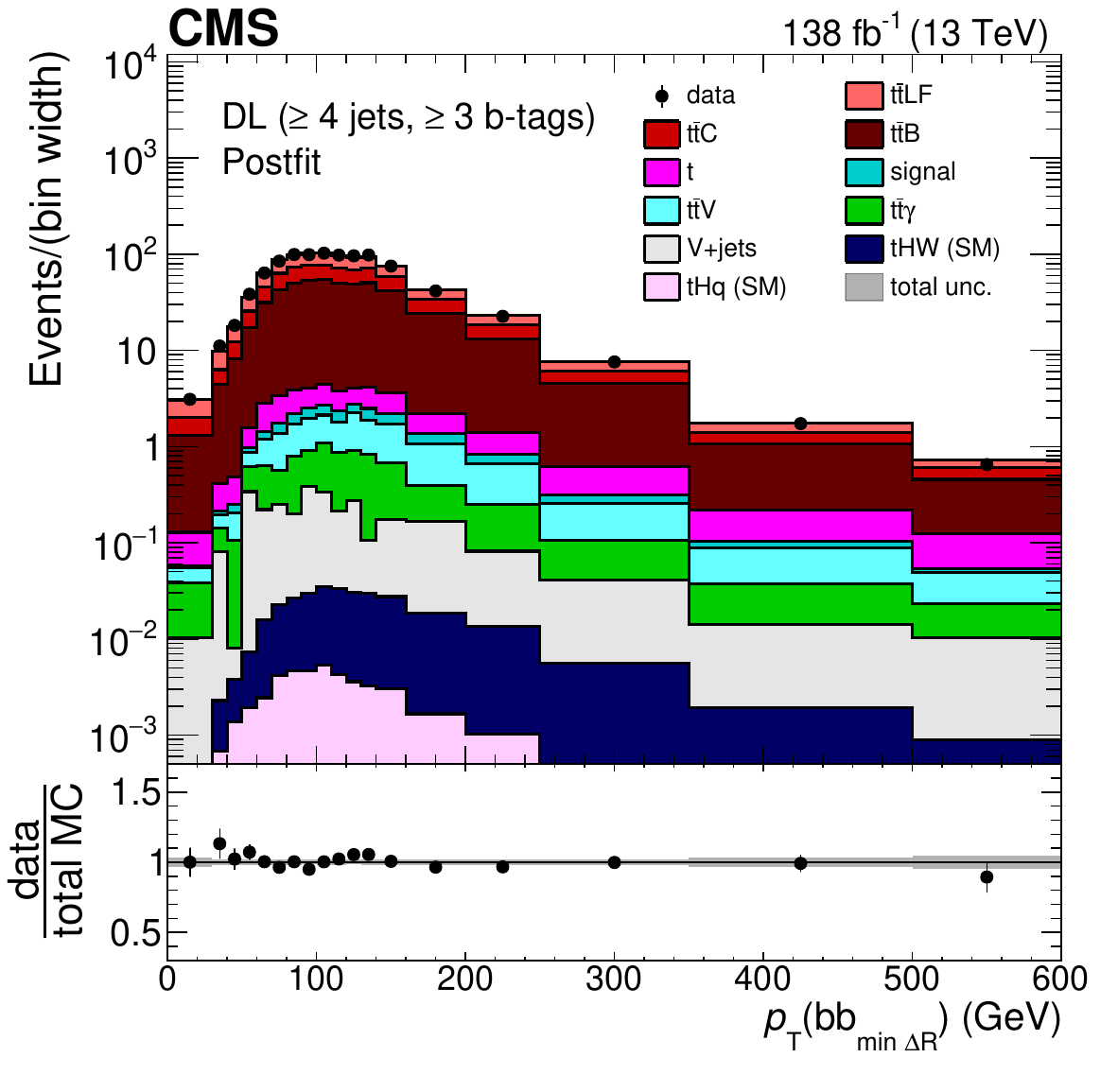} 
  \caption{
    Minimum $\Delta R$ between any two \PQb-tagged jets (upper), MEM discriminant output (middle), and \pt of the Higgs boson candidate identified as the pair of \PQb-tagged jets closest in $\Delta R$ (lower) for events passing the baseline selection requirements and additionally $\geq$4 jets in the DL channel prefit (left) and with the postfit background model (right) obtained from the fit to data described in Section~\ref{sec:results}.
    In the prefit case, the \ttH signal contribution, scaled by a factor 50 for better visibility, is also overlayed (line).
    The uncertainty band represents the total (statistical and systematic) uncertainty.
    Where applicable, the last bin in each distribution includes the overflow events.
  }
  \label{fig:background:controlvariables:dl}
\end{figure}

The modelling of the \ttB background has been validated in data in various observables and relevant phase space regions.
Only observables for which also the correlation to other observables is well-modelled were used further in the analysis, as described in Section~\ref{sec:strategy}.
For example, the shape of the predicted jet multiplicity distributions in the SL and DL channels, shown in Fig.~\ref{fig:selection:njets} after the baseline selection and prior to the fit to data, agrees well with the one observed in data.
The overall normalisation difference is attributed to the underprediction of the \ttB component in the simulation.
Further example distributions are shown in Figs.~\ref{fig:background:controlvariables:sl} and~\ref{fig:background:controlvariables:dl} for events in signal-enriched regions of high jet and \PQb-tagged jet multiplicity:
the average $\Delta\eta$ and the minimum $\Delta R$ between any two \PQb-tagged jets, which are sensitive to the \ttB modelling details;
the MEM discriminant output, which is one of the most powerful single variables used in the analysis to separate the \ttH signal from the \ttB background;
and the \pt of the Higgs boson candidate identified either using the aforementioned reconstruction BDT or as the pair of \PQb-tagged jets closest in $\Delta R$, which is relevant specifically for the STXS measurement.
Prior to the fit to data, the shapes of the data distributions are well described by the simulation.
Differences in the normalisation are attributed to the expected under-prediction of the \ttB yield.
With the postfit background model obtained from the final fit to data described in Section~\ref{sec:results}, the predicted distributions agree well with the ones observed in data.
The effect of the fit on these observables mostly changes the relative normalisation of each component, reduces the corresponding uncertainties, and slightly modifies the shapes of the distributions within the associated systematic uncertainties.

\section{Analysis strategy and classification}
\label{sec:strategy}
In each analysis channel, the selected events are divided into categories depending on the jet and \PQb-tagged jet multiplicity.
In each category, dedicated ANNs are trained and optimised to separate specific signal or background processes.
The ANN output is used, in the SL and DL channels, to categorise further the events and, in all channels, to construct the final discriminating observables.
The events are categorised independently for the 2016, 2017, and 2018 data-taking periods.

The signal is extracted in a simultaneous binned profile likelihood fit to the data of the expected signal and background distributions of the discriminating observable or of the event yield, depending on the channel and category~\cite{CMS:2024onh}.
The fit is performed across all categories and channels.
The systematic uncertainties discussed in Section~\ref{sec:systematics} are taken into account using nuisance parameters with appropriate constraints~\cite{CMS:2024onh,Conway:2011in}, and are correlated among the processes, categories, channels, and data-taking periods, as described in Section~\ref{sec:systematics}.
The different background composition in the categories helps to constrain the uncertainties and, thus, to increase the overall sensitivity.
The signal distributions include contributions from all SM Higgs boson decays to take into account contamination by other decay channels than \Htobb.
The other decay channels contribute approximately 7\% of the signal events, dominated by $\PH\to\PGt\PGt$ and $\PH\to\PW\PW$ decays.

The categorisation of the signal events differs depending on the interpretation.
As a first step, the signal categories are optimised for the inclusive \ttH and \tH production rate measurements, as well as for the coupling and \CP measurement.
For the STXS measurement, the \ttH signal categories in the different channels are split further into five subcategories (STXS categories) that target different ranges of \ptH (with lower boundaries at 0, 60, 120, 200, and 300\GeV).
Independent signal templates are constructed per particle-level Higgs boson \pt bin and fitted simultaneously.
In each subcategory, the contributions from all \pt bins are considered, thereby taking into account migration of events across bins due to reconstruction and resolution effects.

\begin{figure}[!htpb]
  \centering
  \includegraphics[width=\textwidth]{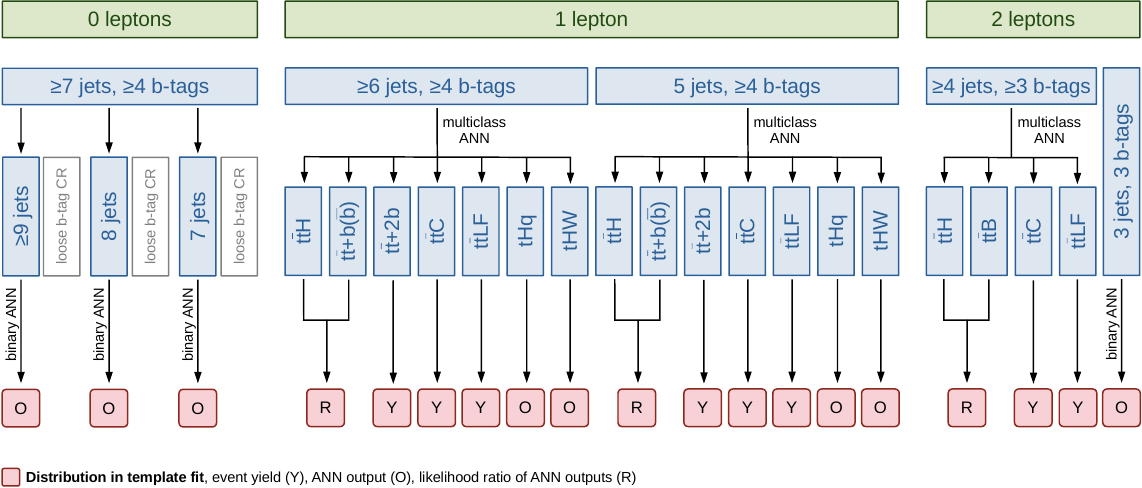} \\[50pt]
  \includegraphics[width=\textwidth]{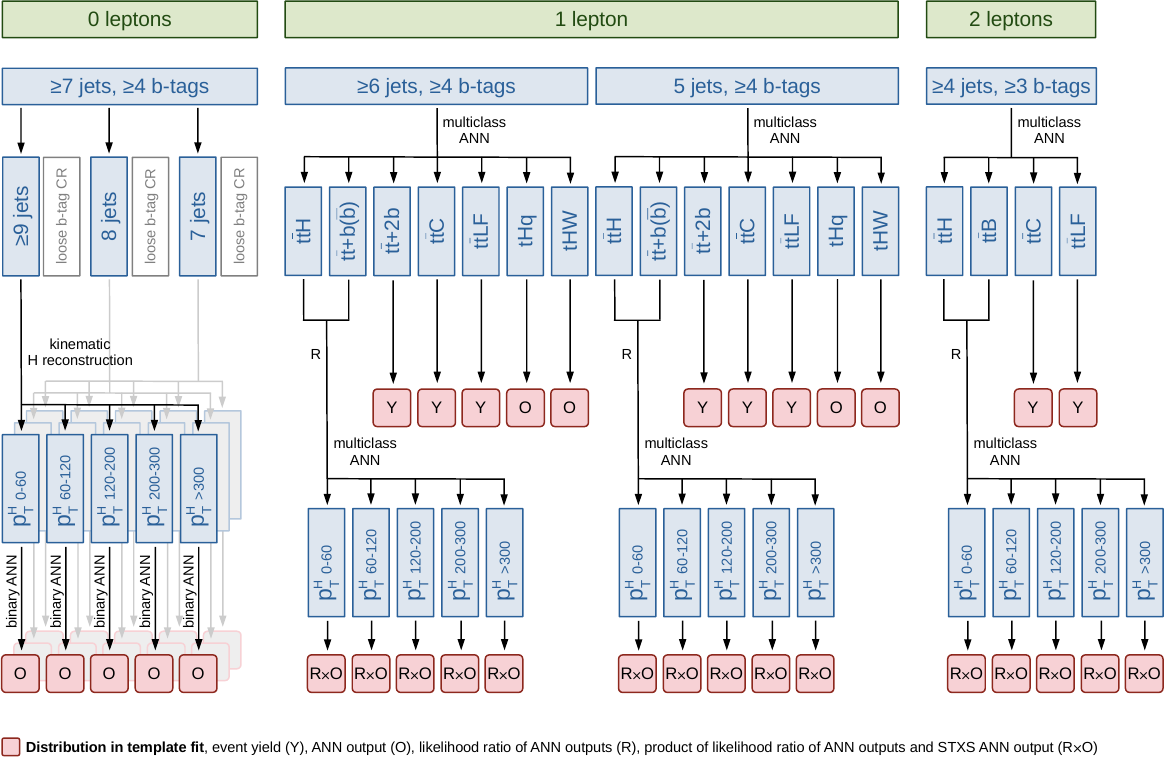}
  \caption{
    Illustration of the analysis strategy for the inclusive \ttH and \tH production rate, coupling, and \CP measurements (upper), and for the \ttH STXS measurement (lower).
    Details of the ANNs and the definition of the likelihood ratio are given in the text.
    The procedure is applied separately for the three years of data taking.
  }
  \label{fig:analysisstrategy}
\end{figure}

The analysis strategy is illustrated in Fig.~\ref{fig:analysisstrategy} and detailed in the following.
It has been optimised separately in each channel, based on both the quality of the background modelling and the expected sensitivity to an SM signal evaluated with simulated events and, in the case of the FH channel, data for the QCD multijet background determination.
Common aspects of the ANNs are described in Section~\ref{sec:strategy:ann}, followed by a description of the analysis strategy in the FH channel in Section~\ref{sec:strategy:fh}, and in the SL and DL channels in Section~\ref{sec:strategy:sldl}.

\subsection{Neural network architecture and training}\label{sec:strategy:ann}

In all analysis channels, dedicated ANNs are utilised for event categorisation or classification, as detailed further below.
The ANNs are implemented in \textsc{Keras}~\cite{chollet2015keras} as feed-forward neural networks.
The architecture consists of three or four hidden layers with between 100 to 2048 nodes, depending on the channel and category.
The activation function used in the nodes of the hidden layers are the ``ReLu'' function in the FH channel and the ``LeakyReLu'' function in the SL and DL channels~\cite{ml:general}.
The hyperparameters have been optimised using a ``Tree Parzen Estimator''~\cite{snoek2012practical,pmlr-v70-franceschi17a}, which uses information from past trials when testing the next set of hyperparameters.
The cost function that is minimised during the training is the ``categorical cross entropy'' in case of multiclassification ANNs, and the ``squared hinge'' function or the ``binary cross entropy'' in case of binary classification ANNs~\cite{ml:general}.
Potential overtraining is minimised using dropout and L2 regularisation or dropout and simultaneously L1 and L2 regularisation, depending on the channel~\cite{hinton2012improving}.
The ANNs have been found to provide the best analysis sensitivity compared to other types of ANNs or BDTs, evaluated using simulated data.

In each category, one single ANN is trained that is valid for all three data-taking periods.
Simulated data and, in case of the FH channel, data of all three periods, weighted to reflect the different integrated luminosities, are used in the training, thereby reducing statistical fluctuations in the trained ANN parameter values.
When evaluating the ANNs to obtain the final discriminant distributions in simulation and data, separate distributions are constructed per data-taking period in order to better control systematic uncertainties specific to the different periods, such as components of the jet energy scale and the \PQb tagging uncertainties discussed in Section~\ref{sec:systematics}.
It has been validated that the sensitivity of the analysis does not degrade compared to the case of training ANNs separately for each data-taking period or when using information about the year of data taking as an input feature to the ANN.

Depending on the category, between 0.5 and 1.5 million events are used for the training procedure, with at least 16\,000 events for the \ttB class in all cases.
These events are further split into three independent subsamples used for the actual training (60\%), for the optimisation of the hyperparameters (20\%), and for validating the performance of the ANNs (20\%).
Additional weights are applied such that the effective number of events per class is the same in order to avoid that the ANN classification decision is biased by the relative frequency of the different processes or, in case of the STXS classification, the \ptH bins.
The training is terminated once either the performance obtained with the validation sample does not improve after 50 full passes over the training data (``epoch'') or the relative difference between the performance on the training and validation samples diverges (``early stopping'').
The aforementioned events are used exclusively for training, optimisation, and validation, and another statistically independent set of events is used for the final analysis of the data to avoid biases due to potential overtraining.

The input variables of the ANNs are listed in Table~\ref{tab:anninputvariables}.
The variables have been chosen out of a larger set of possible variables as the ones that are most important for the performance of the ANNs, based on several ranking procedures including a Taylor expansion of the ANN output as a function of the input variables~\cite{Wunsch:2018oxb} and a ``data Shapley'' metric~\cite{ghorbani2019data}.
\begin{table}
  \centering
  \topcaption{
    Observables used as input variables to the primary ANNs ($\times$) and STXS ANNs ($\circ$) per channel.
    Categories are labelled as ``$<$(min.) number jets$>$-$<$min.\ number \PQb-tagged jets$>$'', \eg the FH \fhNineSR category is labelled as ``9-4''.
    The $\dagger$ indicates that the observable is constructed using information from the BDT-based event reconstruction.
  }
  \renewcommand{\arraystretch}{1.4}
  \begin{scriptsize}
    \begin{tabular}{lp{0.45\textwidth}ccccccc}
      && \multicolumn{3}{c}{FH} & \multicolumn{2}{c}{SL} & \multicolumn{2}{c}{DL} \\
      \multicolumn{2}{c}{Observable} & 9-4 & 8-4 & 7-4 & 6-4 & 5-4 & 4-3 & 3-3 \\
      \hline
      MEM                              & matrix element method discriminant                                         & $\times$ & $\times$ & $\times$ & $\times$, $\circ$ & $\times$, $\circ$ & $\times$, $\circ$ & \\
      BLR                              & \PQb tagging likelihood ratio discriminant                                & & & & & & $\times$, $\circ$ & \\
      $\ln\left(\frac{\text{BLR}}{1 - \text{BLR}}\right)$ & transformed \PQb tagging likelihood ratio discriminant & & & & $\times$ & $\times$ & & \\
      $\pt(\text{j}^{2})$               & \pt of second leading jet, ranked in \pt                                   & & & & & & $\times$, $\circ$ & \\
      $\pt(\text{j}^{3})$               & \pt of third leading jet, ranked in \pt                                    & & & & & & & $\times$, $\circ$ \\
      $\pt(\text{j}^{7})$               & \pt of seventh leading jet, ranked in \pt                                  & $\times$ & & & & & & \\ 
      $\pt(\PQb^{i})$                  & \pt of $i^{\text{th}}$, $i=$1--4, leading \PQb-tagged jet, ranked in \pt    & & & & & & $\times$, $\circ$ & \\
      $\eta(\text{j}^{i})$              & $\eta$ of $i^{\text{th}}$, $i=$1--2, leading jet, ranked in \PQb tag.\ discr. value & $\times$ & $\times$ & $\times$ & & & & \\ 
      
      $\mean{d_{\PQb}(\text{j})}$          & average \PQb tagging discriminant value of all jets                       & & & & $\times$ & $\times$ & & \\
      $\mean{d_{\PQb}(\PQb)}$             & average \PQb tagging discriminant value of all \PQb-tagged jets          & & & & $\times$ & $\times$ & & \\
      $d^{3}_{\PQb}(\text{j})$              & third highest \PQb tagging discriminant value of all jets                 & & & & $\times$ & $\times$ & & \\
      $\text{Var}(d_{\PQb}(\text{j}))$     & variance of \PQb tagging discriminant values of all jets                  & & & & $\times$ & $\times$ & & \\
      $\mean{\Delta R(\PQb\PQb)}$    & average of $\Delta R$ between two \PQb-tagged jets                         & & & & $\circ$ & $\circ$ & $\times$, $\circ$ & \\
      $\mean{\Delta R(\text{jj})}$     & average of $\Delta R$ between two jets                                      & $\times$ & $\times$ & & & & & \\
      $\text{min}\;\Delta R(\text{jj})$ & minimum of $\Delta R$ between two jets                                     & & $\times$ & $\times$ & & & & $\times$, $\circ$ \\
      $\text{max}\;\Delta R(\text{jj})$ & maximum of $\Delta R$ between two jets                                     & $\times$ & $\times$ & $\times$ & & & & \\
      $\mean{\Delta\eta(\PQb\PQb)}$  & average of $\Delta\eta$ between two \PQb-tagged jets                       & & & & $\times$ & $\times$ & & \\
      $\mean{\Delta\eta(\text{jj})}$   & average of $\Delta\eta$ between two jets                                    & $\times$ & $\times$ & $\times$ & $\times$ & $\times$ & & \\
      $\mean{m(\PQb)}$                & average invariant mass of all \PQb-tagged jets                             & & & & $\times$, $\circ$ & $\times$, $\circ$ & & \\
      $\mean{m(\text{j})}$             & average invariant mass of all jets                                          & & & & $\times$ & $\times$ & & \\
      $m(\PQb\PQb_{\text{min}\;\Delta R})$ & invariant mass of pair of \PQb-tagged jets closest in $\Delta R$          & & & & $\times$ & $\times$ & $\times$, $\circ$ & \\
      $m(\text{j}\PQb_{\text{min}\;\Delta R})$ & invariant mass of pair of jet and \PQb-tagged jet closest in $\Delta R$  & & & & & & $\times$, $\circ$ & \\
      $m(\text{jj}_{125\GeV})$           & invariant mass of pair of jets with mass closest to 125\GeV                & $\times$ & & & & & $\circ$ & $\circ$ \\
      $m(\PQb\PQb_{\text{max}\;m})$     & maximum invariant mass of pairs of \PQb-tagged jets                        & $\times$ & $\times$ & & & & $\times$, $\circ$ & $\times$, $\circ$ \\
      $m(\text{j}\PQb\PQb_{\text{max}\;\pt})$ & inv.\ mass of jet and pair of \PQb-tagged jets with highest \pt       & & & & & & $\times$, $\circ$ & \\
      $\mean{\pt(\text{j})}$           & average \pt of all jets                                                     & & & & $\times$, $\circ$ & $\times$, $\circ$ & & \\
      $\mean{\pt(\PQb)}$              & average \pt of all \PQb-tagged jets                                        & & & & $\times$, $\circ$ & $\times$, $\circ$ & & \\
      $\pt(\PQb\PQb_{\text{min}\;\Delta R})$ & \pt of pair of \PQb-tagged jets closest in $\Delta R$                   & & & & $\times$, $\circ$ & $\times$, $\circ$ & $\times$, $\circ$ & $\times$, $\circ$ \\
      $\pt(\text{jj}_{\text{min}\;\Delta R})$ & \pt of pair of jets closest in $\Delta R$                                 & & & & & & & $\times$, $\circ$ \\
      $\pt(\text{j}\PQb_{\text{min}\;\Delta R})$ & \pt of pair of jet and \PQb-tagged jet closest in $\Delta R$         & & & & & & & $\times$, $\circ$ \\
      $\HT(\text{j})$                  & scalar sum of \pt of all jets                                            & & & & $\times$ & $\times$ & $\times$, $\circ$ & \\
      $\HT(\PQb)$                     & scalar sum of \pt of all \PQb-tagged jets                                  & & & & $\times$ & $\times$ & $\times$, $\circ$ & \\
      
      $N(\text{j})$                    & number of jets                                                              & & & & $\times$ & & & \\
      $N(\PQb^{\text{loose}})$           & number of jets with loose \PQb tag                      & & & & & & $\times$, $\circ$ & \\
      $d_{\PQb}(\PQb^{\tHW}_{\PQt})  {}^{\dagger}$   & \PQb tagging discr.\ value of \PQb jet from \PQt quark from \tHW reco. & & & & $\times$ & $\times$ & & \\
      $\pt(\text{\PH}^{i}) {}^{\dagger}$    & \pt of Higgs boson from \ttH, \tHq, \tHW reconstruction & & & & $\circ$ & $\circ$ & & \\
      $\ln(\text{min}\ \PQb_{\text{\PH}}^{\tHq}) {}^{\dagger}$    & log.\ of min.\ \pt of \PQb jets from Higgs boson from \tHq reco. & & & & $\circ$ & $\circ$ & & \\
      $\abs{\eta(\PQq^{\tHq})} {}^{\dagger}$    & $\abs{\eta}$ of light-quark jet from \tHq reconstruction & & & & $\times$ & $\times$ & & \\
      $\Delta R(\PQb\PQb_{\text{\PH}}^{i}) {}^{\dagger}$    & $\Delta R$ of \PQb jets from Higgs boson from \ttH, \tHq, \tHW reco. & & & & $\circ$ & $\circ$ & & \\
      $m(\text{\PH}^{i}) {}^{\dagger}$    & inv.\ mass of Higgs boson from \ttH reconstruction & & & & $\circ$ & $\circ$ & & \\
      $m(\PQt^{\ttH}_{\text{lep}}) {}^{\dagger}$      & inv.\ mass of leptonically decaying \PQt quark from \ttH reco. & & & & $\times$, $\circ$ & $\times$, $\circ$ & & \\
      $\text{BDT}^{i} {}^{\dagger}$      & reconstruction BDT output for \tHq, \ttH, \ttbar hypotheses & & & & $\times$ & $\times$ & & \\
      $A$, $S$ & event aplanarity and sphericity~\cite{Bjorken:1969wi} & $\times$ & $\times$ & $\times$ & & & & \\
      $H^{\text{FW}}_{i}$                & $i^{\text{th}}$, $i=$0--5, Fox--Wolfram moment~\cite{tagkey1979543}             & $\times$ & $\times$ & $\times$ & & & & \\ 
      $H^{\text{FW}}_{i}/H^{\text{FW}}_{0}$ & ratio of Fox--Wolfram moments, $i=$1--4                                       & $\times$ & $\times$ & $\times$ & & & & \\
    \end{tabular}
  \end{scriptsize}
  \renewcommand{\arraystretch}{1.0}
  \label{tab:anninputvariables}
\end{table}

\subsection{Fully hadronic channel}\label{sec:strategy:fh}

In the FH channel, events are categorised depending on the jet and \PQb-tagged jet multiplicity, as well as the invariant mass of the pair of non-\PQb-tagged jets closest to the nominal \PW boson mass, \mqq.
Three categories, referred to as SRs, with \fhSevenSR, \fhEightSR, and \fhNineSR are used for the signal extraction and enter the final fit.
The events in the SR are in addition required to fulfil $60<\mqq<100\GeV$ for events with 7 or 8 jets, and $72<\mqq<90\GeV$ for events with $\geq$9 jets, where the different \mqq ranges reflect the dependency of the signal efficiency of the \mqq criterion on the jet multiplicity.
Dedicated binary classification ANNs are trained in each SR to separate the \ttH signal from the background, and the ANN classifier output distribution is used as the final discriminating observable.

For each SR, two mutually exclusive CRs with looser \PQb tagging requirements, referred to as ``evaluation regions'' (ERs) and ``training regions'' (TRs), are defined, which are used to estimate the QCD multijet background and to train the ANNs, respectively.
Events in the ERs (TRs) are required to contain at least four jets
fulfilling the loose \PQb tag requirement, out of which exactly three
(two) are also \PQb tagged.
Otherwise, events in the ERs and TRs must pass the same selection criteria as events in the SRs.
For each SR, ER, and TR, further mutually exclusive regions, referred to as ``validation regions'' (VR-SR, VR-ER, VR-TR), are defined by inverting the criterion on \mqq.
The VRs are signal depleted, with an expected \ttH contribution below 0.6\%, and used to test the background estimation and ANN training procedures described in the following.
The categorisation scheme is summarised in Table~\ref{tab:categorisation:fh}.
\begin{table}[!htpb]
  \centering
  \topcaption{
    Categorisation scheme in the FH channel, applied independently in each jet-multiplicity category.
    The \mqq selection criteria refer to events with 7 or 8 ($\geq$9) jets.
  }
  \renewcommand{\arraystretch}{1.1}
  \begin{tabular}{lcc}
    & & $30 < \mqq < 60 (72)\GeV$ \\
    & $60 (72) < \mqq < 100 (90)\GeV$ & or \\
    & & $100 (90) < \mqq < 250\GeV$ \\
    \hline
    2 \PQb tags & {\itshape training region (TR)} & {\itshape validation region (VR-TR)} \\
    $\geq$4 loose \PQb tags & QCD events for ANN training & input variable validation \\[\cmsTabSkip]
    3 \PQb tags & {\itshape evaluation region (ER)} & {\itshape validation region (VR-ER)} \\
    $\geq$4 loose \PQb tags & discriminant shape for QCD & discriminant shape for QCD \\[\cmsTabSkip]
    $\geq$4 \PQb tags & {\itshape signal region (SR)} & {\itshape validation region (VR-SR)} \\
    & analysis region & comparison of QCD shape with data \\    
  \end{tabular}
  \renewcommand{\arraystretch}{1.0}
  \label{tab:categorisation:fh}
\end{table}

The dominant background contribution (approximately 72--80\%) in the SR originates from QCD multijet production.
The contribution is determined from the data to avoid effects from inaccurate modelling and insufficient number of events in the relevant phase space regions of the QCD simulation.
The procedure is as follows:
the ANNs are trained using the events from the TRs, which are estimated to originate from QCD multijet production more than 80\% of the time, followed by \ttLF events, as background together with simulated \ttH events, where only Higgs boson decays to \bbbar are considered.
Differences of the expected distributions of the other, non-QCD backgrounds between the prefit and the postfit configurations are negligible for the procedure, given their small size compared to the QCD background in the TRs.
The shape of the discriminant distribution is then determined from the events in the ERs by evaluating the ANN output from these events, and then subtracting the remaining contributions from non-QCD backgrounds (mainly \ttjets) using simulation.
The yield of the QCD events is left as free parameter in the final fit, independently in each SR.
The subtracted non-QCD background contribution amounts to 20--29\% in the ERs, depending on the category.
The subtraction is performed in-situ in the final fit, and the full set of systematic uncertainties, discussed in Section~\ref{sec:systematics}, in the non-QCD backgrounds is taken into account in the procedure.

Due to differing heavy-flavour compositions, the kinematic properties
of \PQb-tagged and untagged jets vary across the TR, ER, and SR.
To align the kinematic properties of loosely \PQb-tagged jets with
those of \PQb-tagged jets, event weights are applied to events in the
TR and ER.
These weights are calculated using per-jet correction factors (\tflm).

In the ER, the \tflm correction factor for the leading loosely
\PQb-tagged jet that is not also \PQb tagged is used as the event
weight.
In the TR, \tflm correction factors for the two leading loosely
\PQb-tagged jets that are not also \PQb tagged are multiplied to
determine the event weight.
The \tflm corrections are derived from jets in events that pass the
baseline selection, excluding the first two jets, based on the jet
\pt, $\eta$, and the minimum distance between the jet and the first
two \PQb-tagged jets, following the procedure in
Ref.~\cite{Sirunyan:2018ygk}.
The jet ranking is based on the \PQb tagging discriminant value.

Once the event weights are applied, the ANN output distribution shape
in the ER is expected to accurately represent the QCD multijet
background shape in the SR.
This is confirmed by applying the same procedure to events in the
VR-ER and comparing the predicted shape to the observed one in the
VR-SR.
The method is further validated by dividing the ER into two regions
based on \PQb tagging discriminant values and applying the procedure
analogously.

The ANNs are constructed to separate \ttH events from QCD multijet background events.
Taking into account explicit separation against \ttjets events has been found to not improve the results further.
The ANNs utilise input variables related to the kinematic properties of individual objects and event shape information, as well as the MEM discriminant output, as listed in Table~\ref{tab:anninputvariables}.
It has been verified that the distributions of the chosen variables agree between each SR, ER, and TR after application of the \tflm correction, using the data in the VR and simulated \ttH events.
For example, variables that are directly correlated with the \PQb tagging discriminant value do not fulfil this criterion and are not used as input.
The observed distributions of the chosen input variables have been further verified to be well modelled based on a $\chi^{2}$ criterion, with corresponding $p$-values well above 5\%.

\begin{figure}[!htp]
  \centering

    \includegraphics[width=0.45\textwidth]{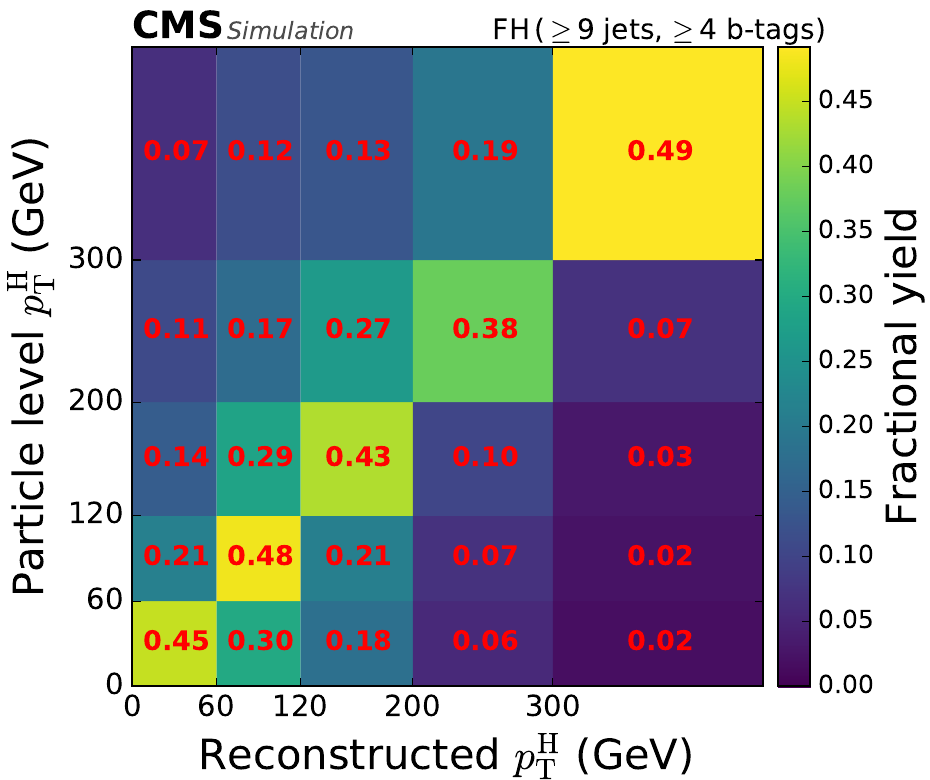}
    \includegraphics[width=0.45\textwidth]{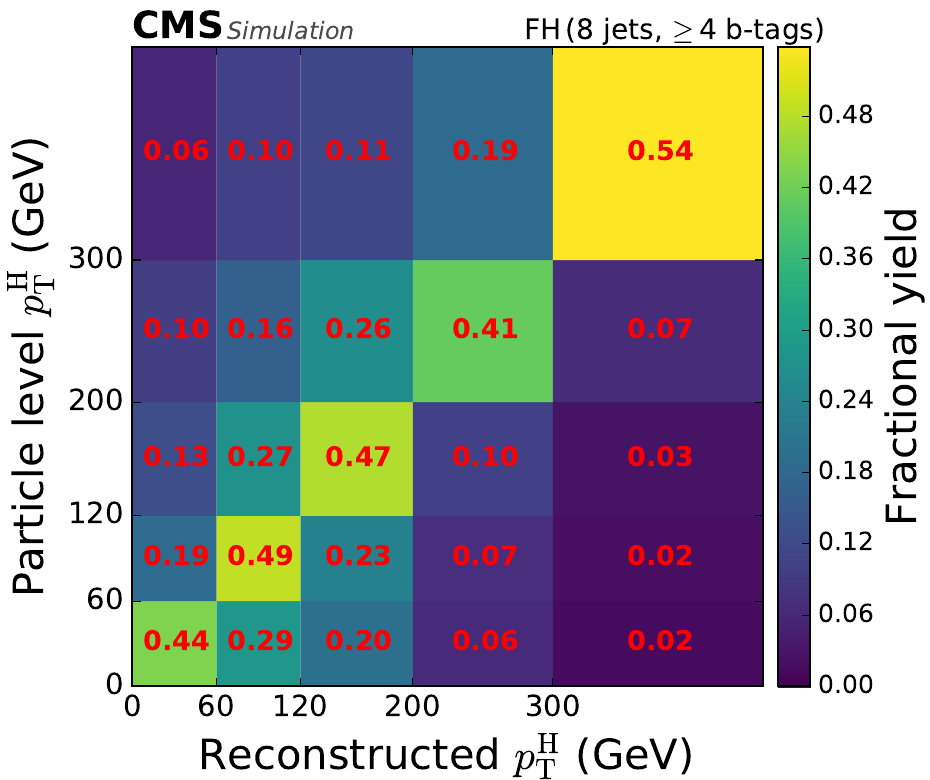}
    \includegraphics[width=0.45\textwidth]{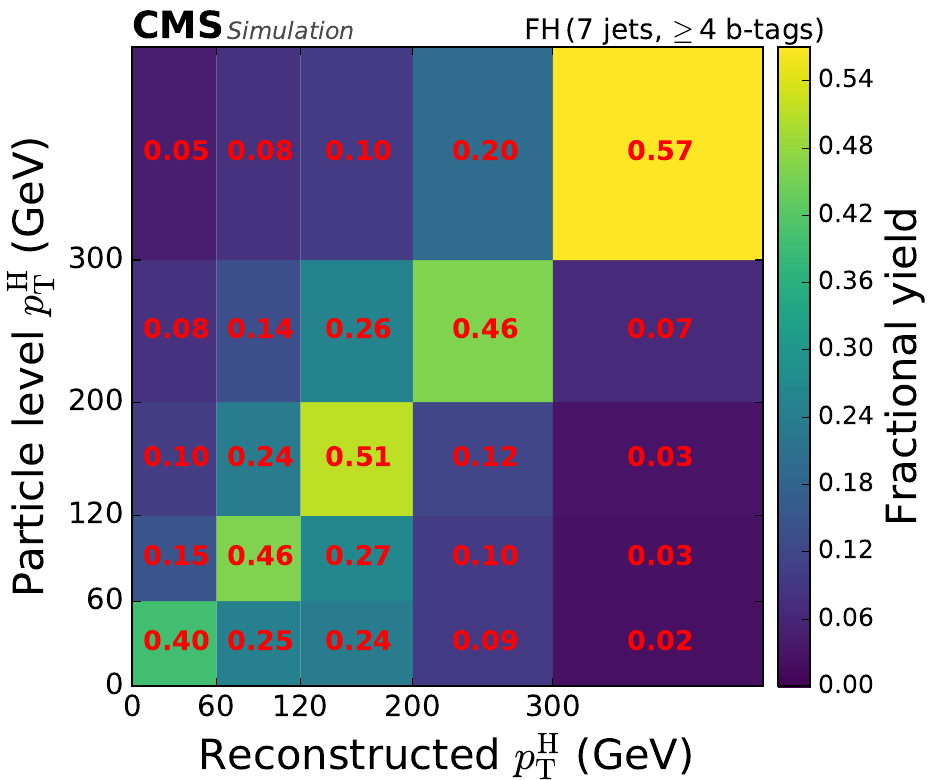}
    \includegraphics[width=0.45\textwidth]{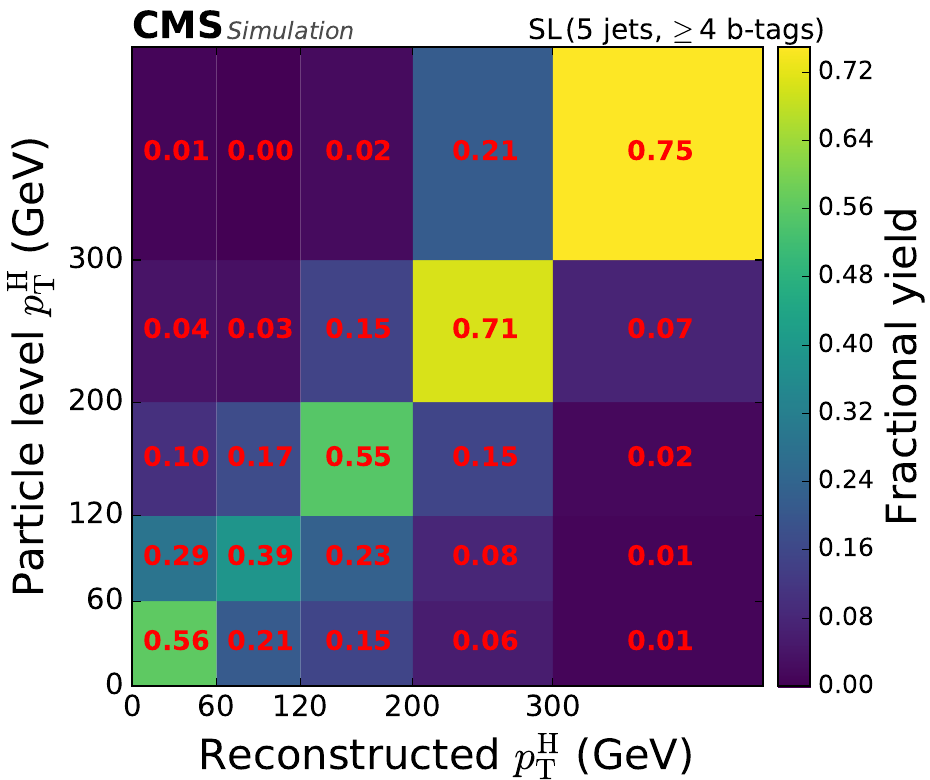} 
    \includegraphics[width=0.45\textwidth]{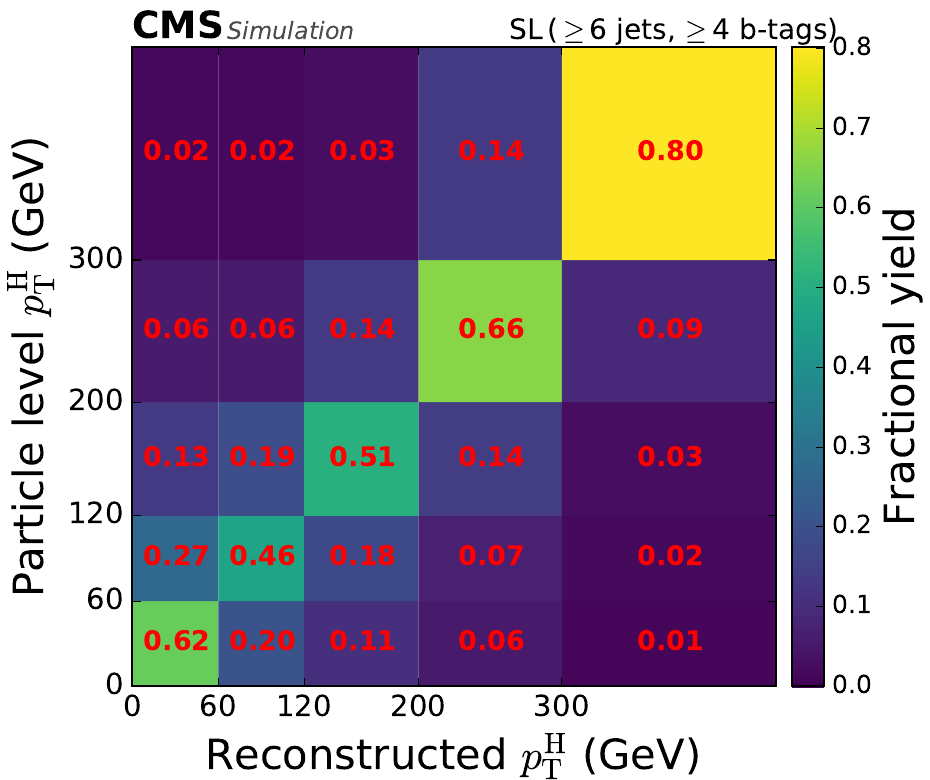} 
    \includegraphics[width=0.45\textwidth]{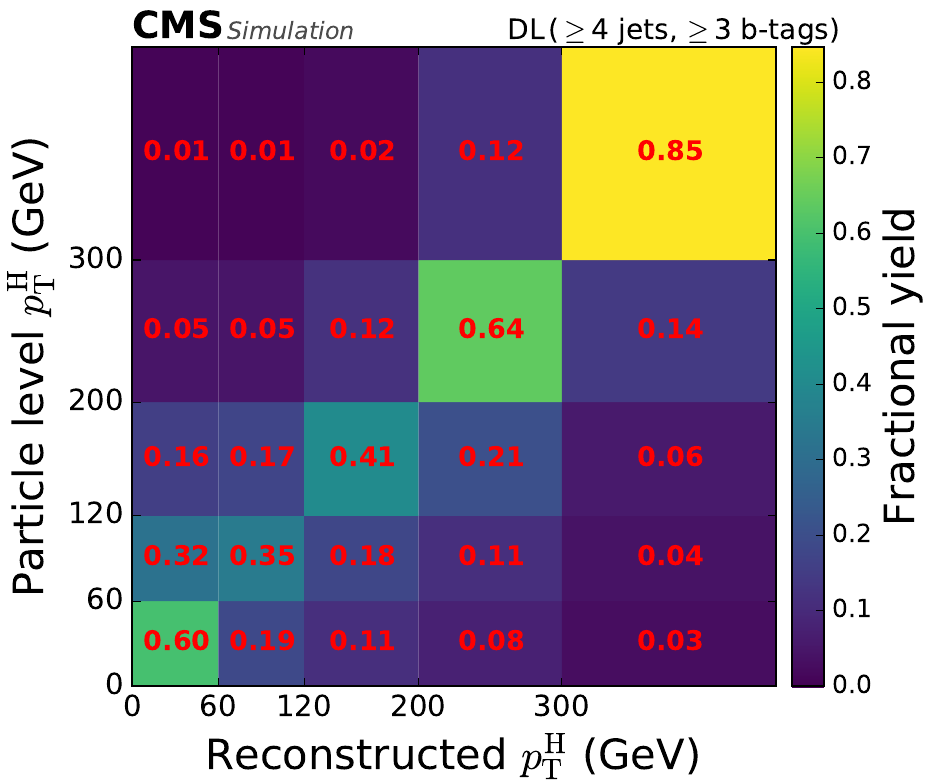} 
  \caption{
    Categorisation efficiency of the \ttH signal events in the STXS analysis in the different categories of the FH channel (upper row, middle row left), the SL channel (middle row right, lower row left), and the DL channel (lower row right).
  }
  \label{fig:stxsmigration}
\end{figure}

For the STXS measurement, a Higgs boson candidate is reconstructed from that pair of jets, out of the four jets with the highest \PQb tagging discriminant value, whose invariant dijet mass is most compatible with that of a 125\GeV Higgs boson, based on a $\chi^{2}$ criterion taking into account the experimental resolution of the jet \pt.
Depending on the \pt of the reconstructed Higgs boson candidate, the SR events are categorised further and assigned to one of the five \ptH ranges, resulting in 38--57\% correct assignments for \ttH events, as presented in Fig.~\ref{fig:stxsmigration}.
Thus, there are in total 15 STXS categories in the FH channel for each of the three data-taking periods.
In each category, the ANN output distribution is used as final discriminant in the fit to extract the STXS.

\subsection{Single-lepton and dilepton channels}\label{sec:strategy:sldl}

In the channels with leptons, events are separated based on the jet and \PQb-tagged jet multiplicity.
In the SL channel, two categories with \slSixFour and \slFiveFour are considered, and in the DL channel two categories with \dlFourThree and \dlThreeThree are used.

In each of the two SL categories and in the \dlFourThree DL category, events are further categorised based on the output of multiclassification ANNs, which are designed to separate between different signal and background processes.
The values obtained in the output nodes of the ANNs are normalised to unity using a ``soft-max'' function~\cite{Goodfellow-et-al-2016}, and, as a result, the output values \OANN{i} can be interpreted as probabilities describing the likelihood of the event being of a certain process $i$.
Events are assigned to a category corresponding to the most probable process according to this ANN multiclassification.
In the SL channel, the three signal processes \ttH, \tHq, and \tHW, and the four \ttjets background processes \ttLF, \ttC, \tttwob, and \ttborbb are considered in the ANN multiclassification.
The dedicated class for \tttwob events out of the \ttB background is designed in order to constrain the uncertainty related to collinear gluon splitting.
In the DL channel, the \ttH signal process and the three \ttjets background processes \ttLF, \ttC, and \ttB are considered in the ANN multiclassification.
No dedicated classes for \tHq, \tHW, or \tttwob events are included in the DL channel, motivated by the reduced statistical precision due to the lower \tH and \ttjets event rate in this channel. 

The output values of the ANNs are subsequently used to compute the final discriminant observables.
In the SL channel, for events in the \ttH and \ttborbb process categories, a likelihood ratio discriminant is computed from the ANN output values $\text{O}$ as
\begin{equation}
  \text{R}_{\text{SL}} = \frac{\OANN{\ttH}}{\OANN{\ttH} + \OANN{\ttborbb} + \OANN{\tttwob}},
\end{equation}
and used as the discriminating observable.
This allows exploiting more information from the ANN multiclassification: \eg, for an event that is categorised as \ttH, not only the output value of the \ttH node is used but also the output values of the \ttborbb and \tttwob nodes, which provides further information about the likelihood of the categorisation.
With the likelihood ratio discriminant, an improvement of the expected sensitivity by 18\% is achieved compared to fitting directly the ANN output values \OANN{\ttH} and \OANN{\ttborbb}.
In the \tHq and \tHW process categories, the \OANN{\tHq} and \OANN{\tHW} distributions are used as the final discriminating observables, respectively.
The binning of the discriminant distributions has been chosen to
optimise the sensitivity, while at the same time avoiding artificial
constraints of the uncertainties that can arise if their
parametrisation is not flexible enough to describe all shape
variations of the distributions.
The main purpose of the remaining process categories \ttLF, \ttC, and \tttwob is to constrain the normalisation of the corresponding background processes; therefore, in these categories only the event yield is considered instead of the full distribution.
Similarly, in the DL \dlFourThree category, a likelihood ratio discriminant is computed from the events in the \ttH and \ttB process categories as
\begin{equation}
  \text{R}_{\text{DL}} = \frac{\OANN{\ttH}}{\OANN{\ttH} + \OANN{\ttB}},
\end{equation}
and used as the discriminating observable, while for the \ttLF and \ttC process categories, the event yield is used.

In the DL \dlThreeThree category, a binary classification ANN is designed to separate the \ttH signal from the inclusive \ttjets background, and the ANN classifier output distribution is used as final discriminating observable.

In total, this leads to 12 categories in the SL channel and four categories in the DL channel per data-taking period, as shown Fig.~\ref{fig:analysisstrategy}.

As for the FH channel, the ANN input variables are related to the kinematic properties of individual objects, the event shape, as well as the MEM discriminant outputs, and in addition the jet \PQb tagging and the BLR discriminants, as listed in Table~\ref{tab:anninputvariables}.
In the SL channel, variables constructed using information from the BDT-based event reconstruction are also used (marked with a dagger symbol (${}^{\dagger}$) in Table~\ref{tab:anninputvariables}).
The input variables and their correlations have been verified to be well described in data, based on a goodness-of-fit test that takes into account the full uncertainty model, \ie including statistical and systematic uncertainties, using the ``saturated model'' method~\cite{CMS:2024onh,Lindsey1996-nh}.
For each variable and each pair of variables under scrutiny, the one- and two-dimensional distributions, respectively, have been fitted to the data in the analysis categories, confirming they are well modelled with $p$-values greater than 5\%.
The effect of the signal contribution on the validation procedure has been found to be negligible, given the limited sensitivity to the signal of each of the input variables alone.

The ANNs are trained using simulated \ttH, \tH, and \ttjets events.
Only events with Higgs boson decays to \bbbar are considered, and the \ttbar system is required to have one or two leptons in the final state for the SL and the DL channels, respectively.
For the \tH events, only those with leptonic decays of the top quark are considered.
The achieved classification accuracy ranges from 35--65\% for \ttH events to 40--70\% for \tH events, as well as 30--40\% and 60--70\% for \ttB and \ttLF events, respectively, depending on the channel.

For the STXS measurement, the most signal-like events are categorised further, targeting the five regions in \ptH.
Specifically, these are the events in the \ttH and \ttborbb process categories of the SL \slSixFour and \slFiveFour categories, and the \ttH and \ttB process categories of the DL \dlFourThree category.
The STXS categorisation is performed by additional multiclassification ANNs (STXS ANNs) designed to classify \ttH events by \ptH.
Each event is assigned to the \ptH bin with the highest probability according to the STXS ANN multiclassification; the STXS ANN output value in that node, multiplied by the observable value of the inclusive \ttH measurement, \ie the likelihood ratio value, is used as final discriminating observable.
Events that are categorised with high certainty by the STXS ANN thus contribute with a higher weight to the final discriminating observable.
This approach has been found to achieve superior sensitivity compared to other classification approaches studied, including approaches based on a kinematic reconstruction of \ptH.
Thus, in total, there are ten STXS categories in the SL channel, corresponding to the five Higgs boson \pt ranges in each of the jet multiplicity categories, and five STXS categories in the DL channel.
The categorisation of all other events remains the same as for the inclusive \ttH measurement, with the exception of the events in the DL \dlThreeThree category, which are not used since they do not improve the sensitivity in the STXS analysis.

A dedicated STXS ANN is trained in each of the \slSixFour and \slFiveFour categories of the SL channel and in the \dlFourThree category of the DL channel, using the input variables listed in Table~\ref{tab:anninputvariables}.
For the training, only simulated \ttH events are used, and from these only those where both \PQb quarks are within the experimental acceptance defined by the event selection described in Section~\ref{sec:selection}.
The achieved categorisation efficiency is presented in Fig.~\ref{fig:stxsmigration}.

\section{Systematic uncertainties}
\label{sec:systematics}
Several sources of systematic uncertainties are considered in the analysis.
The uncertainties are taken into account via nuisance parameters in the final profile likelihood fit described in Section~\ref{sec:results} and alter either the rate or both the rate and the discriminant shape of the signal or background processes~\cite{CMS:2024onh}.
The effects from the same source are treated as fully correlated among the different categories.
In general, and unless stated otherwise hereafter, theoretical uncertainties are treated as fully correlated among the different data-taking periods, while experimental uncertainties are treated as uncorrelated for the data recorded in 2016, 2017, and 2018.
The latter is justified since the experimental uncertainties are mainly of statistical origin related to the limited size of the data and simulation samples used in auxiliary measurements, which are independent between the data-taking periods.

The theoretical uncertainties in the cross sections used to predict the rates of the signal and background processes, which arise primarily from the factorisation and renormalisation scale choices and the PDFs, are propagated to the yield estimates.
The cross section uncertainties are each separated into their scale (renorm./fact. scales) and PDF components, and are correlated where appropriate among processes.
In addition, the normalisation of the \ttB and \ttC background processes is left unconstrained in the final fit.
To take into account additional uncertainties in the modelling of collinear gluon splitting, an additional 100\% log-normal constrained rate uncertainty is assigned to the \tttwob component relative to the normalisation of the overall \ttB process.
Since the shapes of the relevant distributions are different for the \tttwob component and the overall \ttB background, this leads effectively to a rate and shape variation of the \ttB distributions.
The prior size of 100\% has been chosen conservatively to give the fit the freedom to adjust the distributions to the data, and it is constrained in the fit.

Uncertainties arising from missing higher-order terms in the \POWHEG \ttbbsample and \ttsample simulations at ME level are evaluated by independent variations of the renormalisation and factorisation scales by factors of two up and down with respect to the nominal values, and two independent nuisance parameters are assigned in the fit (\muR/\muF scale).
In addition, the uncertainties are treated as independent among the signal, the \ttB, and the other \ttbar processes.
The uncertainty arising from the PDF set is determined from the PDF variations provided with the NNPDF set~\cite{Ball:2014uwa}, correlating processes for which the same FS and order in \alpS are used in the PDF set.
The corresponding uncertainty in the \PYTHIA PS is determined by varying the parameters controlling the amount of initial- and final-state radiation independently by factors of two up and down~\cite{Skands:2014pea} (PS scale ISR/FSR), separately for signal, \ttB, \ttC, and \ttLF.
These variations are applied using event weights obtained directly from the generators.
Uncertainties related to the ME-PS matching scheme and the underlying event tune are evaluated by comparing the reference \ttbbsample and \ttsample simulation with samples with varied parton-shower matching scale (\texttt{hdamp} parameter) and varied tune parameters, respectively.
The event count in these additional samples was small and induced changes to the discriminant distributions comparable in size to the statistical fluctuations of the additional samples and compatible with a pure rate variation.
For this reason, the uncertainties have been estimated as the changes in the rates of the different \ttjets subprocesses independently for each category.
The derived rate variations amount typically to 5--10\% and the uncertainties are treated as uncorrelated among the \ttjets subprocesses.
This approach has been verified with an independent estimation of the variations obtained from the same additional samples and applying the same analysis selection, except for using events with fewer \PQb-tagged jets, thus reducing the statistical fluctuations.
In this cross-check, the effect of the selection with fewer \PQb-tagged jets is accounted for via event weights that act as transfer factors, that depend on the \pt, $\eta$, and flavour of the jets in the event and encode the probability to observe jets originating from \PQb quarks among the non-\PQb-tagged jets.
The impact of the mismodelling of the top quark \pt spectrum in the \ttbar simulation~\cite{Sirunyan:2018wem} has been found to be negligible.

The flexibility of the background model and its robustness against potential mismodelling of the \ttB component have been confirmed using pseudo-experiments.
A large number of pseudo-experiments have been sampled, both from the nominal and from alternative \ttB predictions, and fitted using the nominal background model, taking  into account the systematic uncertainties.
The fits have also been performed including only the theoretical \ttjets modelling uncertainties.
Several alternative \ttB predictions have been used to sample the pseudo-data, including scaling the \ttB component by a factor of 1.2 in order to verify the ability of the background model to compensate for the expected larger rate, or using the \ttB component of the \POWHEG~\ttsample sample, which is not otherwise used in the analysis, or combinations thereof.
The average fitted signal strength deviated from the injected value by at most 3\% in case the \ttB background was sampled from the nominal model, and by at most 6\% when it was sampled from the \ttsample sample, which is well within the systematic uncertainty of the measurement, and the average fitted \ttB normalisation corresponded to the applied scale of the \ttB background.

For the STXS measurement, the \ttjets background uncertainty model is extended to provide further flexibility towards potential effects that depend on the reconstructed \ptH.
The extension is a conservative choice, intentionally made because the STXS measurement is inherently sensitive to effects that depend on \ptH.
The \ttB normalisation parameter, as well as the nuisance parameters associated with the uncertainties due to the modelling of collinear gluon splitting, the ISR and FSR PS scale, and the ME-PS matching scheme are partially decorrelated between each of the five STXS categories and the other categories.
The decorrelation is implemented by assigning, per uncertainty, one parameter that acts on all categories and five additional parameters that act each on one of the STXS categories.
This effectively allows an overall variation in all categories simultaneously and additional independent variations per STXS category.

Furthermore, the renormalisation and factorisation scale uncertainty model of the \ttH signal process is extended, following Refs.~\cite{Stewart:2011cf,stxs_ttHbb_migration_uncertainties}.
The extended model comprises an uncertainty component in the inclusive cross section prediction as before, as well as uncertainties in the migration and acceptance effects of signal events between the STXS bins.
The uncertainties are constructed by studying the effect of all combinations of factorisation and renormalisation scale variations across each \ptH boundary on the total cross section above this boundary.
The largest effect is taken as the absolute uncertainty, which is propagated by increasing the process normalisation in the STXS bins above the \ptH boundary, and decreasing the process normalisation in the bin directly below the boundary under consideration of potential double-counting effects.
The migration uncertainties are implemented as rate uncertainties per STXS bin and amount to 5--10\% per parameter.
Acceptance effects within each STXS bin are taken into account by varying the renormalisation and factorisation scales, taking into account a normalisation factor to ensure that the variations do not change the overall \ttH cross section.
The resulting shape variation is typically 1--3\% and as large as up to 7\% depending on the classifier bin.
In addition, for the simulated \ttH signal events, the renormalisation and factorisation scale variations (\muR/\muF scale) are performed simultaneously, and, as well as the parton-shower uncertainties (PS scale ISR/FSR), are treated as uncorrelated for events with \ptH below and above 300\GeV.

The integrated luminosities for the 2016, 2017, and 2018 data-taking years have 1.2--2.5\% individual uncertainties~\cite{CMS-LUM-17-003,CMS-PAS-LUM-17-004,CMS-PAS-LUM-18-002}, which are partially correlated to account for common sources of uncertainty in the luminosity measurement. They amount to an overall uncertainty of 1.6\% for the 2016--2018 period.
The trigger efficiency uncertainty in the FH channel is determined from the bin-by-bin uncertainties in the ratio of efficiency in data relative to simulation, and are 1--2\% on average, with some being as large as 9\%.
The efficiencies of the single-electron and dilepton triggers are measured in data using reference triggers based on single-muon and \ptmiss requirements, respectively, that are uncorrelated with those used in the analysis; the uncertainties range up to 8\%, dominated by statistical fluctuations in the data samples used in the auxiliary measurement.
Uncertainties in the electron and muon reconstruction, identification, and isolation efficiency corrections are found to be small, typically at the 1\% level.
The uncertainty in the L1 trigger prefiring correction is determined from the uncertainty in the prefiring probability estimate and amounts to approximately 0.5\%.

Effects of the uncertainty in the distribution of the number of pileup interactions are evaluated by varying the total inelastic cross section used to predict the number of pileup interactions in the simulated events by $\pm4.6\%$ from its nominal value~\cite{CMS:2020ebo}.

The uncertainty related to the jet energy scale (resolution) is determined by varying the energy scale (resolution) correction of all jets in the signal and background predictions by one standard deviation.
The jet energy scale uncertainty is divided into 11 independent sources, which include uncertainties owing to the extrapolation between samples of different jet-flavour composition and the presence of pileup collisions in the derivation of the corrections, and which are treated as fully uncorrelated in the fit.
While most of the sources are dominated by statistical fluctuations in auxiliary measurements and are treated as uncorrelated among the data-taking periods, some sources are related to theoretical predictions in the MC simulation used, \eg to extrapolate between samples of different jet-flavour composition, and are thus treated as correlated among the data-taking periods~\cite{CMS:2016lmd}.

The \PQb tagging discriminant corrections receive uncertainties due to the contamination of background processes in the data samples used in the correction factor measurements, the jet energy scale uncertainty, and the statistical uncertainty in the correction factor evaluation~\cite{BTV-16-002}.
The impact of the statistical uncertainty is parameterised as the sum of two contributions:
one term with linear dependence on the \PQb tagging discriminant value, allowing an overall tilt of the discriminant distribution, and another term with quadratic dependence, allowing an overall shift of the discriminant distribution.
Each source of \PQb tagging uncertainty is considered separately per jet flavour.
The uncertainty related to the background contamination is treated as correlated among the data-taking periods of 2017 and 2018, and as uncorrelated with 2016 to allow for effects due to the upgraded pixel detector~\cite{Phase1Pixel}.
The statistical component is treated as uncorrelated among the data-taking periods.

Many uncertainties that are related to the MC simulation of the QCD multijet background in the FH channel are avoided by estimating this contribution from data.
Small uncertainties remain in the \tflm correction applied to the loosely \PQb-tagged jets, which is estimated by applying an additional $\eta$-dependent correction to \tflm to account for small effects of missing higher-order iterations in the correction procedure.
The total QCD background normalisation in each category is left unconstrained in the final fit.

The impact of statistical fluctuations in the signal and background prediction due to the limited sample size is accounted for using the Barlow--Beeston lite approach~\cite{Barlow:1993dm}.

The described sources of uncertainty are summarised in Table~\ref{tab:systematics:summary} and their impact on the final result is discussed in Section~\ref{sec:results}.
\begin{table}[!hbtp]
  \topcaption{
    Systematic uncertainties considered in the analysis.
    ``Type'' refers to rate (R) or rate and shape (S) altering uncertainties.
    ``Correlation'' indicates whether the uncertainty is treated as correlated, partially correlated (as detailed in the text), or uncorrelated across the years 2016--18.
    Uncertainties for \ttjets events marked with a ${}^{\dagger}$ are treated as partially correlated between each of the STXS categories and the other categories in the STXS analysis.
  }
  \label{tab:systematics:summary}
  \centering
  \cmsTable{
  \renewcommand{\arraystretch}{1.2}
    \begin{tabular}{lccp{0.43\textwidth}}
      Source                           & Type & Correlation        & Remarks \\
      \hline
      Renorm./fact. scales             & R & correlated & Scale uncertainty of (N)NLO prediction, independent for \ttH, \tHq, \tHW, \ttbar, t, \Vjets, \diboson \\
      PDF+\alpS ($\Pg\Pg$)                   & R & correlated & PDF uncertainty for $\Pg\Pg$ initiated processes, independent for \ttH, \tHq, \tHW, and others \\
      PDF+\alpS ($\PQq\PAQq$)                 & R & correlated & PDF uncertainty for $\PQq\PAQq$ initiated processes (\ttW,\PW,\PZ) except \tHq \\
      PDF+\alpS ($\PQq\Pg$)                  & R & correlated & PDF uncertainty for $\PQq\Pg$ initiated processes (single t) except \tHW \\[\cmsTabSkip]
      Collinear gluon splitting${}^{\dagger}$ & S & correlated & Additional 100\% rate uncertainty on \tttwob component of \ttB background \\[\cmsTabSkip]
      \muR scale       & S & correlated & Renormalisation scale uncertainty of the ME generator, independent for \ttH, \tHq, \tHW, \ttB (\ttbbsample sample), other \ttbar (\ttsample sample) \\
      \muF scale       & S & correlated & Factorisation scale uncertainty of the ME generator, independent for \ttH, \tHq, \tHW, \ttB (\ttbbsample sample), other \ttbar (\ttsample sample) \\
      PDF shape         & S & correlated & From NNPDF variations, independent for \tHq, \tHW, \ttB (\ttbbsample sample), other \ttbar (\ttsample sample) and \ttH \\
      PS scale ISR${}^{\dagger}$              & S & correlated & Initial state radiation uncertainty of the PS (\PYTHIA), independent for \ttH, \ttB, \ttC, \ttLF \\
      PS scale FSR${}^{\dagger}$             & S & correlated & Final state radiation uncertainty of the PS (\PYTHIA), independent for \ttH, \ttB, \ttC, \ttLF \\[\cmsTabSkip]
      ME-PS matching (\ttbar)${}^{\dagger}$       & R & correlated        & NLO ME-PS matching (for \ttjets events), independent for \ttB, \ttC, \ttLF \\
      Underlying event (\ttbar)   & R & correlated        & Underlying event (for all \ttjets events) \\[\cmsTabSkip]
      STXS migration  & R & correlated & Signal, only in STXS measurement \\
      STXS acceptance & S & correlated & Signal, only in STXS measurement \\[\cmsTabSkip]
      Integrated luminosity  & R & partially    & Signal and all backgrounds \\
      Lepton ID/Iso (2 sources)      & S & uncorrelated & Signal and all backgrounds \\
      Trigger efficiency (4 sources) & S & uncorrelated & Signal and all backgrounds \\
      L1 prefiring correction & S & uncorrelated & Signal and all backgrounds \\
      Pileup                           & S & correlated   & Signal and all backgrounds \\[\cmsTabSkip]
      Jet energy scale (11 sources)  & S & partially    & Signal, \ttjets and single~t \\
      Jet energy resolution & S & uncorrelated & Signal, \ttjets and single~t \\[\cmsTabSkip]
      \PQb tag bkg.\ contam.\ (2 sources)          & S & partially   & Signal and all backgrounds \\
      \PQb tag bkg.\ contam.\ stat.\ (4 sources) & S & uncorrelated & Signal and all backgrounds \\
      \PQb tag charm (2 sources)      & S & partially & Signal and all backgrounds \\[\cmsTabSkip]
      \tflm correction                 & S & uncorrelated & QCD multijet background estimate \\[\cmsTabSkip]
      Size of the MC samples & S & uncorrelated & Statistical uncertainty of signal and background prediction due to limited sample size \\
    \end{tabular}
  }
\end{table}

\section{Statistical analysis and results}
\label{sec:results}
The production rates of the \ttH and \tH signal processes are
determined in a simultaneous binned profile likelihood fit to the
final discriminant distributions in all channels, categories, and
data-taking periods, using the techniques detailed
in Ref.~\cite{CMS:2024onh}.
The rates of the \ttB, \ttC, as well as the QCD multijet background, are separately left unconstrained in the fit.
Several signal interpretations are performed and described below.

\subsection{The \texorpdfstring{\ttH}{ttH} production rate}
First, the \ttH production rate is measured.
For this interpretation, the \tH contribution is assumed to conform to
the SM expectation and is treated as background.

\begin{figure}[!hbtp]
  \centering
  \includegraphics[width=0.9\textwidth]{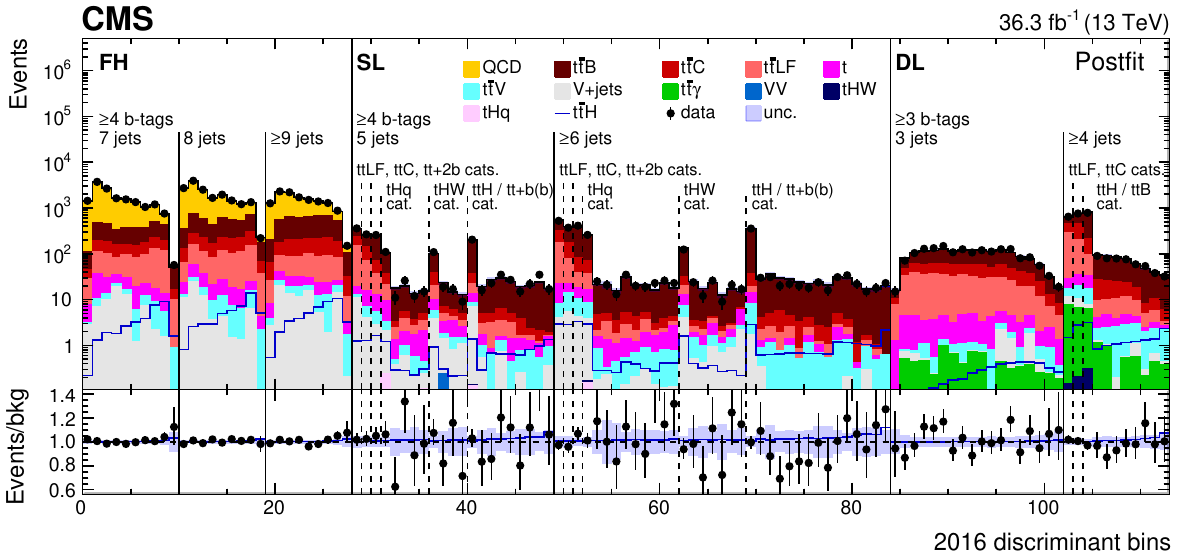}
  \includegraphics[width=0.9\textwidth]{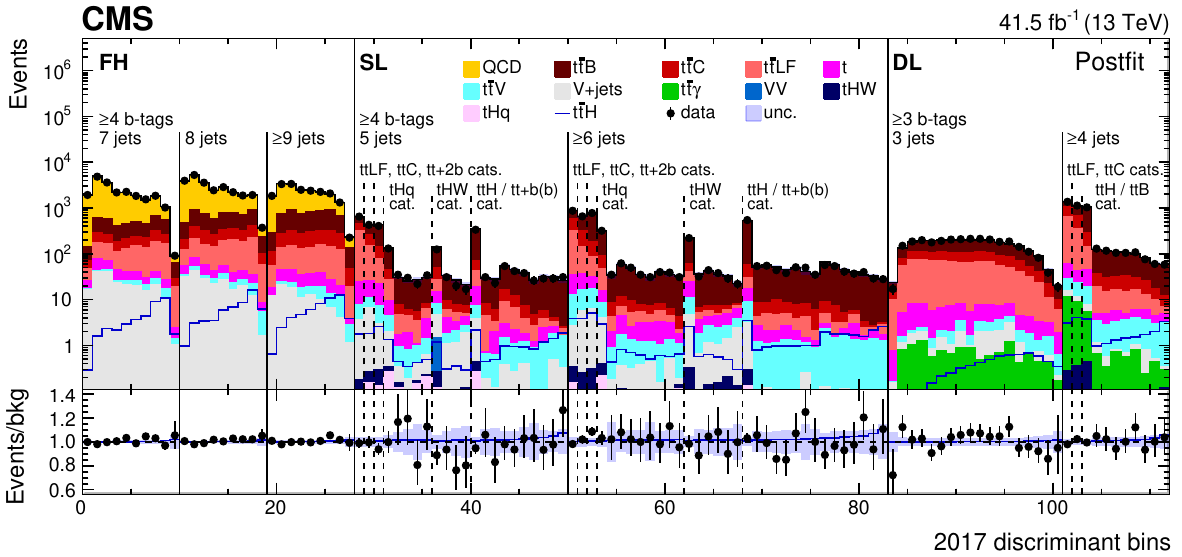}
  \includegraphics[width=0.9\textwidth]{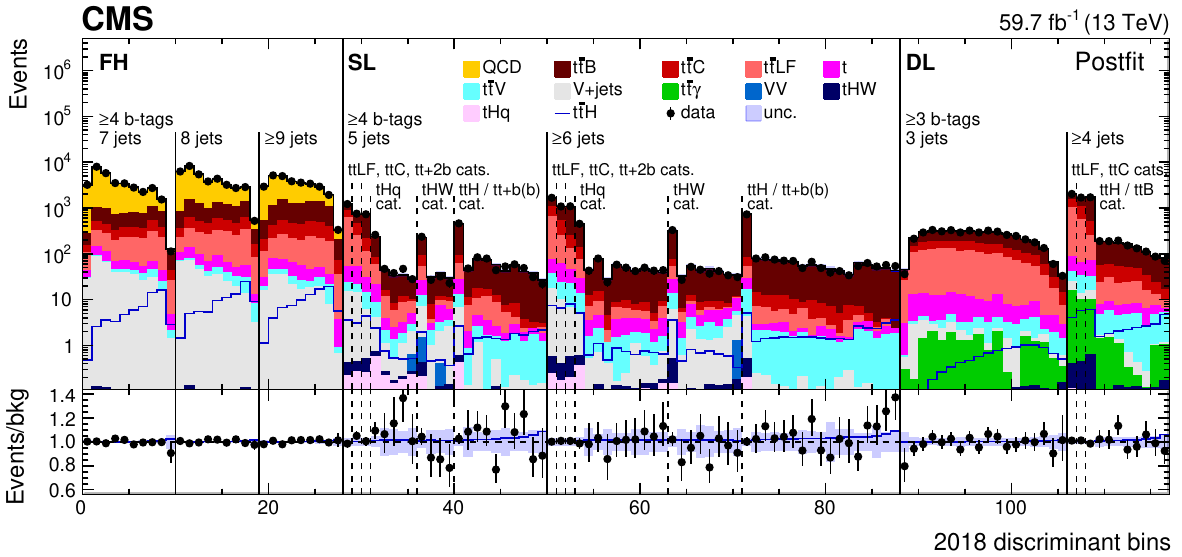}
  \caption{
    Observed (points) and postfit expected (filled histograms) yields in each discriminant (category yield, ANN score, or ratio of ANN scores) bin for the 2016 (upper), 2017 (middle), and 2018 (lower) data-taking periods.
    The uncertainty bands include the total uncertainty of the fit model.
    The lower pads show the ratio of the data to the background (points) and of the postfit expected signal+background to the background-only contribution (line).
  }
  \label{fig:discriminants:tth}
\end{figure}

\begin{figure}[!htbp]
  \centering
  \includegraphics[width=0.55\textwidth]{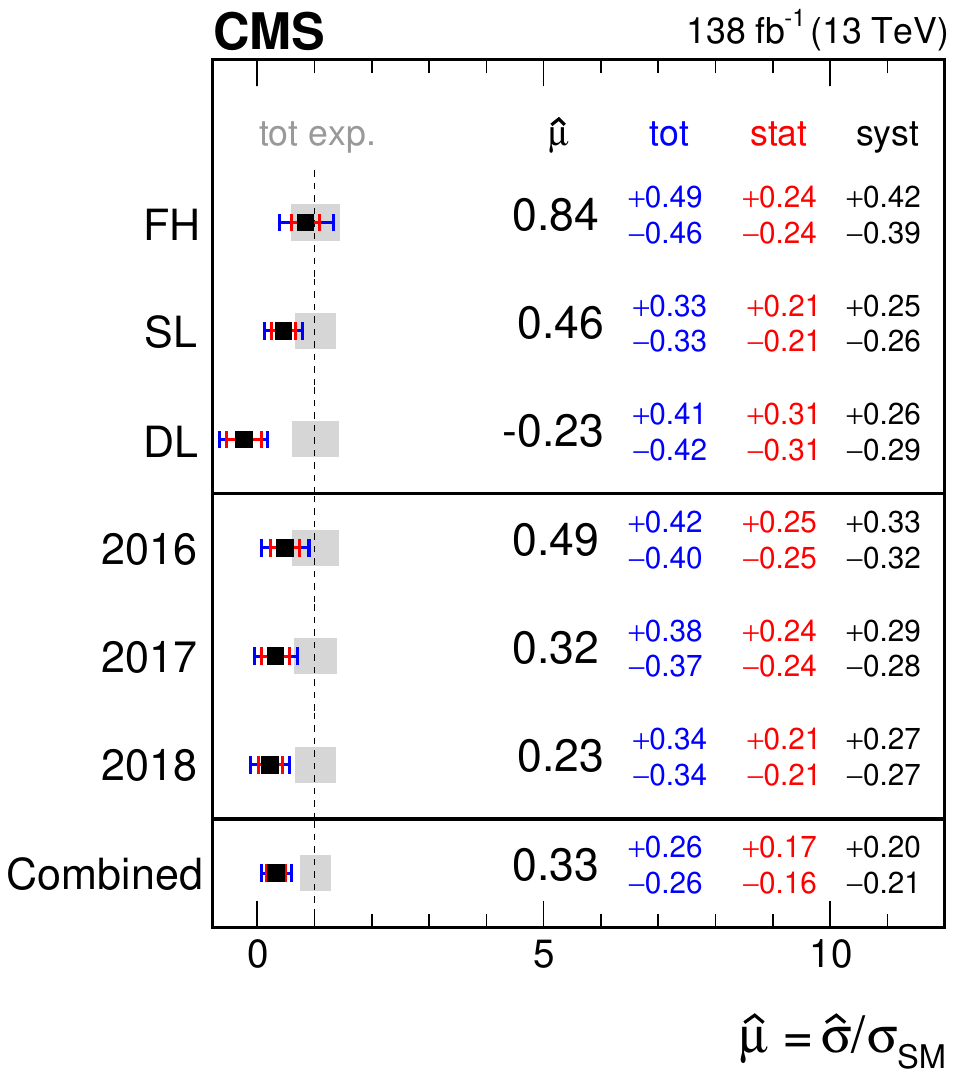}
  \caption{
    Best fit results \muhat of the \ttH signal-strength modifier \muttH in
    each channel (upper three rows), in each year (middle three rows),
    and in the combination of all channels and years (lower
    row). Uncertainties are correlated between the channels and
    years.
  }
  \label{fig:mutth}
\end{figure}

The event yields observed in data in the different analysis channels and categories, together with the signal and background yields after the fit to data, are listed in Table~\ref{tab:postfityields}.
\begin{table}[!htp]
  \topcaption{
    Event yields observed in data and expected for the \ttH signal and total background after the fit to data in the different analysis categories for all data-taking periods together.
    The uncertainties denote the total uncertainty of the fit model.
    Contributions from \ttjets and \tH background processes are also listed individually.
  }\label{tab:postfityields}
  \centering
  \renewcommand{\arraystretch}{1.4}
  \begin{tabular}{llrrrrr@{$\,\pm\,$}rr@{$\,\pm\,$}rr}
  \multicolumn{2}{l}{Categories} & \ttLF & \ttC & \ttB & \tH & \multicolumn{2}{r}{ Total bkg.\ } & \multicolumn{2}{r}{ \ttH } & Data \\
  \hline
  \multicolumn{11}{l}{{\bfseries FH} $\geq$4\,{\PQb}\,tags} \\
  & 7\,jets       & $      2980$ & $      2790$ & $      8100$ & $         2$ & $     69950$ & $  300$ & $       134$ & $   99$ & $     70070$\\ 
  & 8\,jets       & $      3510$ & $      3970$ & $     11850$ & $         2$ & $     78960$ & $  360$ & $       190$ & $  140$ & $     79179$\\ 
  & $\geq$9\,jets & $      2460$ & $      3850$ & $     13200$ & $      0.98$ & $     60950$ & $  410$ & $       180$ & $  140$ & $     61193$\\[\cmsTabSkip]
  \multicolumn{11}{l}{{\bfseries SL} 5\,jets, $\geq$4\,{\PQb}\,tags} \\
  & \ttLF         & $       996$ & $       494$ & $       667$ & $     0.689$ & $      2254$ & $   88$ & $         7$ & $    5$ & $      2235$\\ 
  & \ttC          & $       377$ & $       374$ & $       564$ & $      0.80$ & $      1408$ & $   56$ & $         6$ & $    5$ & $      1449$\\ 
  & \tttwob       & $       161$ & $       196$ & $       984$ & $     0.885$ & $      1406$ & $   44$ & $         9$ & $    7$ & $      1396$\\ 
  & \tHq          & $       102$ & $        83$ & $       536$ & $         3$ & $       790$ & $   26$ & $        10$ & $    8$ & $       846$\\ 
  & \tHW          & $        64$ & $       100$ & $       446$ & $         1$ & $       697$ & $   29$ & $         7$ & $    5$ & $       665$\\ 
  & \ttH/\ttborbb & $       125$ & $       139$ & $      1555$ & $         3$ & $      1930$ & $   63$ & $        35$ & $   26$ & $      1965$\\[\cmsTabSkip]
  \multicolumn{11}{l}{{\bfseries SL} $\geq$6\,jets, $\geq$4\,{\PQb}\,tags} \\
  & \ttLF         & $      1020$ & $       820$ & $      1133$ & $         1$ & $      3100$ & $  120$ & $        14$ & $   11$ & $      3074$\\ 
  & \ttC          & $       347$ & $       562$ & $      1078$ & $      0.71$ & $      2108$ & $   78$ & $        13$ & $   10$ & $      2114$\\ 
  & \tttwob       & $       161$ & $       252$ & $      1680$ & $      0.97$ & $      2188$ & $   75$ & $        17$ & $   13$ & $      2284$\\ 
  & \tHq          & $       146$ & $       201$ & $      1471$ & $         3$ & $      1946$ & $   55$ & $        25$ & $   19$ & $      1992$\\ 
  & \tHW          & $        65$ & $        82$ & $       952$ & $         3$ & $      1244$ & $   39$ & $        16$ & $   13$ & $      1217$\\ 
  & \ttH/\ttborbb & $       101$ & $       197$ & $      3020$ & $         3$ & $      3460$ & $  110$ & $        74$ & $   56$ & $      3507$\\[\cmsTabSkip]
  \multicolumn{11}{l}{{\bfseries DL} 3\,jets, 3\,{\PQb}\,tags} \\
  & & $      2500$ & $      1700$ & $      3980$ & $         3$ & $      8510$ & $  160$ & $        20$ & $   16$ & $      8594$\\[\cmsTabSkip]
  \multicolumn{11}{l}{{\bfseries DL} $\geq$4\,jets, $\geq$3\,{\PQb}\,tags} \\
  & \ttLF     & $      1760$ & $      1090$ & $      1017$ & $     0.816$ & $      4045$ & $   79$ & $         9$ & $    6$ & $      4050$\\ 
  & \ttC      & $       970$ & $      1100$ & $      1325$ & $     0.995$ & $      3534$ & $   74$ & $        13$ & $   10$ & $      3586$\\ 
  & \ttH/\ttB & $       538$ & $       890$ & $      4270$ & $         4$ & $      5980$ & $  120$ & $        57$ & $   41$ & $      5931$\\ 
\end{tabular}
\renewcommand{\arraystretch}{1}
\end{table}
The observed yields in each bin of the final discriminant
distributions in all channels and categories entering the fit are
shown in Fig.~\ref{fig:discriminants:tth}, together with the fitted
signal and background yields.
The best fit values of the inclusive \ttH production rate relative to
the SM expectation, denoted as the signal strength modifier \muttH,
are presented in Fig.~\ref{fig:mutth}.
Results are shown for fits performed simultaneously in all channels and years
using either one signal-strength modifier per channel or per year and correlating
the uncertainties, or using one overall signal-strength modifier.
For the overall signal-strength modifier, a best fit value of $\muttH
  = \resttHMu\pm\resttHMuUpDn =
  \resttHMu\,{}^{+\resttHMuStatUp}_{-\resttHMuStatDn}\stat{}^{+\resttHMuSystUp}_{-\resttHMuSystDn}\syst$
is obtained, with an expected uncertainty of $\pm\resttHMuExpStatUpDn\stat{}^{+\resttHMuExpSystUp}_{-\resttHMuExpSystDn}\syst$.
The observed signal has a significance compared to the background-only
hypothesis corresponding to \resttHSigObs\,\SD, with an
expectation of \resttHSigExp\,\SD.
The goodness-of-fit is quantified using a $p$-value that takes into account the postfit uncertainty model~\cite{CMS:2024onh,Lindsey1996-nh} and amounts to $p=\resttHgof$, indicating good description of the data by the fit model.

\begin{figure}[!btp]
  \centering
  \includegraphics[width=0.51\textwidth]{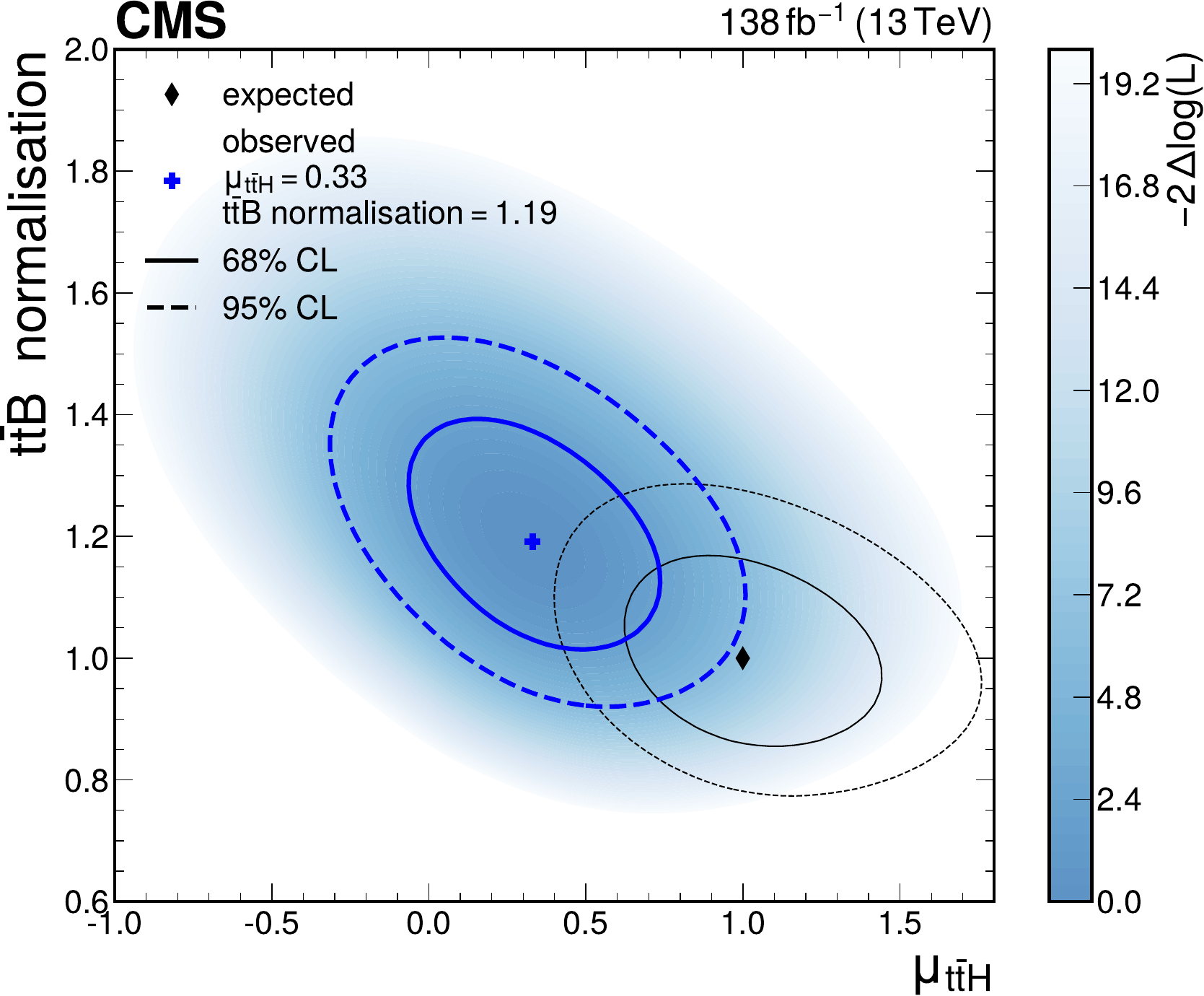}
  \caption{
    Observed likelihood-ratio test statistic (blue shading) as a
    function of the \ttH signal-strength modifier \muttH and the \ttB
    background normalisation, together with the observed (blue) and SM
    expected (black) best fit points (cross and diamond markers) as
    well as the 68\% (solid lines) and 95\% (dashed lines) \CL
    regions. The \ttC background normalisation and all other nuisance
    parameters are profiled such that the likelihood attains its
    minimum at each point in the plane.
  }
  \label{fig:tth:nllcontour}
\end{figure}

The postfit values and uncertainties of the \ttB and \ttC background
normalisation parameters obtained in the combined fit of all channels
are \mbox{$\resttHttB\,{}_{-\resttHttBDn}^{+\resttHttBUp}$} and
\mbox{$\resttHttC\,{}_{-\resttHttCDn}^{+\resttHttCUp}$}, respectively. 
These values are consistent with the results of dedicated inclusive
$\ttbar\bbbar$ and $\ttbar\ccbar$ cross section measurements in
similar phase space regions, which are at the level of 6--20\% relative
precision depending on the exact phase space and the size of the analysed dataset~\cite{CMS:2023xjh,Sirunyan:2020kga,Sirunyan:2019jud,CMS:2020utv}, and
are consistent with the known underprediction of the $\ttbar\bbbar$ cross
section by the simulation.
The anticorrelation between the \ttH signal strength and the \ttB
background normalisation is 48\% and
is visible in Fig.~\ref{fig:tth:nllcontour}.

\begin{figure}[hbtp]
  \centering
 \includegraphics[width=0.75\textwidth]{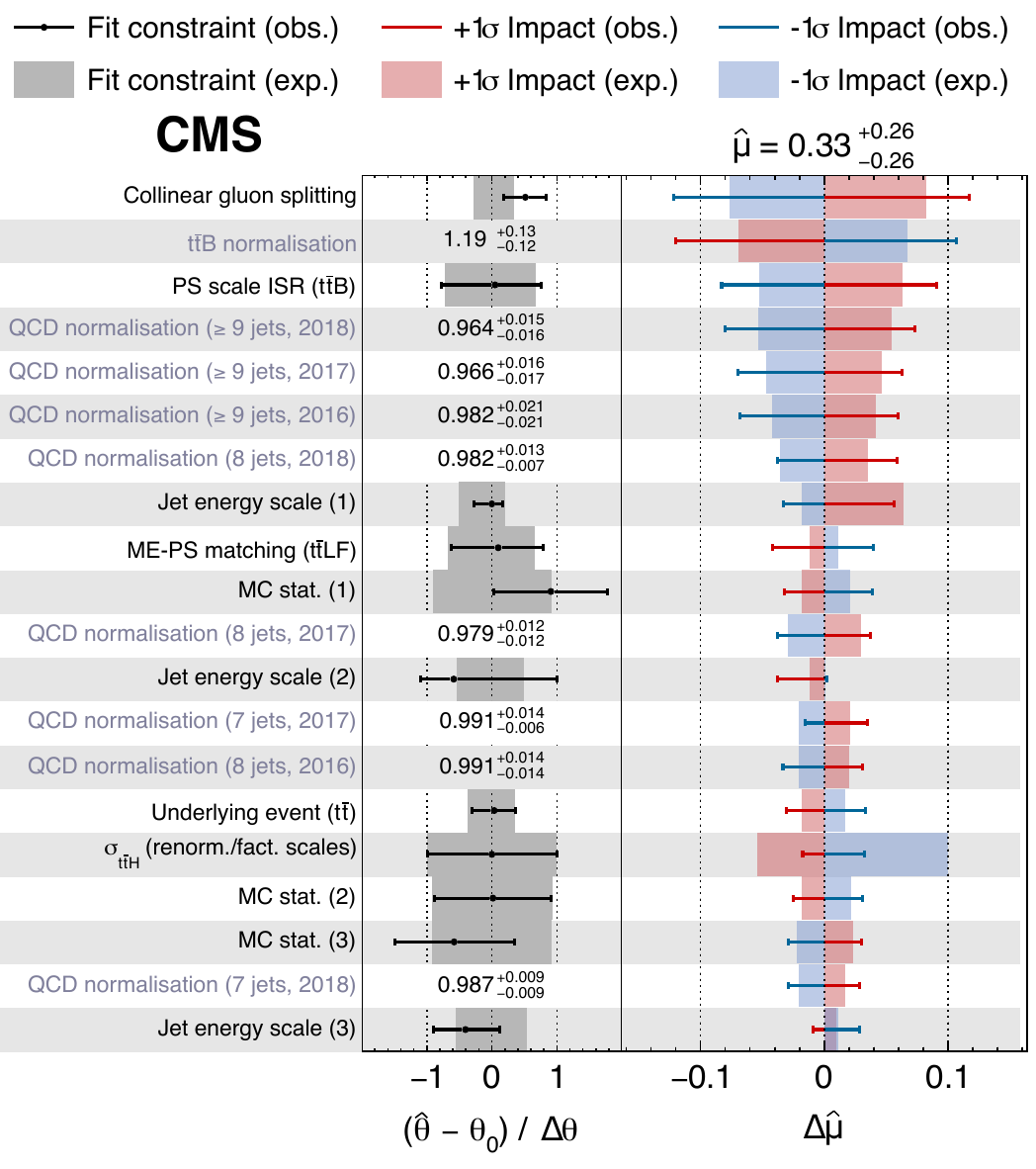}
  \caption{
    Postfit values of the nuisance parameters (black markers), shown
    as the difference of their best fit values, $\hat{\theta}$, and
    prefit values, $\theta_{0}$, relative to the prefit uncertainties
    $\Delta\theta$. The impact $\Delta\hat{\mu}$ of the nuisance
    parameters on the signal strength \muttH is computed as the
    difference of the nominal best fit value of $\mu$ and the best fit
    value obtained when fixing the nuisance parameter under scrutiny
    to its best fit value $\hat{\theta}$ plus/minus its postfit
    uncertainty (coloured areas). The nuisance parameters are ordered
    by their impact, and only the 20 highest ranked parameters are
    shown.
  }
  \label{fig:impacts:tth}
\end{figure}

The best fit values and the impacts of the 20 nuisance parameters ranked highest in impact are presented in Fig.~\ref{fig:impacts:tth}.
The impact of each nuisance parameter is evaluated as the difference of the nominal best fit value of $\mu$ and the best fit value obtained when fixing the nuisance parameter under scrutiny to its best fit value plus/minus its postfit uncertainty.
The nuisance parameters with the highest impact are related to the
\ttB background modelling, followed by the QCD multijet background
normalisation and, to a lesser extent, the jet energy scale, the
uncertainty in the signal cross section ($\sigma_{\ttH}$), and the
statistical uncertainty in the signal and background prediction due to
the number of simulated events (MC stat.).
The three highest ranked nuisance parameters of the jet energy scale uncertainty are related to the \pt dependence of the jet energy scale and resolution of the forward jets (1) as well as to the uncertainty in the response obtained from the in-situ calibration of the absolute jet energy scale (2, 3).

The best fit values of the nuisance parameters are within 1\,\SD of
the prior uncertainty for more than 93\% of the total number of
nuisance parameters.
As expected, significant shifts from the nominal value are observed for nuisance parameters related to the ttB background modelling, such as the \ttB normalisation and the gluon-splitting uncertainty, since the a-priori knowledge does not reflect the data distributions.
Furthermore, the values of the nuisance parameters related to the \muR scale uncertainty of the \ttsample sample used to model the \ttLF and \ttC processes as well as the uncertainty in the PS modelling (PS scale FSR) of the \ttLF process are shifted by $+1\,\SD$ and $-1\,\SD$, respectively; the nuisance parameters are not shown in Fig.~\ref{fig:impacts:tth} since their impact is lower than those of the shown parameters.
Consequently, the fit constrains these nuisance parameters
relative to their prior values. 
Several other nuisance parameters, in particular those related to jet
energy scale and \PQb tagging uncertainties, are constrained.
This is attributed to the following: events are selected
according to different, large multiplicities of jets and \PQb-tagged
jets, thus increasing the sensitivity of the analysis to changes of
the jet energy scale and \PQb tagging efficiency, for example by
their effect on the event yield per analysis category; and in
several cases the prior uncertainties are large.
No large constraints are observed in the measurement of the remaining nuisance parameters.

The contributions of the statistical and various groups of systematic
uncertainties to the uncertainty in \muttH are listed in
Table~\ref{tab:uncertaintygroups}.
The statistical uncertainty is evaluated by fixing all nuisance
parameters to their postfit values and repeating the fit.
The contribution by a group of systematic uncertainties is evaluated by
repeating the fit fixing only the nuisance parameters related to the
uncertainty under scrutiny to their postfit values, and subtracting
the uncertainty obtained in quadrature from the total uncertainty of
the fit in which no parameters are fixed.
The uncertainties obtained in the fit to data agree with the
expectation from simulation.
The total uncertainty of the full fit on the signal strength ($\pm\resttHMuUpDn$) is different from the quadratic
sum of the listed contributions because of correlations between the
nuisance parameters.
The statistical uncertainty also includes components
from the background normalisations.
The total contribution from the systematic uncertainties ($+\resttHMuSystUp$/$-\resttHMuSystDn$) is larger than
from the statistical uncertainties ($+\resttHMuStatUp$/$-\resttHMuStatDn$), which include the uncertainties due to the freely-floating background normalisations, albeit at a similar level.
The theoretical uncertainties amount to $\pm0.16$ and are dominated by the uncertainties of the \ttjets background modelling.
Experimental uncertainties amount to $\pm0.10$, dominated by the
jet energy scale and resolution as well as \PQb
tagging related uncertainties.
Systematic uncertainties due to the size of the various simulated samples used to model the background and signal templates are at the same order and amount to $+0.13$/$-0.12$.
\begin{table}[!htp]
  \centering
  \topcaption{
    Contributions of different sources of uncertainty to the result
    for the fit to the data (observed) and to the expectation from
    simulation (expected).
    The quoted uncertainties $\Delta\muttH$ in \muttH are obtained by
    fixing the listed sources of uncertainties to their postfit
    values in the fit and subtracting the obtained result in
    quadrature from the result of the full fit.
    The statistical uncertainty is evaluated by fixing all nuisance
    parameters to their postfit values and repeating the fit.
    The quadratic sum of the contributions is different from the total uncertainty because of correlations between the nuisance parameters.
  }
  \label{tab:uncertaintygroups}
  \renewcommand{\arraystretch}{1.4}
  \begin{tabular}{lcc}
    Uncertainty source & $\Delta\muttH$ (observed) & $\Delta\muttH$ (expected) \\
    \hline
    Total experimental                       & $+0.10/-0.10$ & $+0.11/-0.10$ \\
    \hspace{15pt}jet energy scale and resolution & $+0.08/-0.07$ & $+0.09/-0.09$ \\
    \hspace{15pt}\PQb tagging                & $+0.07/-0.06$ & $+0.06/-0.02$ \\
    \hspace{15pt}integrated luminosity       & $+0.02/-0.02$ & $+0.01/-0.01$ \\[\cmsTabSkip]
    Total theory                             & $+0.16/-0.16$ & $+0.18/-0.14$ \\
    \hspace{15pt}\ttjets background          & $+0.15/-0.16$ & $+0.12/-0.11$ \\
    \hspace{15pt}signal modelling            & $+0.06/-0.01$ & $+0.13/-0.06$ \\[\cmsTabSkip]
    Size of the simulated event samples      & $+0.13/-0.12$ & $+0.10/-0.10$ \\[\cmsTabSkip]
    Total systematic                         & $+0.20/-0.21$ & $+0.23/-0.19$ \\[\cmsTabSkip]
    Statistical                              & $+0.17/-0.16$ & $+0.17/-0.17$ \\
    \hspace{15pt}background normalisation    & $+0.13/-0.13$ & $+0.13/-0.13$ \\
    \hspace{15pt}\ttB and \ttC normalisation & $+0.12/-0.12$ & $+0.12/-0.12$ \\
    \hspace{15pt}QCD normalisation           & $+0.01/-0.01$ & $+0.01/-0.01$ \\[\cmsTabSkip]
    Total                                    & $+0.26/-0.26$ & $+0.28/-0.25$ \\
  \end{tabular}
  \renewcommand{\arraystretch}{1.0}
\end{table}

The single-channel best fit results for \muttH are compatible with the combined result at
a level corresponding to a $p$-value of \pvalueInclToComb (\sigmaInclToComb\,\SD), and the
$p$-value compatibility of the combined result with the SM expectation
of $\muttH=1$ is \pvalueInclToSM (\sigmaInclToSM\,\SD).
The result obtained in the combination of the leptonic channels with
the 2016 data only is compatible to the central value of the earlier result in
Ref.~\cite{Sirunyan:2018mvw} obtained from a similar dataset at a $p$
value of \pvalueLeptonicToPaper (\sigmaLeptonicToPaper\,\SD).
For the $p$-value computation, the postfit uncertainty model was taken into account in each case.
The result was extensively validated through several methods, including fitting different observables like MEM and \HT, testing the stability of the background model with various combinations of categories, decorrelating uncertainties between different \ttB components (such as those with one or two \PQb jets within acceptance), and using background predictions derived from data.
The performed cross checks confirm the observed result, albeit at a lower precision.
The value is very close to the result reported by the ATLAS Collaboration in Ref.~\cite{ATLAS:2021qou}, but lower than the one reported in Ref.~\cite{ATLAS:2024gth}.
The latter includes several changes with respect to Ref.~\cite{ATLAS:2021qou}, most notably, an updated treatment of the \ttjets background with a revised \ttbbsample simulation and corresponding systematic uncertainty model.

\begin{figure}[!hbtp]
  \centering
  \includegraphics[width=0.9\textwidth]{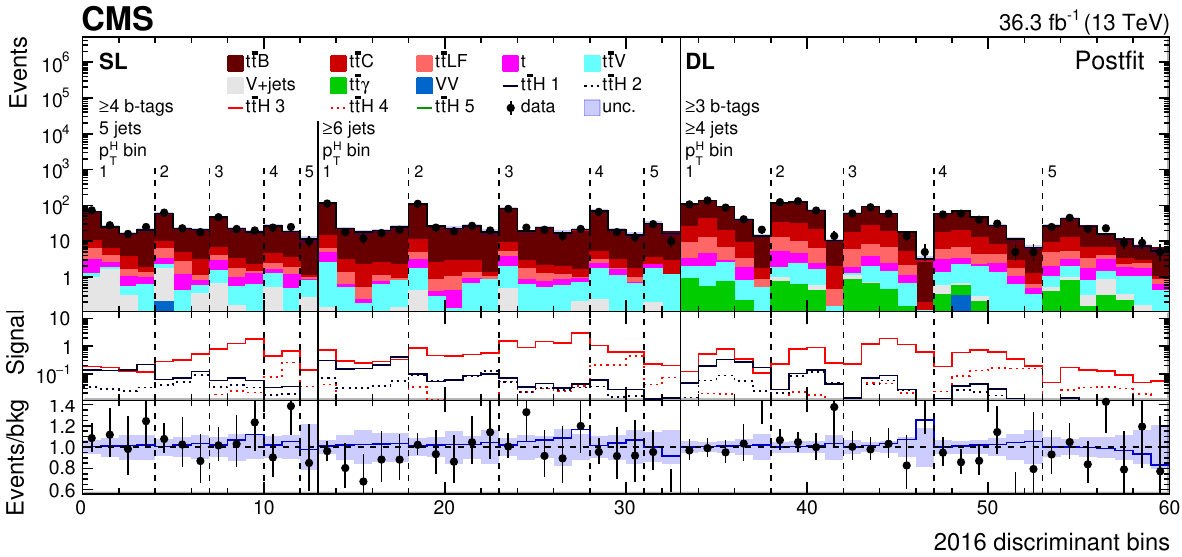}
  \includegraphics[width=0.9\textwidth]{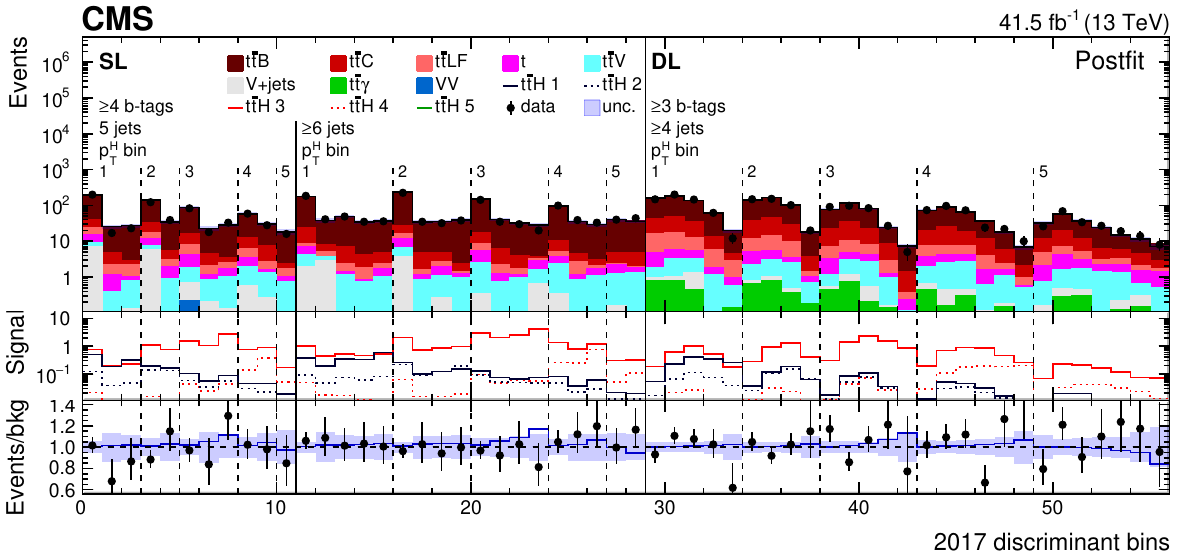}
  \includegraphics[width=0.9\textwidth]{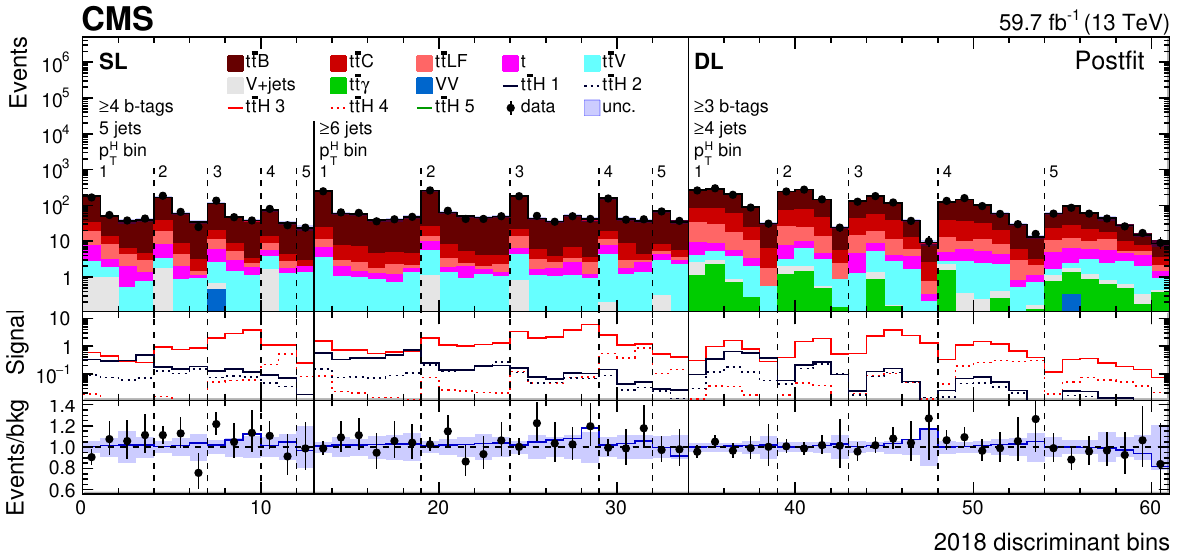}
  \caption{
    Observed (points) and postfit expected (filled histograms) yields
    in each STXS analysis discriminant bin in the signal regions of
    the SL and DL channels for the 2016 (upper), 2017 (middle), and 2018
    (lower) data-taking periods.
    The vertical dashed lines separate the STXS categories (labelled 1
    to 5).
    The fitted signal distributions (lines labelled \ttH 1 to 5) in
    each \ptH bin are shown in the middle pads.
    The lower pads show the ratio of the data to the background
    (points) and of the postfit expected total signal+background to the
    background-only contribution (line).
    The uncertainty bands include the total uncertainty of the fit model.
  }
  \label{fig:discriminants:stxs}
\end{figure}

\begin{figure}[!htbp]
  \centering
  \includegraphics[width=0.50\textwidth]{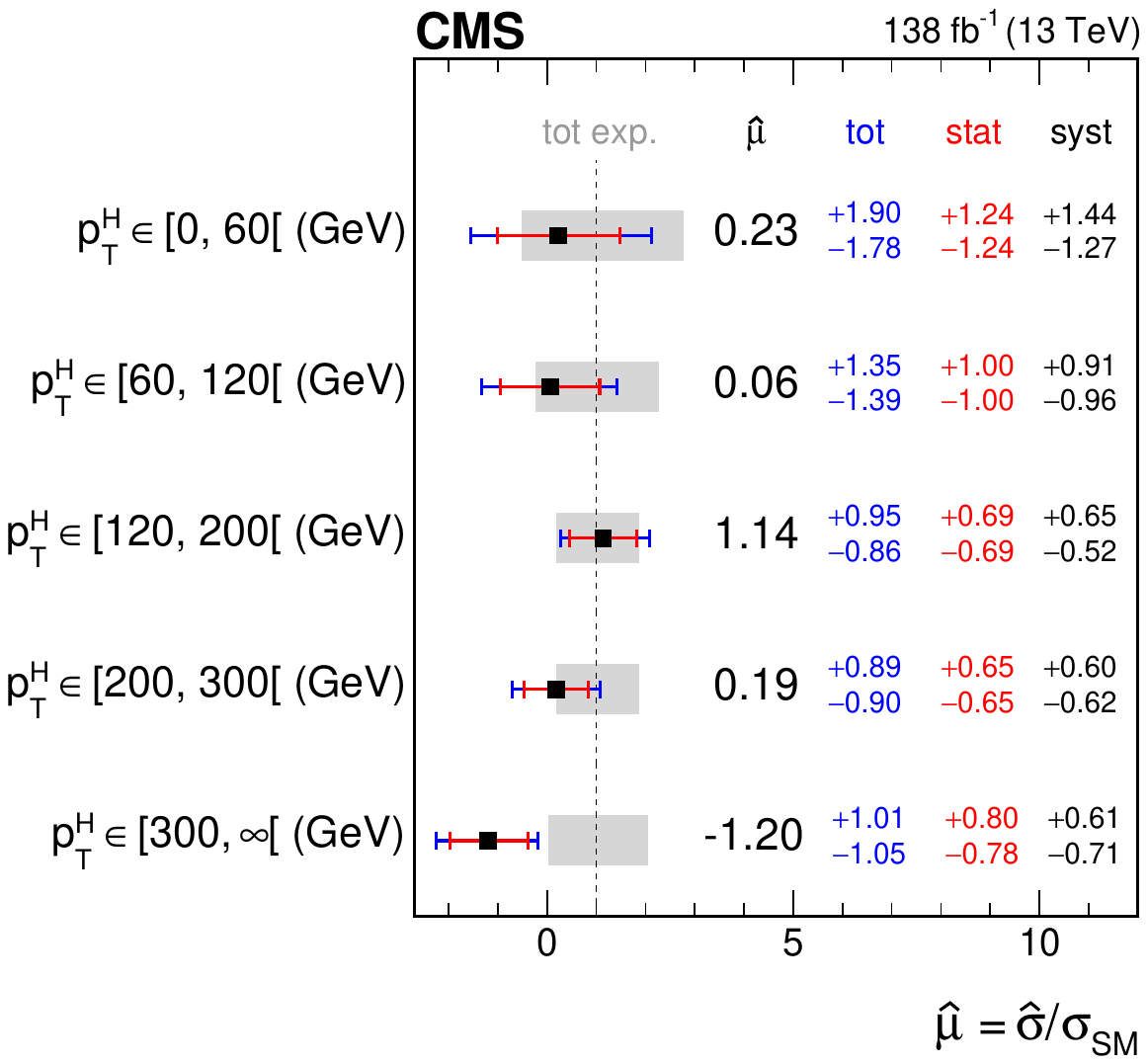}
  \includegraphics[width=0.47\textwidth]{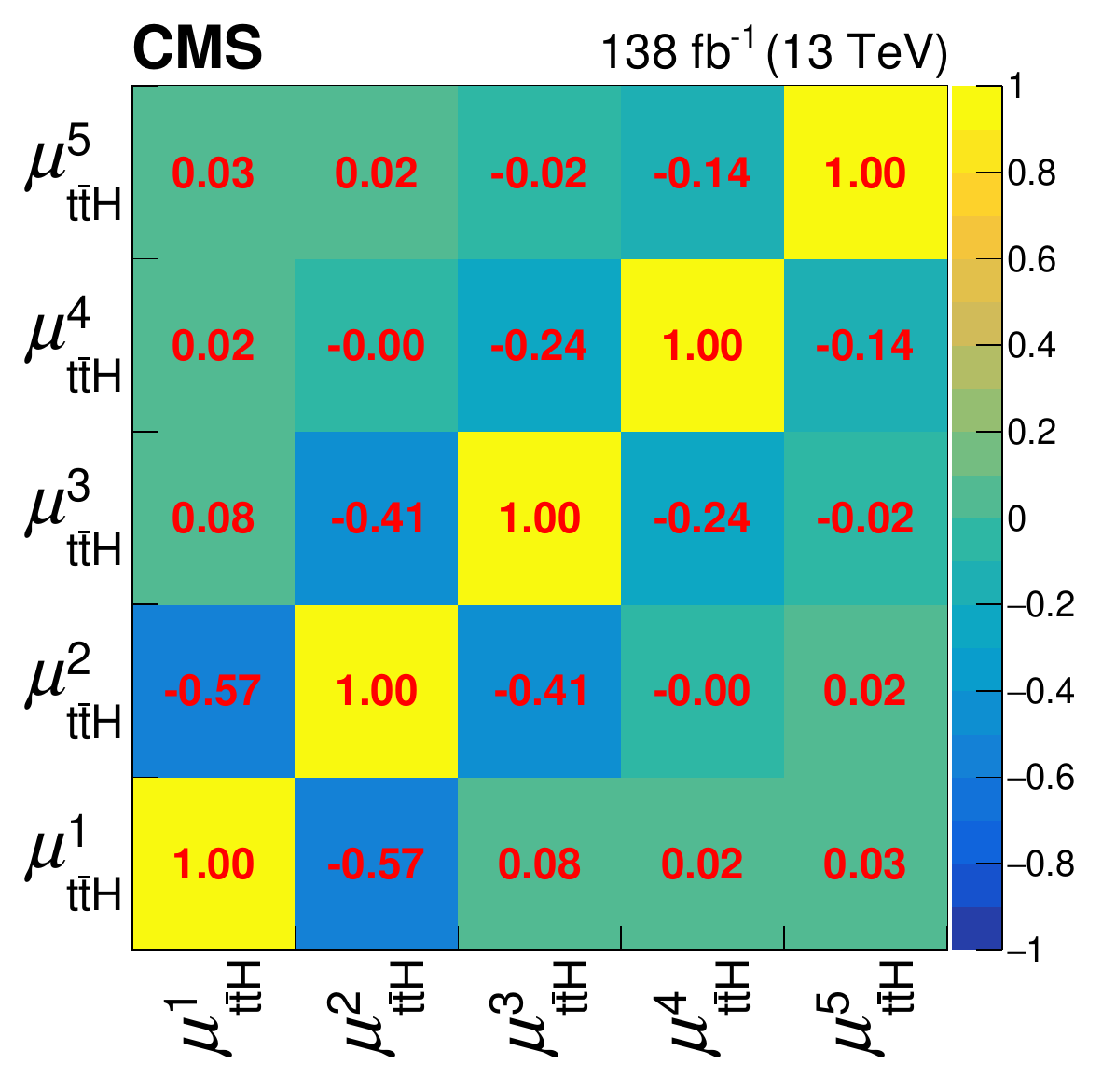}
  \caption{
    Best fit results \muhat of the \ttH signal-strength modifiers \muttH in
    the different \ptH bins (left) and their correlations (right) of the STXS measurement.
  }
  \label{fig:stxs}
\end{figure}

The observed yields in each bin of the final discriminant
distributions of the STXS measurement in the signal regions of the SL
and DL channels, which drive the signal sensitivity, are shown in Fig.~\ref{fig:discriminants:stxs}, together with the fitted signal and background yields.
The results of the STXS measurement are presented in
Fig.~\ref{fig:stxs}, which shows the best fit values of the \ttH signal-strength
modifier per region of \ptH and their correlations, obtained
in the combination of all channels and years.
The goodness-of-fit $p$-value is \resSTXSgof, indicating good description of the data by the fit model.
The highest expected sensitivity is reached in the medium-\ptH range
between 120 and 300\GeV.

The measured best fit values are compatible with the central result of the
inclusive \ttH production rate measurement, corresponding to an overall $p$-value of
\pvalueSTXStoIncl (\sigmaSTXStoIncl\,\SD).
This $p$-value takes into account the correlations between
the \pt bins shown in Fig.~\ref{fig:stxs}.
The post fit values of the \ttB normalisation parameters are compatible with the value obtained in the inclusive result.
As an additional compatibility test, a combined fit of the STXS signal
templates with a single parameter of interest was performed, resulting
in a best fit value that agrees within 3\% to that of the inclusive
result with uncertainties that are larger by 20\%.
The larger uncertainty is expected and an effect of the additional \ptH dependent uncertainties of the \ttB background model in
the STXS fit described in Section~\ref{sec:systematics}.

The compatibility with the SM expectation has a $p$-value of \pvalueSTXStoSM (\sigmaSTXStoSM\,\SD).
No significant trend in \ptH compared to the SM expectation is observed.
The largest deviation from the SM expectation is observed for
$\pt>300\GeV$ with a local significance of approximately 2\,\SD.

\subsection{The \texorpdfstring{\tH}{tH} production rate}
\begin{figure}[!htbp]
  \centering
  \includegraphics[width=0.6\textwidth]{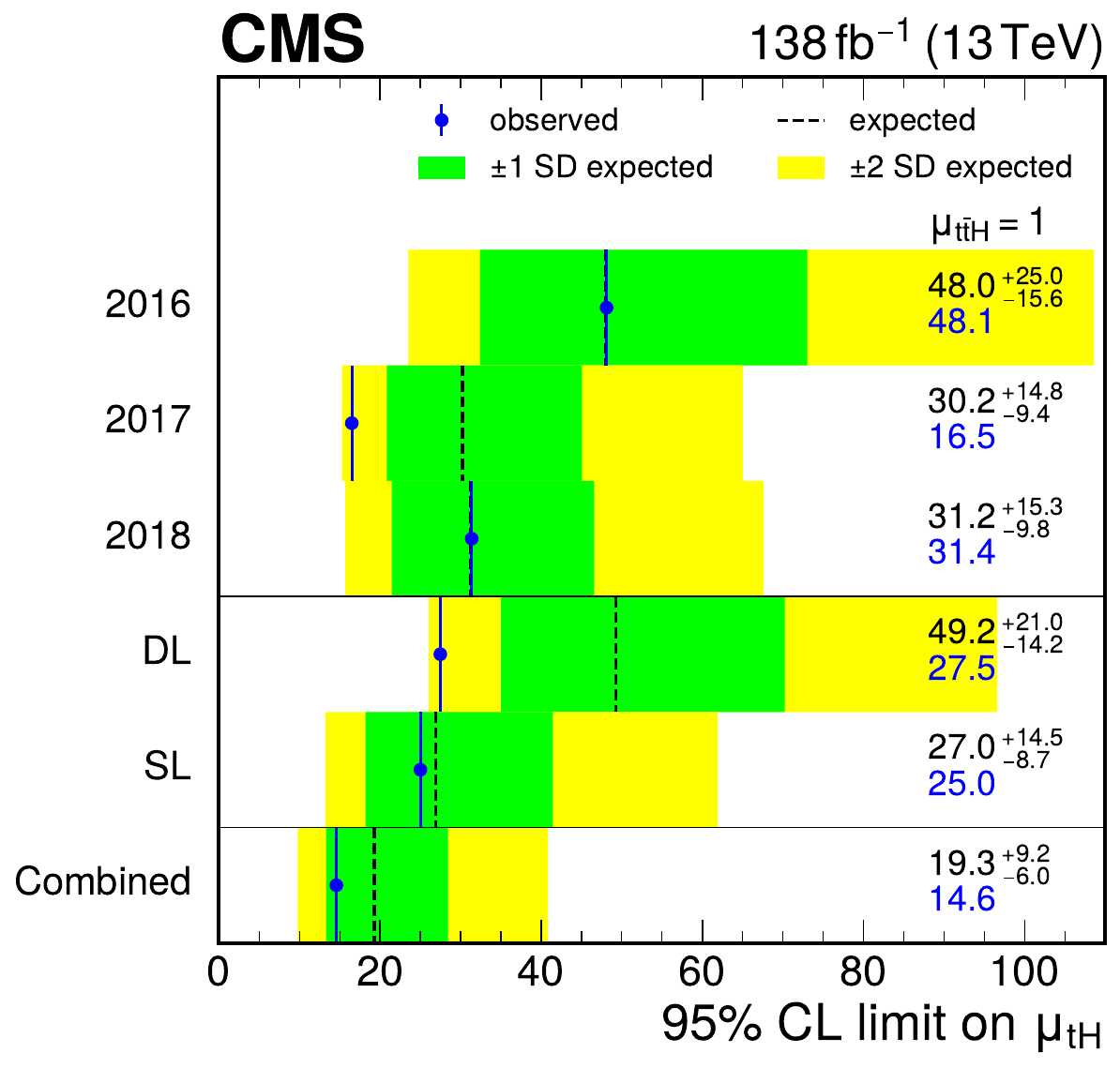}
  \caption{
    Observed (solid vertical line) and expected (dashed vertical line)
    upper 95\%~\CL limit on the \tH signal strength modifier \mutH for
    different channels and years, where the uncertainties are
    uncorrelated between the channels and years, and in their
    combination.
    The green (yellow) areas indicate the one (two) standard deviation
    confidence intervals on the expected limit.
  }
  \label{fig:results:tH:limit}
\end{figure}

Second, the \tH signal process is targeted.
Here, the \ttH contribution is assumed to conform to the SM
expectation ($\muttH=1$) and is treated as background. 
An upper limit at 95\%~\CL on the \tH production rate relative to the
SM expectation, denoted as the signal strength modifier \mutH, of
\restHLimit is obtained, with an expectation of
$\restHExpLimit^{+\restHExpLimitUp}_{-\restHExpLimitDn}$.
The limits per channel and year and in their combination are
shown in Fig.~\ref{fig:results:tH:limit}.
The expected limit for 2017 is slightly lower than the one for 2018
despite the higher luminosity of the 2018 dataset, driven by a
slightly larger selection efficiency for \ttH and \ttjets background
processes in 2018.
The systematic uncertainties with the highest impact are related to the \ttB
background modelling, the jet energy scale, and the uncertainty in the
\tH signal cross section.
When fixing \muttH to the measured value of \resttHMu, an upper limit on \mutH of 16.1 is obtained.

\begin{figure}[!htbp]
  \centering
  \includegraphics[width=0.51\textwidth]{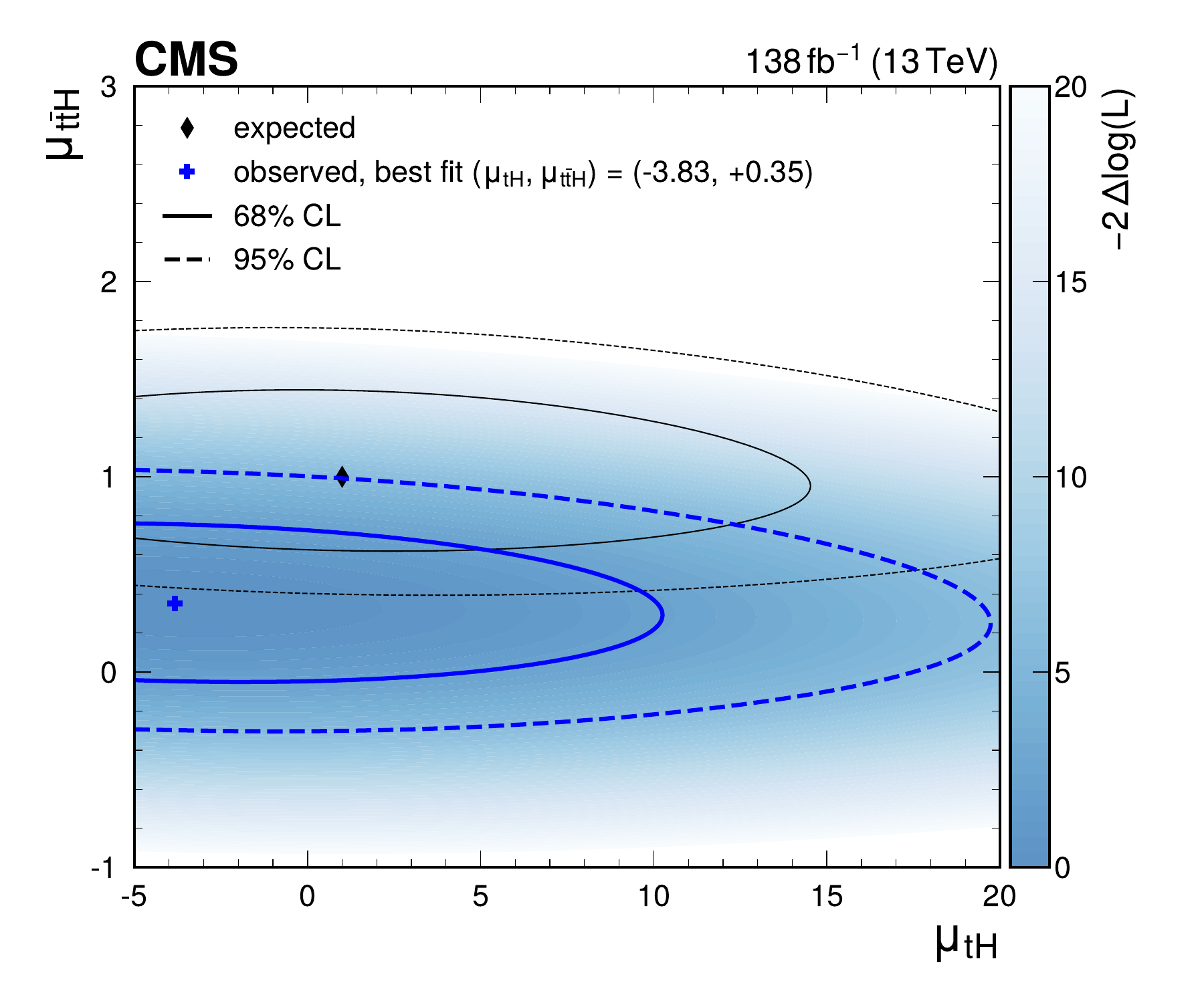}
  \caption{
    Observed likelihood-ratio test statistic (blue shading) as a
    function of the \ttH and \tH signal strength modifiers \muttH and
    \mutH, together with the observed (blue) and SM expected (black)
    best fit points (cross and diamond markers) as well as the 68\% (solid lines) and
    95\% (dashed lines) \CL regions. 
  }
  \label{fig:results:correlation:ttHtH}
\end{figure}

Furthermore, a simultaneous fit of the \ttH and \tH signal strength
modifiers is performed. The observed and expected values of the
likelihood-ratio test statistic are shown in
Fig.~\ref{fig:results:correlation:ttHtH}, with best fit values of
$(\muttH,\mutH)$ of $(\resTwoDttH,\resTwoDtH)$.
The correlation between the \ttH and \tH signal strength modifiers is
of moderate size ($-15\%$), which demonstrates the
discrimination between the two signal processes achieved in this
analysis.

\subsection{Coupling measurement}
Third, the Higgs boson coupling is analysed in different models, where
both \ttH and \tH are treated as signal.
Variations in the kinematic properties of the signal events are taken into account via event weights in the simulated samples, as described in Section~\ref{sec:mcsamples}.

Assuming SM Higgs boson coupling structure, the coupling strength
of the Higgs boson to top quarks and to vector bosons is allowed to
vary.
This is parameterised in terms of the coupling strength modifiers
\kappat and \kappaV, which denote the coupling strengths relative to
the SM expectation following Ref.~\cite{Heinemeyer:2013tqa}.
While the \ttH production rate is proportional to $\kappat^{2}$, for
\tH production interference occurs between processes in which the
Higgs boson couples to the top quark or to the \PW boson, as shown in
Fig.~\ref{fig:intro:productiondiagrams}.
As such, the \tH production cross section $\sigma_{\tHq/\tHW}$ is
sensitive to the relative sign of \kappat and \kappaV:
\begin{equation}\label{eq:tHxs}
  \begin{aligned}
    \sigma_{\tHq} &= \left( 2.63\kappat^2 +3.58\kappaV^2 -5.21\kappat\kappaV\right) \sigma^{\text{SM}}_{\tHq},\\
    \sigma_{\tHW} &= \left( 2.91\kappat^2 +2.31\kappaV^2 -4.22\kappat\kappaV\right)  \sigma^{\text{SM}}_{\tHW}.
  \end{aligned}
\end{equation}

\begin{figure}[!htbp]
  \centering
    \includegraphics[width=0.51\textwidth]{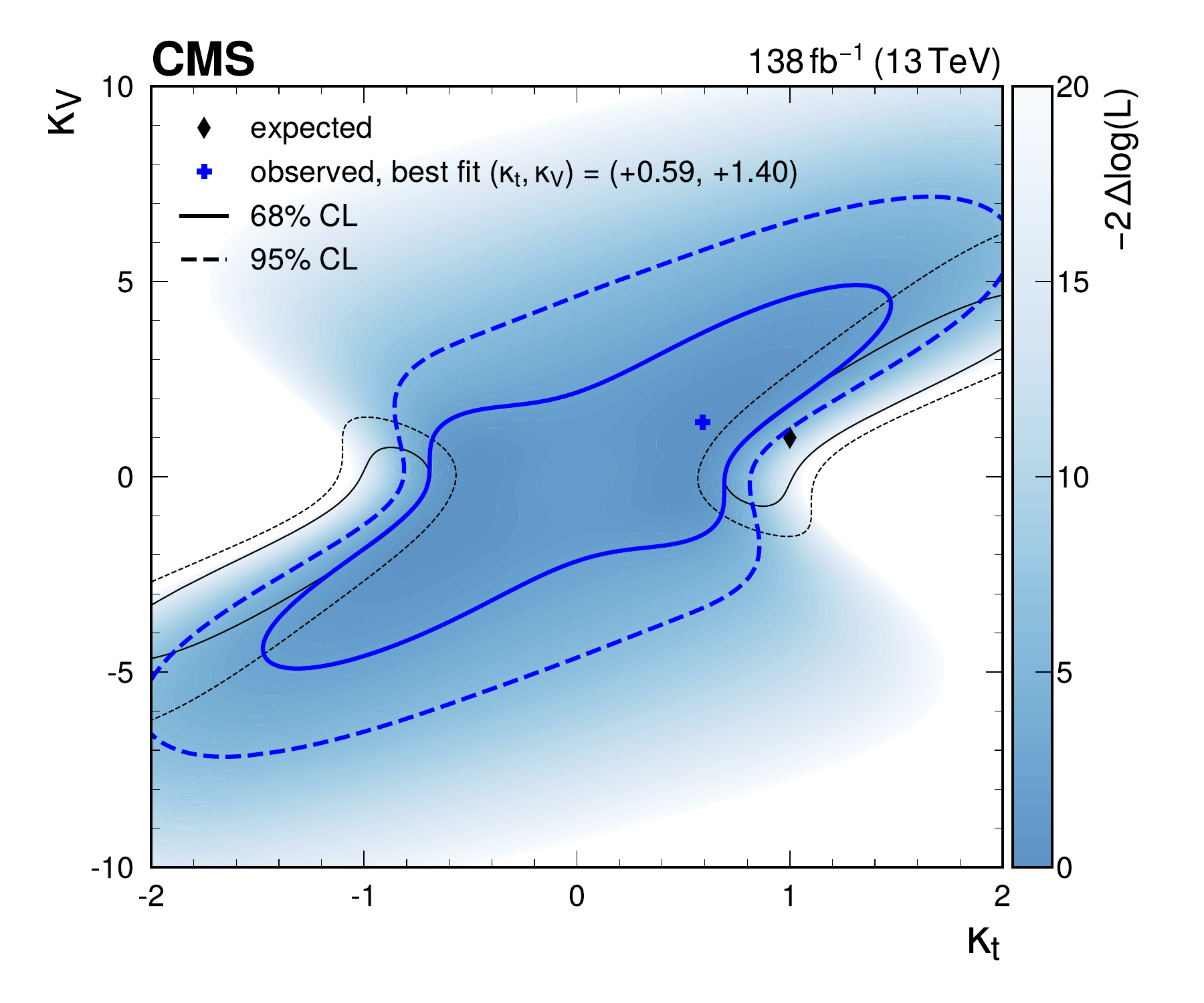}
    \includegraphics[width=0.43\textwidth]{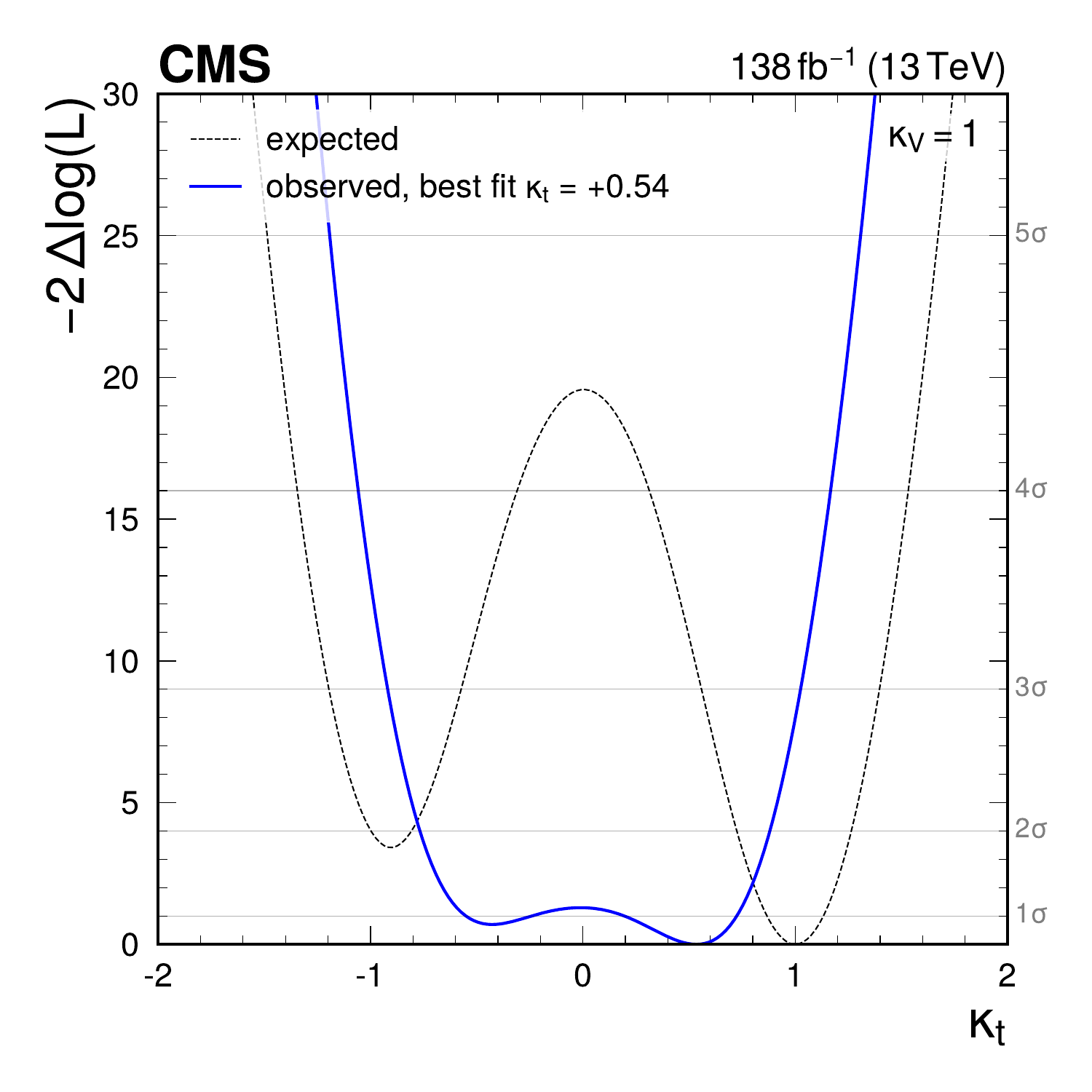}
  \caption{
    Observed likelihood ratio test statistic (blue shading) as a
    function of \kappat and \kappaV, together with the observed
    (blue) and SM expected (black) best fit points (cross and diamond markers) as well
    as the 68\% (solid lines) and 95\% (dashed lines) \CL regions (left).
    The observed (solid blue line) and expected (dotted black line)
    values of the likelihood ratio for $\kappaV=1$ are also shown
    (right). 
  }
  \label{fig:results:kappa:SMcoupling}
\end{figure}

The observed and expected values of the likelihood ratio test statistic for
different values of \kappat and \kappaV are shown in
Fig.~\ref{fig:results:kappa:SMcoupling}.
Best fit values of $(\kappat,\kappaV)$ of $(\resktkVkt,\resktkVkV)$ are
observed, compatible with the SM expectation at the level of 2.4\,\SD.
Assuming $\kappaV=1$, a best fit value of $\kappat=\reskt$ is obtained with the 68\% \CL region ranging from \resktNegLo to \resktNegHi and \resktPosLo to \resktPosHi.

Furthermore, the \CP structure of the top-Higgs coupling is probed for
potential non-SM contributions.
For this, the amplitude of the top-Higgs interaction is parameterised
as in Ref.~\cite{Gritsan:2016hjl} as     
\begin{equation}
\mathcal{A}(\PH\PQt\PAQt) = - \frac{m_{\PQt}}{v} \overline{\psi}_{\PQt} \left ( \kappa_{\PQt}  + \text{i}  \kappatcp  \gamma_{5} \right ) {\psi}_{\PQt},
\end{equation}
where $\psi_{\PQt}$ and $\overline{\psi}_{\PQt}$ are a Dirac spinor
and its adjoint, respectively, $m_{\PQt}$ is the top quark mass,
\kappat and \kappatcp denote the coupling strength modifiers to a
purely \CP-even and a purely \CP-odd component, respectively, and $v$ is
the vacuum expectation value of the Higgs field.
In the SM, $\kappat = 1$ and $\kappatcp = 0$.
The same interference model as in Eq.~\eqref{eq:tHxs} is taken into account for the \CP-even top-Higgs coupling.

\begin{figure}[!htbp]
  \centering
  \includegraphics[width=0.51\textwidth]{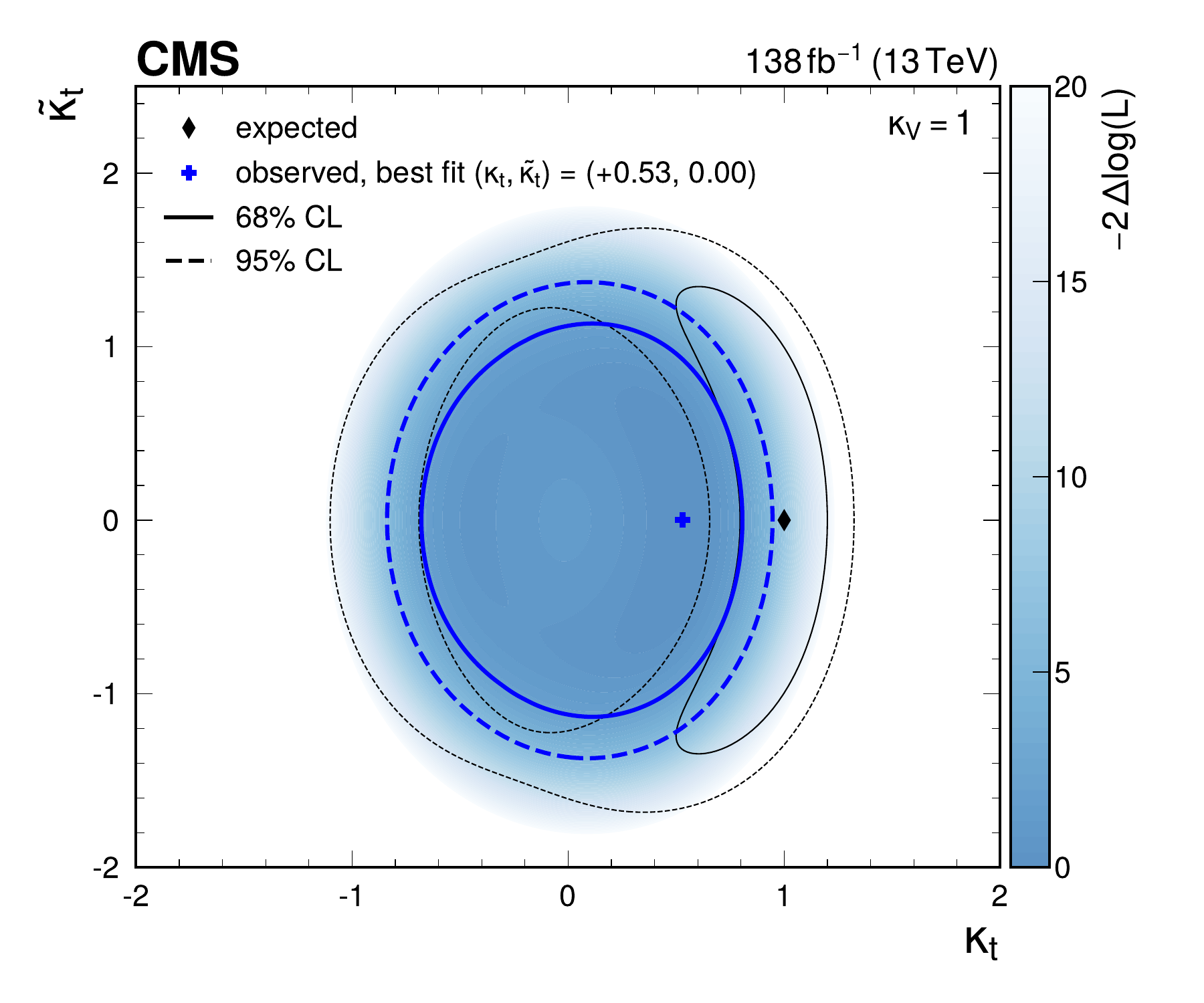}
  \caption{
    Observed likelihood ratio test statistic (blue shading) as a
    function of \kappat and \kappatcp, where $\kappaV=1$, together with the observed
    (blue) and SM expected (black) best fit points (cross and diamond markers) as well
    as the 68\% (solid lines) and 95\% (region between dashed lines) \CL regions.
  }
  \label{fig:results:kappa:NonSMcoupling}
\end{figure}

\begin{figure}[!htbp]
    \includegraphics[width=0.48\textwidth]{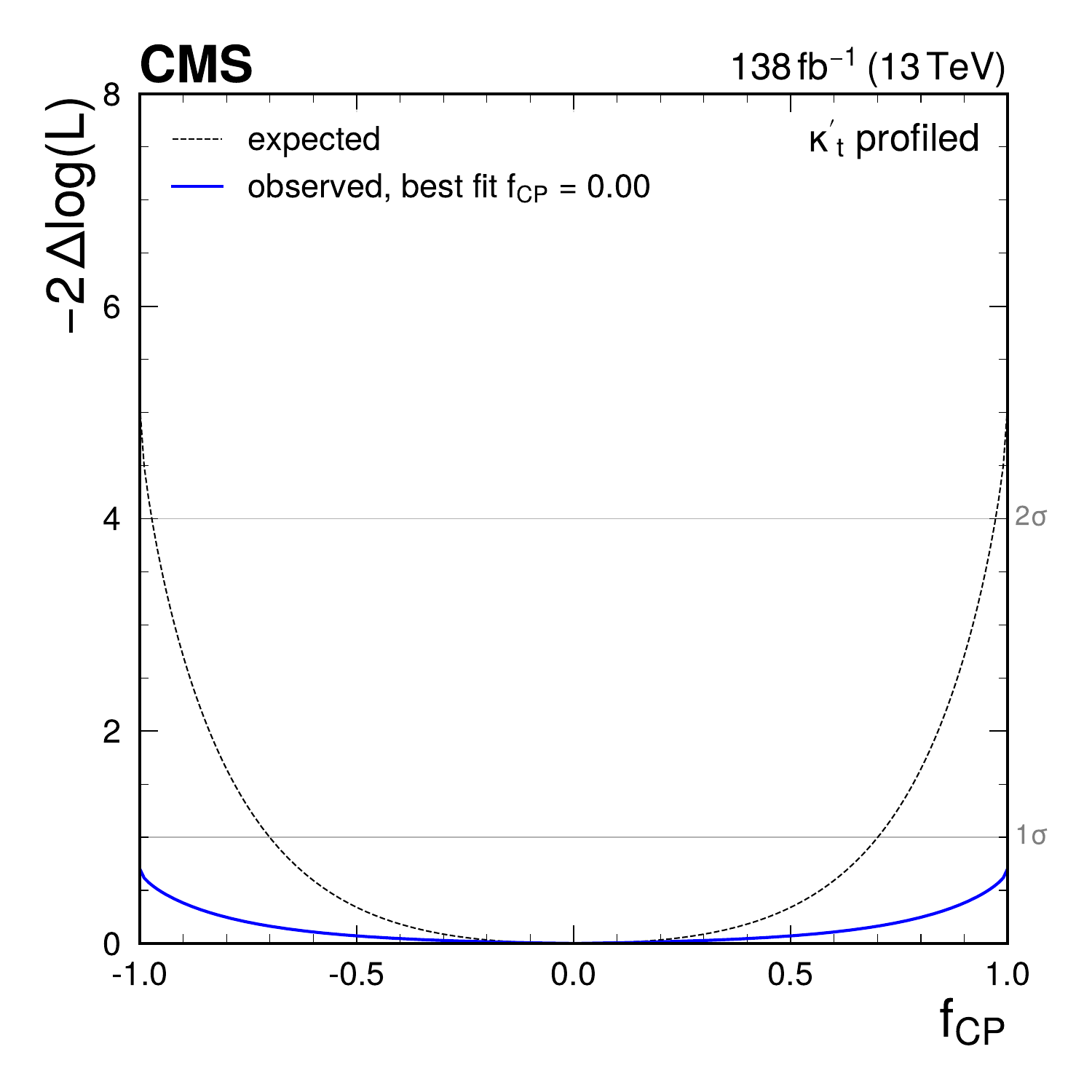} 
    \includegraphics[width=0.48\textwidth]{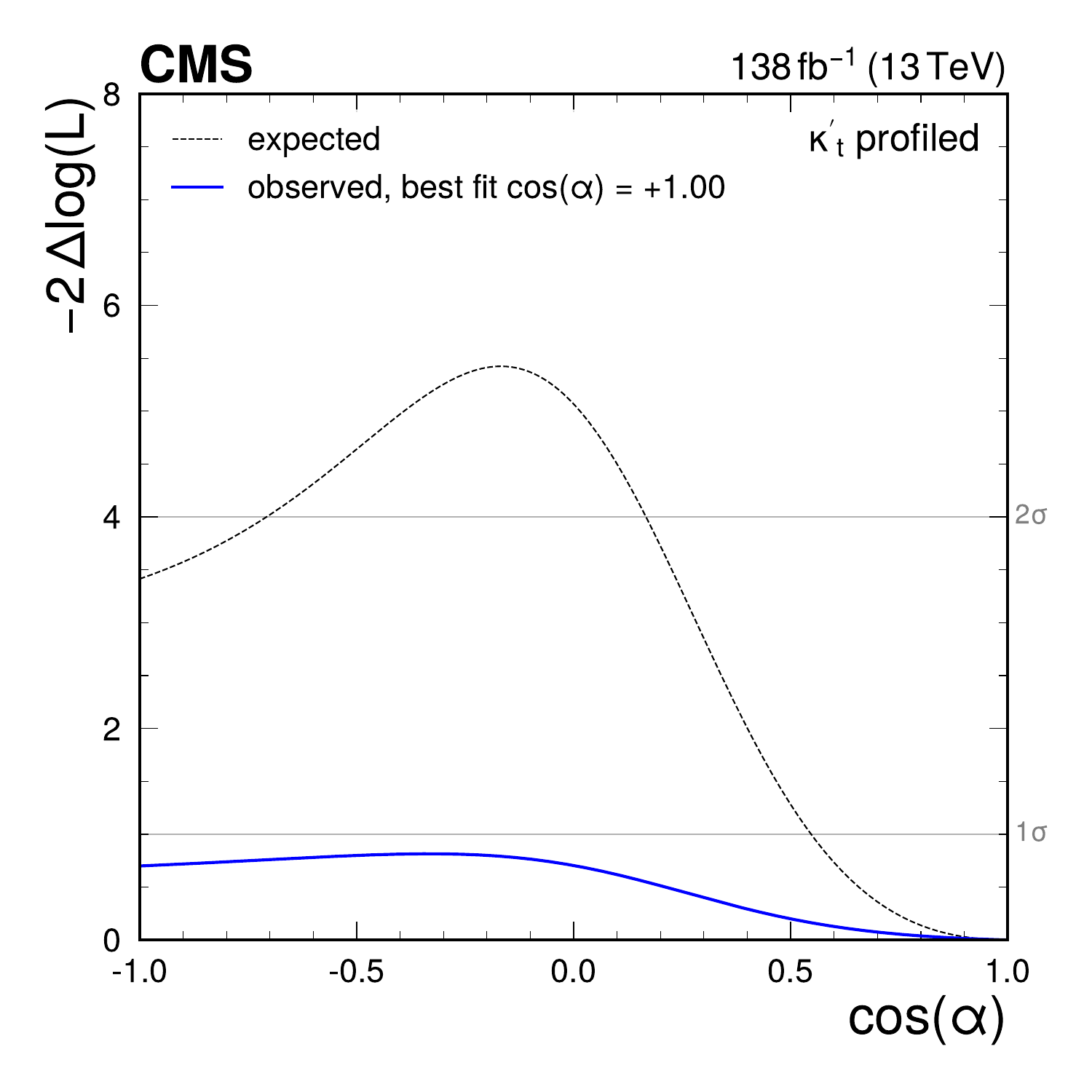} 
  \caption{
    Observed (solid blue line) and expected (dotted black line)
    likelihood ratio test statistic as a function of \fCP (left) and
    \cosa (right), where \kappaV is 1 and
    $\kappatprime=\sqrt{\smash[b]{\kappatcp^{2}+\kappat^{2}}}$, the overall
    modifier of the top-Higgs coupling strength, is profiled.  
  }
  \label{fig:results:kappa:fCPcosa}
\end{figure}

Figure~\ref{fig:results:kappa:NonSMcoupling} shows the observed and expected values
of the likelihood ratio test statistic as a function of \kappat and
\kappatcp, where \kappaV is fixed to the SM value of 1.
Best fit values of $(\kappat,\kappatcp)$ of $(\resktcpeven,\resktcpodd)$ are
observed, compatible with the SM expectation at the level of 2\,\SD.
The results are also expressed in terms of the \CP-odd fraction~\cite{Gritsan:2016hjl}
\begin{equation}
\fCP =
       \frac{\kappatcp^{2}}{\kappatcp^{2}+\kappat^{2}}\sign\left(\kappatcp/\kappat\right)
\end{equation}
as well as the \CP mixing angle~\cite{Demartin:2014fia}
\begin{equation}
\cosa = \frac{\kappat}{\sqrt{\smash[b]{\kappatcp^{2}+\kappat^{2}}}}
\end{equation}
shown in Fig.~\ref{fig:results:kappa:fCPcosa}.
The reduction in observed sensitivity relative to the expectation is a
consequence of the best fit value of \kappat being smaller than 1,
which leads to a shallower likelihood contour along a circle in
$(\kappat,\kappatcp)$ space.

\subsection{Combination of coupling measurement with results in other channels}
The results for \kappat and \kappatcp are combined with the results of
\ttH measurements by the CMS Collaboration in the
$\PH\to\PGg\PGg$~\cite{CMS:2020cga} and $\PH\to\PZ\PZ$~\cite{CMS:2021nnc}, and the $\PH\to\PW\PW$ and $\PH\to\PGt\PGt$~\cite{CMS:2022dbt} decay channels.
Potential overlap of events selected in the different analyses is considered negligible, due to the lepton and \PQb-tagged jets selection requirements, as well as the used multivariate classifiers.
Systematic uncertainties are correlated between the decay channels
following the scheme described in Ref.~\cite{CMS:2022dbt}, where also a
combination of the results in the aforementioned non-\bbbar channels is presented.
In particular,
theoretical uncertainties on the rates of signal and background processes estimated from simulation as well as uncertainties in the acceptance due to missing higher orders are treated as correlated;
uncertainties in the background rates determined from dedicated control regions in data are taken as uncorrelated;
and uncertainies in the luminosity estimate and the pileup modelling are correlated.
Several of the components of the experimental uncertainties related to the \PQb tagging and the jet energy scale are common between the channels and are correlated, while the other components as well as all other experimental uncertainties are analysis specific and therefore treated as uncorrelated.

\begin{figure}[!htbp]
  \centering
  \includegraphics[width=0.65\textwidth]{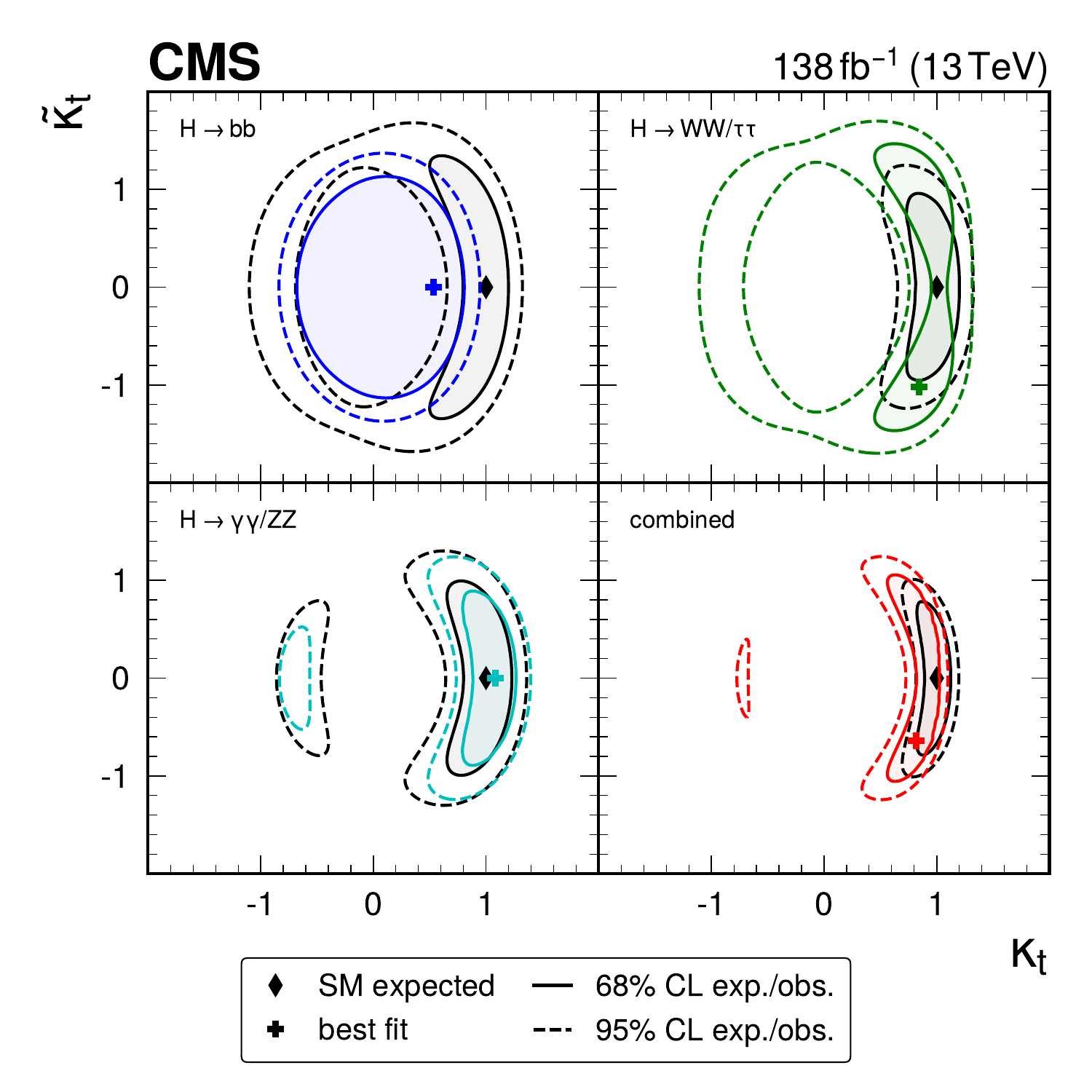}
  \caption{
    Observed 68\% (solid lines) and 95\% (dashed lines) \CL regions of
    the likelihood ratio test statistic as a function of \kappat and
    \kappatcp, where $\kappaV=1$, and best fit values (crosses), for
    the \Htobb decay channel (blue), the $\PH\to\PGg\PGg$ and
    $\PH\to\PZ\PZ$ channels (cyan), the $\PH\to\PW\PW$ and
    $\PH\to\PGt\PGt$ channels (green), and for the combination of all
    channels (red).
    The SM expected \CL regions (black lines) and best fit values
    (black diamonds) are superimposed.    
  }
  \label{fig:results:kappacombination:NonSMcoupling}
\end{figure}

The result of the combination is presented in
Fig.~\ref{fig:results:kappacombination:NonSMcoupling}, which shows the
values of the likelihood ratio test statistic as
a function of \kappat and \kappatcp for the individual channels and
the combination, which yields best fit values of $(\kappat,\kappatcp)$
of $(\resktcpevenComb,\resktcpoddComb)$.
The change of sign of \kappatcp for the result in the $\PH\to\PW\PW$
and $\PH\to\PGt\PGt$ channels compared to Ref.~\cite{CMS:2022dbt} is
not meaningful but reflects the degeneracy of the likelihood with
respect to \kappatcp.
The sensitivity of the combined result is driven by the non-\bbbar
channels, but the addition of the \Htobb channel leads
to a slight improvement in sensitivity, visible in the reduction of
the 68\% and 95\% \CL confidence regions compared to the combination
in Ref.~\cite{CMS:2022dbt}.

\begin{figure}[!htbp]
  \centering
    \includegraphics[width=0.48\textwidth]{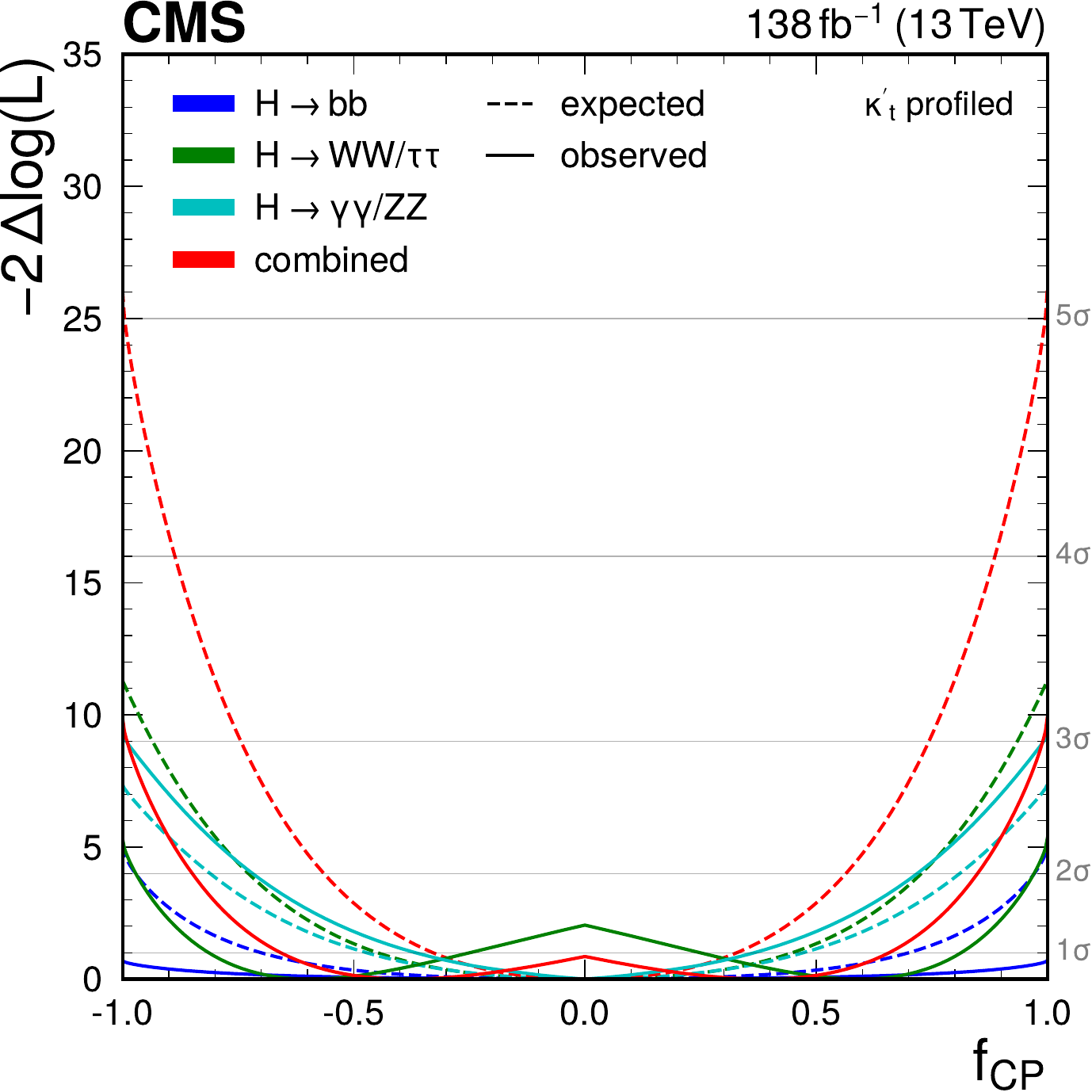}
    \includegraphics[width=0.48\textwidth]{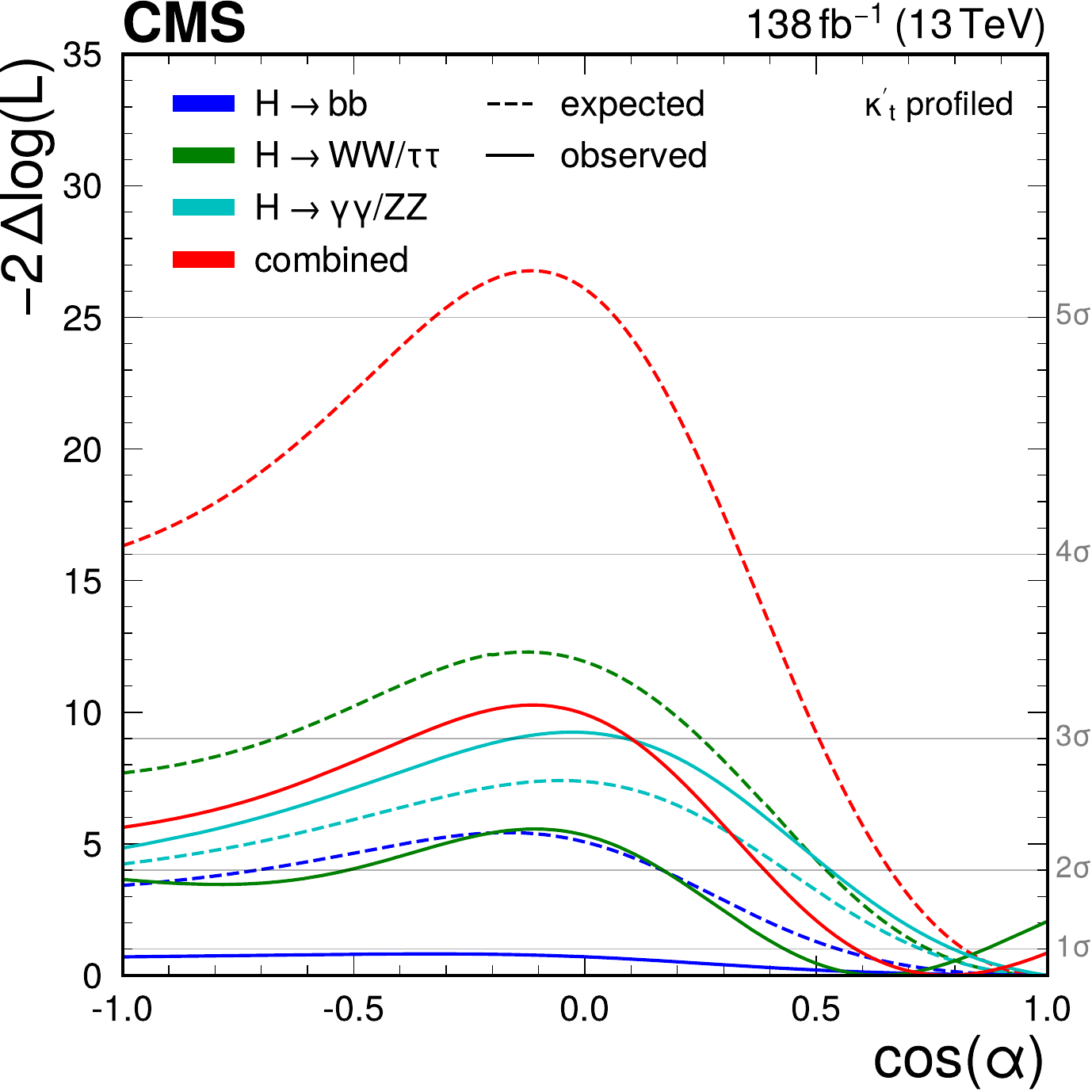} 
  \caption{
    Observed (solid lines) and expected (dotted lines) likelihood
    ratio test statistic as a function of \fCP (left) and \cosa
    (right), where \kappaV is 1 and
    $\kappatprime=\sqrt{\smash[b]{\kappatcp^{2}+\kappat^{2}}}$, the overall
    modifier of the top-Higgs coupling strength, is profiled, for
    the \Htobb decay channel (blue), the $\PH\to\PGg\PGg$ and
    $\PH\to\PZ\PZ$ channels (cyan), the $\PH\to\PW\PW$ and
    $\PH\to\PGt\PGt$ channels (green), and for the combination of all
    channels (red).
  }
  \label{fig:results:kappacombination:fCPcosa}
\end{figure}

Figure~\ref{fig:results:kappacombination:fCPcosa} shows the
corresponding results for \fCP and \cosa.
They are constrained to $|\fCP|<\resfCPLimitComb$ and $\cosa>\rescosaLimitComb$ at 95\% \CL.
While the sensitivity to parameter values close to the SM expectation remains
essentially unchanged, consistent with
Fig.~\ref{fig:results:kappacombination:NonSMcoupling}, the addition of the
\Htobb channel leads to a substantial improvement in the expected
sensitivity to non-SM scenarios with values of $\abs{\fCP}$ close to 1 and
\cosa around 0, corresponding to large values of $\abs{\kappatcp}$ or
small values of $\abs{\kappat}$.
The significance of the observed exclusion is reduced, also with
respect to the previous combination in Ref.~\cite{CMS:2022dbt}.
This is driven by the low observed \ttH signal strength in the \Htobb
channel compared to the other channels.
Given that the sensitivity of the presented analysis to the \ttH signal
strength is at a similar level as in the other channels, the overall
top-Higgs coupling strength in the combination is reduced, leading in turn to
the reduced significance of the observed exclusion in \fCP and \cosa.

\section{Summary}
\label{sec:summary}
A combined analysis of the associated production of a Higgs boson (\PH) with a top quark-antiquark pair (\ttH) or a single top quark (\tH) with the Higgs boson decaying into a bottom quark-antiquark pair has been presented.
The analysis has been performed using proton proton collision data recorded with the CMS detector at a centre-of-mass energy of 13\TeV, corresponding to an integrated luminosity of \lumitotal.
Candidate events are selected in mutually exclusive categories according to the lepton and jet multiplicity, targeting three different final states of the top quark decays.
Neural network discriminants are used to further categorise the events according to the most probable process, targeting the signal and different topologies of the dominant \ttjets background, as well as to separate the signal from the background.
Compared to previous CMS results in this channel, which were obtained with an approximately four times smaller dataset, several refinements of the analysis strategy as well as modelling of the \ttjets background based on state-of-the art \ttbb simulations have been adopted.
Furthermore, an extended set of interpretations is performed, including the first analysis within the simplified template cross section (STXS) framework and the first analysis of the \CP structure of the top-Higgs coupling in this channel by the CMS Collaboration.

A best fit value of the \ttH production cross section relative to the
standard model (SM) expectation of $\resttHMu\pm\resttHMuUpDn =
\resttHMu \pm\resttHMuStatUp\stat\pm\resttHMuSystDn\syst$ is obtained.
The observed rate of the dominant background from \ttbb production is larger than predicted, in agreement with dedicated measurements of the process~\cite{CMS:2023xjh}, and the results motivate further studies of \ttbb production.
The analysis is additionally performed within the STXS framework in five intervals of Higgs boson \pt, probing potential \pt dependent deviations from the SM expectation.
An observed (expected) upper limit on the \tH production cross section
relative to the SM expectation of \restHLimit (\restHExpLimit) at 95\%
confidence level (\CL) is derived.
Information on the Higgs boson coupling strength is furthermore inferred from a simultaneous fit of the \ttH and \tH production rates, probing either the coupling strength of the Higgs boson to top quarks and to heavy vector bosons, or possible \CP-odd admixtures in the coupling between the Higgs boson and top quarks.
The results on the \CP nature of the coupling are combined
with those from measurements in other Higgs boson decay channels,
constraining the \CP-odd fraction \fCP to
$\abs{\fCP}<\resfCPLimitComb$ and the \CP mixing angle \cosa to $\cosa>\rescosaLimitComb$ at 95\% \CL.

\begin{acknowledgments}
  We congratulate our colleagues in the CERN accelerator departments for the excellent performance of the LHC and thank the technical and administrative staffs at CERN and at other CMS institutes for their contributions to the success of the CMS effort. In addition, we gratefully acknowledge the computing centres and personnel of the Worldwide LHC Computing Grid and other centres for delivering so effectively the computing infrastructure essential to our analyses. Finally, we acknowledge the enduring support for the construction and operation of the LHC, the CMS detector, and the supporting computing infrastructure provided by the following funding agencies: SC (Armenia), BMBWF and FWF (Austria); FNRS and FWO (Belgium); CNPq, CAPES, FAPERJ, FAPERGS, and FAPESP (Brazil); MES and BNSF (Bulgaria); CERN; CAS, MoST, and NSFC (China); MINCIENCIAS (Colombia); MSES and CSF (Croatia); RIF (Cyprus); SENESCYT (Ecuador); ERC PRG, RVTT3 and MoER TK202 (Estonia); Academy of Finland, MEC, and HIP (Finland); CEA and CNRS/IN2P3 (France); SRNSF (Georgia); BMBF, DFG, and HGF (Germany); GSRI (Greece); NKFIH (Hungary); DAE and DST (India); IPM (Iran); SFI (Ireland); INFN (Italy); MSIP and NRF (Republic of Korea); MES (Latvia); LMTLT (Lithuania); MOE and UM (Malaysia); BUAP, CINVESTAV, CONACYT, LNS, SEP, and UASLP-FAI (Mexico); MOS (Montenegro); MBIE (New Zealand); PAEC (Pakistan); MES and NSC (Poland); FCT (Portugal); MESTD (Serbia); MCIN/AEI and PCTI (Spain); MOSTR (Sri Lanka); Swiss Funding Agencies (Switzerland); MST (Taipei); MHESI and NSTDA (Thailand); TUBITAK and TENMAK (Turkey); NASU (Ukraine); STFC (United Kingdom); DOE and NSF (USA).
 
  \hyphenation{Rachada-pisek} Individuals have received support from the Marie-Curie programme and the European Research Council and Horizon 2020 Grant, contract Nos.\ 675440, 724704, 752730, 758316, 765710, 824093, 101115353, 101002207, and COST Action CA16108 (European Union); the Leventis Foundation; the Alfred P.\ Sloan Foundation; the Alexander von Humboldt Foundation; the Science Committee, project no. 22rl-037 (Armenia); the Belgian Federal Science Policy Office; the Fonds pour la Formation \`a la Recherche dans l'Industrie et dans l'Agriculture (FRIA-Belgium); the F.R.S.-FNRS and FWO (Belgium) under the ``Excellence of Science -- EOS" -- be.h project n.\ 30820817; the Beijing Municipal Science \& Technology Commission, No. Z191100007219010 and Fundamental Research Funds for the Central Universities (China); the Ministry of Education, Youth and Sports (MEYS) of the Czech Republic; the Shota Rustaveli National Science Foundation, grant FR-22-985 (Georgia); the Deutsche Forschungsgemeinschaft (DFG), among others, under Germany's Excellence Strategy -- EXC 2121 ``Quantum Universe" -- 390833306, and under project number 400140256 - GRK2497; the Hellenic Foundation for Research and Innovation (HFRI), Project Number 2288 (Greece); the Hungarian Academy of Sciences, the New National Excellence Program - \'UNKP, the NKFIH research grants K 131991, K 133046, K 138136, K 143460, K 143477, K 146913, K 146914, K 147048, 2020-2.2.1-ED-2021-00181, and TKP2021-NKTA-64 (Hungary); the Council of Science and Industrial Research, India; ICSC -- National Research Centre for High Performance Computing, Big Data and Quantum Computing and FAIR -- Future Artificial Intelligence Research, funded by the NextGenerationEU program (Italy); the Latvian Council of Science; the Ministry of Education and Science, project no. 2022/WK/14, and the National Science Center, contracts Opus 2021/41/B/ST2/01369 and 2021/43/B/ST2/01552 (Poland); the Funda\c{c}\~ao para a Ci\^encia e a Tecnologia, grant CEECIND/01334/2018 (Portugal); the National Priorities Research Program by Qatar National Research Fund; MCIN/AEI/10.13039/501100011033, ERDF ``a way of making Europe", and the Programa Estatal de Fomento de la Investigaci{\'o}n Cient{\'i}fica y T{\'e}cnica de Excelencia Mar\'{\i}a de Maeztu, grant MDM-2017-0765 and Programa Severo Ochoa del Principado de Asturias (Spain); the Chulalongkorn Academic into Its 2nd Century Project Advancement Project, and the National Science, Research and Innovation Fund via the Program Management Unit for Human Resources \& Institutional Development, Research and Innovation, grant B39G670016 (Thailand); the Kavli Foundation; the Nvidia Corporation; the SuperMicro Corporation; the Welch Foundation, contract C-1845; and the Weston Havens Foundation (USA).  
\end{acknowledgments}

\bibliography{auto_generated}

\providecommand{\href}[2]{#2}\begingroup\raggedright\begin{thebibliography}{100}%
\makeatletter
\providecommand{\hrefCMSnoop }[0]{\@secondoftwo}%
\makeatother
\providecommand{\doi}{\texttt{doi:}\begingroup \urlstyle{tt}\Url}

\bibitem{Alekhin:2012py}
\hrefCMSnoop {}{S.~Alekhin, A.~Djouadi, and S.~Moch, ``The top quark and
  {Higgs} boson masses and the stability of the electroweak vacuum'',} \textit{
  Phys. Lett. B} \textbf{ 716} (2012) 214,
  \href{http://dx.doi.org/10.1016/j.physletb.2012.08.024}{\doi{10.1016/j.physletb.2012.08.024}},
  \href{http://www.arXiv.org/abs/1207.0980}{\texttt{arXiv:1207.0980}}.

\bibitem{Dobrescu:1997nm}
\hrefCMSnoop {}{B.~A. Dobrescu and C.~T. Hill, ``Electroweak symmetry breaking
  via top condensation seesaw'',} \textit{ Phys. Rev. Lett.} \textbf{ 81}
  (1998) 2634,
  \href{http://dx.doi.org/10.1103/PhysRevLett.81.2634}{\doi{10.1103/PhysRevLett.81.2634}},
  \href{http://www.arXiv.org/abs/hep-ph/9712319}{\texttt{arXiv:hep-ph/9712319}}.

\bibitem{Chivukula:1998wd}
\hrefCMSnoop {}{R.~S. Chivukula, B.~A. Dobrescu, H.~Georgi, and C.~T. Hill,
  ``Top quark seesaw theory of electroweak symmetry breaking'',} \textit{ Phys.
  Rev. D} \textbf{ 59} (1999) 075003,
  \href{http://dx.doi.org/10.1103/PhysRevD.59.075003}{\doi{10.1103/PhysRevD.59.075003}},
  \href{http://www.arXiv.org/abs/hep-ph/9809470}{\texttt{arXiv:hep-ph/9809470}}.

\bibitem{Aaboud:2018urx}
\hrefCMSnoop {}{{ATLAS Collaboration}, ``Observation of {Higgs} boson
  production in association with a top quark pair at the {LHC} with the {ATLAS}
  detector'',} \textit{ Phys. Lett. B} \textbf{ 784} (2018) 159,
  \href{http://dx.doi.org/10.1016/j.physletb.2018.07.035}{\doi{10.1016/j.physletb.2018.07.035}},
  \href{http://www.arXiv.org/abs/1806.00425}{\texttt{arXiv:1806.00425}}.

\bibitem{Sirunyan:2018hoz}
\hrefCMSnoop {}{{CMS Collaboration}, ``Observation of {$\PQt\PAQt\PH$}
  production'',} \textit{ Phys. Rev. Lett.} \textbf{ 120} (2018) 231801,
  \href{http://dx.doi.org/10.1103/PhysRevLett.120.231801}{\doi{10.1103/PhysRevLett.120.231801}},
  \href{http://www.arXiv.org/abs/1804.02610}{\texttt{arXiv:1804.02610}}.

\bibitem{ATLAS:2022tnm}
\hrefCMSnoop {}{{ATLAS Collaboration}, ``Measurement of the properties of
  {Higgs} boson production at $\sqrt{s}=13$\,{TeV} in the
  {$\PH\to\gamma\gamma$} channel using 139\,fb{$^{-1}$} of {$\Pp\Pp$} collision
  data with the {ATLAS} experiment'',} \textit{ JHEP} \textbf{ 07} (2023) 088,
  \href{http://dx.doi.org/10.1007/JHEP07(2023)088}{\doi{10.1007/JHEP07(2023)088}},
  \href{http://www.arXiv.org/abs/2207.00348}{\texttt{arXiv:2207.00348}}.

\bibitem{ATLAS:2020ior}
\hrefCMSnoop {}{{ATLAS Collaboration}, ``{$CP$} properties of {Higgs} boson
  interactions with top quarks in the {$\PQt\PAQt\PH$} and {$\PQt\PH$}
  processes using {$\PH\to\gamma\gamma$} with the {ATLAS} detector'',} \textit{
  Phys. Rev. Lett.} \textbf{ 125} (2020) 061802,
  \href{http://dx.doi.org/10.1103/PhysRevLett.125.061802}{\doi{10.1103/PhysRevLett.125.061802}},
  \href{http://www.arXiv.org/abs/2004.04545}{\texttt{arXiv:2004.04545}}.

\bibitem{CMS:2020cga}
\hrefCMSnoop {}{{CMS Collaboration}, ``Measurements of {$\PQt\PAQt\PH$}
  production and the {CP} structure of the {Yukawa} interaction between the
  {Higgs} boson and top quark in the diphoton decay channel'',} \textit{ Phys.
  Rev. Lett.} \textbf{ 125} (2020) 061801,
  \href{http://dx.doi.org/10.1103/PhysRevLett.125.061801}{\doi{10.1103/PhysRevLett.125.061801}},
  \href{http://www.arXiv.org/abs/2003.10866}{\texttt{arXiv:2003.10866}}.

\bibitem{CMS:2021kom}
\hrefCMSnoop {}{{CMS Collaboration}, ``Measurements of {Higgs} boson production
  cross sections and couplings in the diphoton decay channel at
  $\sqrt{s}=13$\,{TeV}'',} \textit{ JHEP} \textbf{ 07} (2021) 027,
  \href{http://dx.doi.org/10.1007/JHEP07(2021)027}{\doi{10.1007/JHEP07(2021)027}},
  \href{http://www.arXiv.org/abs/2103.06956}{\texttt{arXiv:2103.06956}}.

\bibitem{CMS:2020mpn}
\hrefCMSnoop {}{{CMS Collaboration}, ``Measurement of the {Higgs} boson
  production rate in association with top quarks in final states with
  electrons, muons, and hadronically decaying tau leptons at
  $\sqrt{s}=13$\,{TeV}'',} \textit{ Eur. Phys. J. C} \textbf{ 81} (2021) 378,
  \href{http://dx.doi.org/10.1140/epjc/s10052-021-09014-x}{\doi{10.1140/epjc/s10052-021-09014-x}},
  \href{http://www.arXiv.org/abs/2011.03652}{\texttt{arXiv:2011.03652}}.

\bibitem{ATLAS:2021qou}
\hrefCMSnoop {}{{ATLAS Collaboration}, ``Measurement of {Higgs} boson decay
  into {\PQb}-quarks in associated production with a top-quark pair in
  {$\Pp\Pp$} collisions at $\sqrt{s}=13$\,{TeV} with the {ATLAS} detector'',}
  \textit{ JHEP} \textbf{ 06} (2022) 097,
  \href{http://dx.doi.org/10.1007/JHEP06(2022)097}{\doi{10.1007/JHEP06(2022)097}},
  \href{http://www.arXiv.org/abs/2111.06712}{\texttt{arXiv:2111.06712}}.

\bibitem{Sirunyan:2018mvw}
\hrefCMSnoop {}{{CMS Collaboration}, ``Search for {$\PQt\PAQt\PH$} production
  in the {$\PH\to\PQb\PAQb$} decay channel with leptonic {$\PQt\PAQt$} decays
  in proton-proton collisions at $\sqrt{s}=13$\,{TeV}'',} \textit{ JHEP}
  \textbf{ 03} (2019) 026,
  \href{http://dx.doi.org/10.1007/JHEP03(2019)026}{\doi{10.1007/JHEP03(2019)026}},
  \href{http://www.arXiv.org/abs/1804.03682}{\texttt{arXiv:1804.03682}}.

\bibitem{Sirunyan:2018ygk}
\hrefCMSnoop {}{{CMS Collaboration}, ``Search for {$\PQt\PAQt\PH$} production
  in the all-jet final state in proton-proton collisions at
  $\sqrt{s}=13$\,{TeV}'',} \textit{ JHEP} \textbf{ 06} (2018) 101,
  \href{http://dx.doi.org/10.1007/JHEP06(2018)101}{\doi{10.1007/JHEP06(2018)101}},
  \href{http://www.arXiv.org/abs/1803.06986}{\texttt{arXiv:1803.06986}}.

\bibitem{Maltoni:2001hu}
\hrefCMSnoop {}{F.~Maltoni, K.~Paul, T.~Stelzer, and S.~Willenbrock,
  ``Associated production of {Higgs} and single top at hadron colliders'',}
  \textit{ Phys. Rev. D} \textbf{ 64} (2001) 094023,
  \href{http://dx.doi.org/10.1103/PhysRevD.64.094023}{\doi{10.1103/PhysRevD.64.094023}},
  \href{http://www.arXiv.org/abs/hep-ph/0106293}{\texttt{arXiv:hep-ph/0106293}}.

\bibitem{Farina:2012xp}
M.~Farina\hrefCMSnoop {}{ { et~al.}, ``Lifting degeneracies in {Higgs}
  couplings using single top production in association with a {Higgs} boson'',}
  \textit{ JHEP} \textbf{ 05} (2013) 022,
  \href{http://dx.doi.org/10.1007/JHEP05(2013)022}{\doi{10.1007/JHEP05(2013)022}},
  \href{http://www.arXiv.org/abs/1211.3736}{\texttt{arXiv:1211.3736}}.

\bibitem{Agrawal:2012ga}
\hrefCMSnoop {}{P.~Agrawal, S.~Mitra, and A.~Shivaji, ``Effect of anomalous
  couplings on the associated production of a single top quark and a {Higgs}
  boson at the {LHC}'',} \textit{ JHEP} \textbf{ 12} (2013) 077,
  \href{http://dx.doi.org/10.1007/JHEP12(2013)077}{\doi{10.1007/JHEP12(2013)077}},
  \href{http://www.arXiv.org/abs/1211.4362}{\texttt{arXiv:1211.4362}}.

\bibitem{Demartin:2015uha}
\hrefCMSnoop {}{F.~Demartin, F.~Maltoni, K.~Mawatari, and M.~Zaro, ``Higgs
  production in association with a single top quark at the {LHC}'',} \textit{
  Eur. Phys. J. C} \textbf{ 75} (2015) 267,
  \href{http://dx.doi.org/10.1140/epjc/s10052-015-3475-9}{\doi{10.1140/epjc/s10052-015-3475-9}},
  \href{http://www.arXiv.org/abs/1504.00611}{\texttt{arXiv:1504.00611}}.

\bibitem{Sirunyan:2018lzm}
\hrefCMSnoop {}{{CMS Collaboration}, ``Search for associated production of a
  {Higgs} boson and a single top quark in proton-proton collisions at
  $\sqrt{s}=13$\,{TeV}'',} \textit{ Phys. Rev. D} \textbf{ 99} (2019) 092005,
  \href{http://dx.doi.org/10.1103/PhysRevD.99.092005}{\doi{10.1103/PhysRevD.99.092005}},
  \href{http://www.arXiv.org/abs/1811.09696}{\texttt{arXiv:1811.09696}}.

\bibitem{CMS:2021nnc}
\hrefCMSnoop {}{{CMS Collaboration}, ``Constraints on anomalous {Higgs} boson
  couplings to vector bosons and fermions in its production and decay using the
  four-lepton final state'',} \textit{ Phys. Rev. D} \textbf{ 104} (2021)
  052004,
  \href{http://dx.doi.org/10.1103/PhysRevD.104.052004}{\doi{10.1103/PhysRevD.104.052004}},
  \href{http://www.arXiv.org/abs/2104.12152}{\texttt{arXiv:2104.12152}}.

\bibitem{CMS:2022dbt}
\hrefCMSnoop {}{{CMS Collaboration}, ``Search for {$CP$} violation in
  {$\PQt\PAQt\PH$} and {$\PQt\PH$} production in multilepton channels in
  proton-proton collisions at $\sqrt{s}=13$\,{TeV}'',} \textit{ JHEP} \textbf{
  07} (2023) 092,
  \href{http://dx.doi.org/10.1007/JHEP07(2023)092}{\doi{10.1007/JHEP07(2023)092}},
  \href{http://www.arXiv.org/abs/2208.02686}{\texttt{arXiv:2208.02686}}.

\bibitem{ATLAS:2023cbt}
\hrefCMSnoop {}{{ATLAS Collaboration}, ``Probing the {$CP$} nature of the
  top-{H}iggs {Y}ukawa coupling in {$\PQt\PAQt\PH$} and {$\PQt\PH$} events with
  {$\PH\to\PQb\PAQb$} decays using the {ATLAS} detector at the {LHC}'',}
  \textit{ Phys. Lett. B} \textbf{ 849} (2024) 138469,
  \href{http://dx.doi.org/10.1016/j.physletb.2024.138469}{\doi{10.1016/j.physletb.2024.138469}},
  \href{http://www.arXiv.org/abs/2303.05974}{\texttt{arXiv:2303.05974}}.

\bibitem{deFlorian:2016spz}
\hrefCMSnoop {}{{LHC Higgs Cross Section Working Group}, ``Handbook of {LHC}
  {Higgs} cross sections: 4. deciphering the nature of the {Higgs} sector'',}
  2017. \href{http://www.arXiv.org/abs/1610.07922}{\texttt{arXiv:1610.07922}}.

\bibitem{ATLAS:2024gth}
\hrefCMSnoop {}{{ATLAS Collaboration}, ``Measurement of the associated
  production of a top-antitop-quark pair and a {Higgs} boson decaying into a
  {$\PQb\PAQb$} pair in {$\Pp\Pp$} collisions at $\sqrt{s}=13$\,{TeV} using the
  {ATLAS} detector at the {LHC}'',} 2024.
  \href{http://www.arXiv.org/abs/2407.10904}{\texttt{arXiv:2407.10904}}.
  Submitted to \textit{Eur. Phys. J. C.}

\bibitem{Jezo:2018yaf}
\hrefCMSnoop {}{T.~Je{\v{z}}o, J.~M. Lindert, N.~Moretti, and S.~Pozzorini,
  ``New {NLOPS} predictions for {$\PQt\PAQt+\PQb$}-jet production at the
  {LHC}'',} \textit{ Eur. Phys. J. C} \textbf{ 78} (2018) 502,
  \href{http://dx.doi.org/10.1140/epjc/s10052-018-5956-0}{\doi{10.1140/epjc/s10052-018-5956-0}},
  \href{http://www.arXiv.org/abs/1802.00426}{\texttt{arXiv:1802.00426}}.

\bibitem{Buccioni:2019plc}
\hrefCMSnoop {}{F.~Buccioni, S.~Kallweit, S.~Pozzorini, and M.~F. Zoller,
  ``{NLO} {QCD} predictions for {$\PQt\PAQt\PQb\PAQb$} production in
  association with a light jet at the {LHC}'',} \textit{ JHEP} \textbf{ 12}
  (2019) 015,
  \href{http://dx.doi.org/10.1007/JHEP12(2019)015}{\doi{10.1007/JHEP12(2019)015}},
  \href{http://www.arXiv.org/abs/1907.13624}{\texttt{arXiv:1907.13624}}.

\bibitem{hepdata}
\hrefCMSnoop {}{}{HEPD}ata record for this analysis, 2024.
\newblock
  \href{http://dx.doi.org/10.17182/hepdata.152799}{\doi{10.17182/hepdata.152799}}.

\bibitem{CMS:2008xjf}
\hrefCMSnoop {}{{CMS Collaboration}, ``The {CMS} experiment at the {CERN}
  {LHC}'',} \textit{ JINST} \textbf{ 3} (2008) S08004,
  \href{http://dx.doi.org/10.1088/1748-0221/3/08/S08004}{\doi{10.1088/1748-0221/3/08/S08004}}.

\bibitem{Phase1Pixel}
\hrefCMSnoop {}{{CMS Tracker Group} Collaboration, ``The {CMS} phase-1 pixel
  detector upgrade'',} \textit{ JINST} \textbf{ 16} (2021) P02027,
  \href{http://dx.doi.org/10.1088/1748-0221/16/02/P02027}{\doi{10.1088/1748-0221/16/02/P02027}},
  \href{http://www.arXiv.org/abs/2012.14304}{\texttt{arXiv:2012.14304}}.

\bibitem{CMS:2016ngn}
\hrefCMSnoop {}{{CMS Collaboration}, ``The {CMS} trigger system'',} \textit{
  JINST} \textbf{ 12} (2017) P01020,
  \href{http://dx.doi.org/10.1088/1748-0221/12/01/P01020}{\doi{10.1088/1748-0221/12/01/P01020}},
  \href{http://www.arXiv.org/abs/1609.02366}{\texttt{arXiv:1609.02366}}.

\bibitem{CMS:2020cmk}
\hrefCMSnoop {}{{CMS Collaboration}, ``Performance of the {CMS} level-1 trigger
  in proton-proton collisions at $\sqrt{s}=13$\,{TeV}'',} \textit{ JINST}
  \textbf{ 15} (2020) P10017,
  \href{http://dx.doi.org/10.1088/1748-0221/15/10/P10017}{\doi{10.1088/1748-0221/15/10/P10017}},
  \href{http://www.arXiv.org/abs/2006.10165}{\texttt{arXiv:2006.10165}}.

\bibitem{CMS-LUM-17-003}
\hrefCMSnoop {}{{CMS Collaboration}, ``Precision luminosity measurement in
  proton-proton collisions at $\sqrt{s}=13$\,{TeV} in 2015 and 2016 at
  {CMS}'',} \textit{ Eur. Phys. J. C} \textbf{ 81} (2021) 800,
  \href{http://dx.doi.org/10.1140/epjc/s10052-021-09538-2}{\doi{10.1140/epjc/s10052-021-09538-2}},
  \href{http://www.arXiv.org/abs/2104.01927}{\texttt{arXiv:2104.01927}}.

\bibitem{CMS-PAS-LUM-17-004}
\href {{http://cds.cern.ch/record/2621960}}{{CMS Collaboration}, ``{CMS}
  luminosity measurement for the 2017 data-taking period at
  $\sqrt{s}=13$\,{TeV}'',} {CMS Physics Analysis Summary} CMS-PAS-LUM-17-004,
  2017.

\bibitem{CMS-PAS-LUM-18-002}
\href {{http://cds.cern.ch/record/2676164}}{{CMS Collaboration}, ``{CMS}
  luminosity measurement for the 2018 data-taking period at
  $\sqrt{s}=13$\,{TeV}'',} {CMS Physics Analysis Summary} CMS-PAS-LUM-18-002,
  2018.

\bibitem{CMS:2021yvr}
\hrefCMSnoop {}{{CMS Collaboration}, ``Performance of the {CMS} muon trigger
  system in proton-proton collisions at $\sqrt{s}=13$\,{TeV}'',} \textit{
  JINST} \textbf{ 16} (2021) P07001,
  \href{http://dx.doi.org/10.1088/1748-0221/16/07/P07001}{\doi{10.1088/1748-0221/16/07/P07001}},
  \href{http://www.arXiv.org/abs/2102.04790}{\texttt{arXiv:2102.04790}}.

\bibitem{CMS:2020uim}
\hrefCMSnoop {}{{CMS Collaboration}, ``Electron and photon reconstruction and
  identification with the {CMS} experiment at the {CERN} {LHC}'',} \textit{
  JINST} \textbf{ 16} (2021) P05014,
  \href{http://dx.doi.org/10.1088/1748-0221/16/05/P05014}{\doi{10.1088/1748-0221/16/05/P05014}},
  \href{http://www.arXiv.org/abs/2012.06888}{\texttt{arXiv:2012.06888}}.

\bibitem{Agostinelli:2002hh}
\hrefCMSnoop {}{{GEANT4} Collaboration, ``{\GEANTfour}---a simulation
  toolkit'',} \textit{ Nucl. Instrum. Meth. A} \textbf{ 506} (2003) 250,
  \href{http://dx.doi.org/10.1016/S0168-9002(03)01368-8}{\doi{10.1016/S0168-9002(03)01368-8}}.

\bibitem{Nason:2004rx}
\hrefCMSnoop {}{P.~Nason, ``A new method for combining {NLO} {QCD} with shower
  {Monte} {Carlo} algorithms'',} \textit{ JHEP} \textbf{ 11} (2004) 040,
  \href{http://dx.doi.org/10.1088/1126-6708/2004/11/040}{\doi{10.1088/1126-6708/2004/11/040}},
  \href{http://www.arXiv.org/abs/hep-ph/0409146}{\texttt{arXiv:hep-ph/0409146}}.

\bibitem{Frixione:2007vw}
\hrefCMSnoop {}{S.~Frixione, P.~Nason, and C.~Oleari, ``Matching {NLO} {QCD}
  computations with parton shower simulations: the {\POWHEG} method'',}
  \textit{ JHEP} \textbf{ 11} (2007) 070,
  \href{http://dx.doi.org/10.1088/1126-6708/2007/11/070}{\doi{10.1088/1126-6708/2007/11/070}},
  \href{http://www.arXiv.org/abs/0709.2092}{\texttt{arXiv:0709.2092}}.

\bibitem{Alioli:2010xd}
\hrefCMSnoop {}{S.~Alioli, P.~Nason, C.~Oleari, and E.~Re, ``A general
  framework for implementing {NLO} calculations in shower {Monte} {Carlo}
  programs: the {\textsc{powheg}} {\textsc{box}}'',} \textit{ JHEP} \textbf{
  06} (2010) 043,
  \href{http://dx.doi.org/10.1007/JHEP06(2010)043}{\doi{10.1007/JHEP06(2010)043}},
  \href{http://www.arXiv.org/abs/1002.2581}{\texttt{arXiv:1002.2581}}.

\bibitem{Jezo:2015aia}
\hrefCMSnoop {}{T.~Je{\v{z}}o and P.~Nason, ``On the treatment of resonances in
  next-to-leading order calculations matched to a parton shower'',} \textit{
  JHEP} \textbf{ 12} (2015) 065,
  \href{http://dx.doi.org/10.1007/JHEP12(2015)065}{\doi{10.1007/JHEP12(2015)065}},
  \href{http://www.arXiv.org/abs/1509.09071}{\texttt{arXiv:1509.09071}}.

\bibitem{powheg:tth:1}
\hrefCMSnoop {}{H.~B. Hartanto, B.~Jager, L.~Reina, and D.~Wackeroth, ``Higgs
  boson production in association with top quarks in the {\textsc{powheg}}
  {\textsc{box}}'',} \textit{ Phys. Rev. D} \textbf{ 91} (2015) 094003,
  \href{http://dx.doi.org/10.1103/PhysRevD.91.094003}{\doi{10.1103/PhysRevD.91.094003}},
  \href{http://www.arXiv.org/abs/1501.04498}{\texttt{arXiv:1501.04498}}.

\bibitem{Alwall:2014hca}
J.~Alwall\hrefCMSnoop {}{ { et~al.}, ``The automated computation of tree-level
  and next-to-leading order differential cross sections, and their matching to
  parton shower simulations'',} \textit{ JHEP} \textbf{ 07} (2014) 079,
  \href{http://dx.doi.org/10.1007/JHEP07(2014)079}{\doi{10.1007/JHEP07(2014)079}},
  \href{http://www.arXiv.org/abs/1405.0301}{\texttt{arXiv:1405.0301}}.

\bibitem{Sjostrand:2014zea}
T.~Sj{\"o}strand\hrefCMSnoop {}{ { et~al.}, ``An introduction to {\PYTHIA}
  8.2'',} \textit{ Comput. Phys. Commun.} \textbf{ 191} (2015) 159,
  \href{http://dx.doi.org/10.1016/j.cpc.2015.01.024}{\doi{10.1016/j.cpc.2015.01.024}},
  \href{http://www.arXiv.org/abs/1410.3012}{\texttt{arXiv:1410.3012}}.

\bibitem{Ball:2017nwa}
\hrefCMSnoop {}{{NNPDF} Collaboration, ``Parton distributions from
  high-precision collider data'',} \textit{ Eur. Phys. J. C} \textbf{ 77}
  (2017) 663,
  \href{http://dx.doi.org/10.1140/epjc/s10052-017-5199-5}{\doi{10.1140/epjc/s10052-017-5199-5}},
  \href{http://www.arXiv.org/abs/1706.00428}{\texttt{arXiv:1706.00428}}.

\bibitem{Ball:2014uwa}
\hrefCMSnoop {}{{NNPDF} Collaboration, ``Parton distributions for the {LHC}
  {Run II}'',} \textit{ JHEP} \textbf{ 04} (2015) 040,
  \href{http://dx.doi.org/10.1007/JHEP04(2015)040}{\doi{10.1007/JHEP04(2015)040}},
  \href{http://www.arXiv.org/abs/1410.8849}{\texttt{arXiv:1410.8849}}.

\bibitem{Sirunyan:2019dfx}
\hrefCMSnoop {}{{CMS Collaboration}, ``Extraction and validation of a new set
  of {CMS} {\PYTHIA} $8$ tunes from underlying-event measurements'',} \textit{
  Eur. Phys. J. C} \textbf{ 80} (2020) 4,
  \href{http://dx.doi.org/10.1140/epjc/s10052-019-7499-4}{\doi{10.1140/epjc/s10052-019-7499-4}},
  \href{http://www.arXiv.org/abs/1903.12179}{\texttt{arXiv:1903.12179}}.

\bibitem{bib:CUETP8tune}
\hrefCMSnoop {}{{CMS Collaboration}, ``Event generator tunes obtained from
  underlying event and multiparton scattering measurements'',} \textit{ Eur.
  Phys. J. C} \textbf{ 76} (2016) 155,
  \href{http://dx.doi.org/10.1140/epjc/s10052-016-3988-x}{\doi{10.1140/epjc/s10052-016-3988-x}},
  \href{http://www.arXiv.org/abs/1512.00815}{\texttt{arXiv:1512.00815}}.

\bibitem{Maltoni:2012pa}
\hrefCMSnoop {}{F.~Maltoni, G.~Ridolfi, and M.~Ubiali, ``{\PQb}-initiated
  processes at the {LHC}: a reappraisal'',} \textit{ JHEP} \textbf{ 07} (2012)
  022,
  \href{http://dx.doi.org/10.1007/JHEP07(2012)022}{\doi{10.1007/JHEP07(2012)022}},
  \href{http://www.arXiv.org/abs/1203.6393}{\texttt{arXiv:1203.6393}}.
  [Erratum: \DOI{10.1007/JHEP04(2013)095}].

\bibitem{Demartin:2016axk}
F.~Demartin\hrefCMSnoop {}{ { et~al.}, ``{$\PQt\PW\PH$} associated production
  at the {LHC}'',} \textit{ Eur. Phys. J. C} \textbf{ 77} (2017) 34,
  \href{http://dx.doi.org/10.1140/epjc/s10052-017-4601-7}{\doi{10.1140/epjc/s10052-017-4601-7}},
  \href{http://www.arXiv.org/abs/1607.05862}{\texttt{arXiv:1607.05862}}.

\bibitem{Gainer:2014bta}
J.~S. Gainer\hrefCMSnoop {}{ { et~al.}, ``Exploring theory space with {Monte}
  {Carlo} reweighting'',} \textit{ JHEP} \textbf{ 10} (2014) 078,
  \href{http://dx.doi.org/10.1007/JHEP10(2014)078}{\doi{10.1007/JHEP10(2014)078}},
  \href{http://www.arXiv.org/abs/1404.7129}{\texttt{arXiv:1404.7129}}.

\bibitem{Mattelaer:2016gcx}
\hrefCMSnoop {}{O.~Mattelaer, ``On the maximal use of {Monte} {Carlo} samples:
  re-weighting events at {NLO} accuracy'',} \textit{ Eur. Phys. J. C} \textbf{
  76} (2016) 674,
  \href{http://dx.doi.org/10.1140/epjc/s10052-016-4533-7}{\doi{10.1140/epjc/s10052-016-4533-7}},
  \href{http://www.arXiv.org/abs/1607.00763}{\texttt{arXiv:1607.00763}}.

\bibitem{Buccioni:2019sur}
F.~Buccioni\hrefCMSnoop {}{ { et~al.}, ``{OpenLoops 2}'',} \textit{ Eur. Phys.
  J. C} \textbf{ 79} (2019) 866,
  \href{http://dx.doi.org/10.1140/epjc/s10052-019-7306-2}{\doi{10.1140/epjc/s10052-019-7306-2}},
  \href{http://www.arXiv.org/abs/1907.13071}{\texttt{arXiv:1907.13071}}.

\bibitem{Cascioli:2013era}
F.~Cascioli\hrefCMSnoop {}{ { et~al.}, ``{NLO} matching for
  {$\PQt\PAQt\PQb\PAQb$} production with massive {\PQb}-quarks'',} \textit{
  Phys. Lett. B} \textbf{ 734} (2014) 210,
  \href{http://dx.doi.org/10.1016/j.physletb.2014.05.040}{\doi{10.1016/j.physletb.2014.05.040}},
  \href{http://www.arXiv.org/abs/1309.5912}{\texttt{arXiv:1309.5912}}.

\bibitem{powheg:singlet:st}
\hrefCMSnoop {}{S.~Alioli, P.~Nason, C.~Oleari, and E.~Re, ``{NLO} single-top
  production matched with shower in {\POWHEG}: $s$- and $t$-channel
  contributions'',} \textit{ JHEP} \textbf{ 09} (2009) 111,
  \href{http://dx.doi.org/10.1088/1126-6708/2009/09/111}{\doi{10.1088/1126-6708/2009/09/111}},
  \href{http://www.arXiv.org/abs/0907.4076}{\texttt{arXiv:0907.4076}}.
  [Erratum: \DOI{10.1007/JHEP02(2010)011}].

\bibitem{powheg:singlet:st2}
\hrefCMSnoop {}{R.~Frederix, E.~Re, and P.~Torrielli, ``Single-top $t$-channel
  hadroproduction in the four-flavour scheme with {\POWHEG} and a{\MCATNLO}'',}
  \textit{ JHEP} \textbf{ 09} (2012) 130,
  \href{http://dx.doi.org/10.1007/JHEP09(2012)130}{\doi{10.1007/JHEP09(2012)130}},
  \href{http://www.arXiv.org/abs/1207.5391}{\texttt{arXiv:1207.5391}}.

\bibitem{powheg:singlet:tW}
\hrefCMSnoop {}{E.~Re, ``Single-top ${\PW}{\PQt}$-channel production matched
  with parton showers using the {\POWHEG} method'',} \textit{ Eur. Phys. J. C}
  \textbf{ 71} (2011) 1547,
  \href{http://dx.doi.org/10.1140/epjc/s10052-011-1547-z}{\doi{10.1140/epjc/s10052-011-1547-z}},
  \href{http://www.arXiv.org/abs/1009.2450}{\texttt{arXiv:1009.2450}}.

\bibitem{Frederix:2012ps}
\hrefCMSnoop {}{R.~Frederix and S.~Frixione, ``Merging meets matching in
  {\MCATNLO}'',} \textit{ JHEP} \textbf{ 12} (2012) 061,
  \href{http://dx.doi.org/10.1007/JHEP12(2012)061}{\doi{10.1007/JHEP12(2012)061}},
  \href{http://www.arXiv.org/abs/1209.6215}{\texttt{arXiv:1209.6215}}.

\bibitem{Alwall:2007fs}
J.~Alwall\hrefCMSnoop {}{ { et~al.}, ``Comparative study of various algorithms
  for the merging of parton showers and matrix elements in hadronic
  collisions'',} \textit{ Eur. Phys. J. C} \textbf{ 53} (2008) 473,
  \href{http://dx.doi.org/10.1140/epjc/s10052-007-0490-5}{\doi{10.1140/epjc/s10052-007-0490-5}},
  \href{http://www.arXiv.org/abs/0706.2569}{\texttt{arXiv:0706.2569}}.

\bibitem{CMS:2020xrn}
\hrefCMSnoop {}{{CMS Collaboration}, ``A measurement of the {Higgs} boson mass
  in the diphoton decay channel'',} \textit{ Phys. Lett. B} \textbf{ 805}
  (2020) 135425,
  \href{http://dx.doi.org/10.1016/j.physletb.2020.135425}{\doi{10.1016/j.physletb.2020.135425}},
  \href{http://www.arXiv.org/abs/2002.06398}{\texttt{arXiv:2002.06398}}.

\bibitem{bib:xs1}
M.~Cacciari\hrefCMSnoop {}{ { et~al.}, ``Top-pair production at hadron
  colliders with next-to-next-to-leading logarithmic soft-gluon resummation'',}
  \textit{ Phys. Lett. B} \textbf{ 710} (2012) 612,
  \href{http://dx.doi.org/10.1016/j.physletb.2012.03.013}{\doi{10.1016/j.physletb.2012.03.013}},
  \href{http://www.arXiv.org/abs/1111.5869}{\texttt{arXiv:1111.5869}}.

\bibitem{bib:xs2}
\hrefCMSnoop {}{P.~B{\"a}rnreuther, M.~Czakon, and A.~Mitov,
  ``Percent-level-precision physics at the {Tevatron}: next-to-next-to-leading
  order {QCD} corrections to {$\PQq\PAQq\to\PQt\PAQt\text{+X}$}'',} \textit{
  Phys. Rev. Lett.} \textbf{ 109} (2012) 132001,
  \href{http://dx.doi.org/10.1103/PhysRevLett.109.132001}{\doi{10.1103/PhysRevLett.109.132001}},
  \href{http://www.arXiv.org/abs/1204.5201}{\texttt{arXiv:1204.5201}}.

\bibitem{bib:xs3}
\hrefCMSnoop {}{M.~Czakon and A.~Mitov, ``{NNLO} corrections to top-pair
  production at hadron colliders: the all-fermionic scattering channels'',}
  \textit{ JHEP} \textbf{ 12} (2012) 054,
  \href{http://dx.doi.org/10.1007/JHEP12(2012)054}{\doi{10.1007/JHEP12(2012)054}},
  \href{http://www.arXiv.org/abs/1207.0236}{\texttt{arXiv:1207.0236}}.

\bibitem{bib:xs4}
\hrefCMSnoop {}{M.~Czakon and A.~Mitov, ``{NNLO} corrections to top pair
  production at hadron colliders: the quark-gluon reaction'',} \textit{ JHEP}
  \textbf{ 01} (2013) 080,
  \href{http://dx.doi.org/10.1007/JHEP01(2013)080}{\doi{10.1007/JHEP01(2013)080}},
  \href{http://www.arXiv.org/abs/1210.6832}{\texttt{arXiv:1210.6832}}.

\bibitem{bib:xs7}
\hrefCMSnoop {}{M.~Beneke, P.~Falgari, S.~Klein, and C.~Schwinn, ``Hadronic
  top-quark pair production with {NNLL} threshold resummation'',} \textit{
  Nucl. Phys. B} \textbf{ 855} (2012) 695,
  \href{http://dx.doi.org/10.1016/j.nuclphysb.2011.10.021}{\doi{10.1016/j.nuclphysb.2011.10.021}},
  \href{http://www.arXiv.org/abs/1109.1536}{\texttt{arXiv:1109.1536}}.

\bibitem{Czakon:2013goa}
\hrefCMSnoop {}{M.~Czakon, P.~Fiedler, and A.~Mitov, ``Total top-quark
  pair-production cross section at hadron colliders through
  $\mathcal{O}({\alpha_s}^4)$'',} \textit{ Phys. Rev. Lett.} \textbf{ 110}
  (2013) 252004,
  \href{http://dx.doi.org/10.1103/PhysRevLett.110.252004}{\doi{10.1103/PhysRevLett.110.252004}},
  \href{http://www.arXiv.org/abs/1303.6254}{\texttt{arXiv:1303.6254}}.

\bibitem{Czakon:2011xx}
\hrefCMSnoop {}{M.~Czakon and A.~Mitov, ``{Top++}: a program for the
  calculation of the top-pair cross-section at hadron colliders'',} \textit{
  Comput. Phys. Commun.} \textbf{ 185} (2014) 2930,
  \href{http://dx.doi.org/10.1016/j.cpc.2014.06.021}{\doi{10.1016/j.cpc.2014.06.021}},
  \href{http://www.arXiv.org/abs/1112.5675}{\texttt{arXiv:1112.5675}}.

\bibitem{bib:twchan}
\hrefCMSnoop {}{N.~Kidonakis, ``Two-loop soft anomalous dimensions for single
  top quark associated production with {\PWm} or {$\PH^{-}$}'',} \textit{ Phys.
  Rev. D} \textbf{ 82} (2010) 054018,
  \href{http://dx.doi.org/10.1103/PhysRevD.82.054018}{\doi{10.1103/PhysRevD.82.054018}},
  \href{http://www.arXiv.org/abs/1005.4451}{\texttt{arXiv:1005.4451}}.

\bibitem{Aliev:2010zk}
M.~Aliev\hrefCMSnoop {}{ { et~al.}, ``{HATHOR}: {HAdronic} {Top} and {Heavy}
  quarks {crOss} section {calculatoR}'',} \textit{ Comput. Phys. Commun.}
  \textbf{ 182} (2011) 1034,
  \href{http://dx.doi.org/10.1016/j.cpc.2010.12.040}{\doi{10.1016/j.cpc.2010.12.040}},
  \href{http://www.arXiv.org/abs/1007.1327}{\texttt{arXiv:1007.1327}}.

\bibitem{Kant:2014oha}
P.~Kant\hrefCMSnoop {}{ { et~al.}, ``{HatHor} for single top-quark production:
  Updated predictions and uncertainty estimates for single top-quark production
  in hadronic collisions'',} \textit{ Comput. Phys. Commun.} \textbf{ 191}
  (2015) 74,
  \href{http://dx.doi.org/10.1016/j.cpc.2015.02.001}{\doi{10.1016/j.cpc.2015.02.001}},
  \href{http://www.arXiv.org/abs/1406.4403}{\texttt{arXiv:1406.4403}}.

\bibitem{Maltoni:2015ena}
\hrefCMSnoop {}{F.~Maltoni, D.~Pagani, and I.~Tsinikos, ``Associated production
  of a top-quark pair with vector bosons at {NLO} in {QCD}: impact on
  {$\PQt\PAQt\PH$} searches at the {LHC}'',} \textit{ JHEP} \textbf{ 02} (2016)
  113,
  \href{http://dx.doi.org/10.1007/JHEP02(2016)113}{\doi{10.1007/JHEP02(2016)113}},
  \href{http://www.arXiv.org/abs/1507.05640}{\texttt{arXiv:1507.05640}}.

\bibitem{bib:mcfm:diboson}
\hrefCMSnoop {}{J.~M. Campbell, R.~K. Ellis, and C.~Williams, ``Vector boson
  pair production at the {LHC}'',} \textit{ JHEP} \textbf{ 07} (2011) 018,
  \href{http://dx.doi.org/10.1007/JHEP07(2011)018}{\doi{10.1007/JHEP07(2011)018}},
  \href{http://www.arXiv.org/abs/1105.0020}{\texttt{arXiv:1105.0020}}.

\bibitem{CMS:2020ebo}
\hrefCMSnoop {}{{CMS Collaboration}, ``Pileup mitigation at {CMS} in {13\TeV}
  data'',} \textit{ JINST} \textbf{ 15} (2020) P09018,
  \href{http://dx.doi.org/10.1088/1748-0221/15/09/P09018}{\doi{10.1088/1748-0221/15/09/P09018}},
  \href{http://www.arXiv.org/abs/2003.00503}{\texttt{arXiv:2003.00503}}.

\bibitem{CMS-PRF-14-001}
\hrefCMSnoop {}{{CMS Collaboration}, ``Particle-flow reconstruction and global
  event description with the {CMS} detector'',} \textit{ JINST} \textbf{ 12}
  (2017) P10003,
  \href{http://dx.doi.org/10.1088/1748-0221/12/10/P10003}{\doi{10.1088/1748-0221/12/10/P10003}},
  \href{http://www.arXiv.org/abs/1706.04965}{\texttt{arXiv:1706.04965}}.

\bibitem{CMS-TDR-15-02}
\href {http://cds.cern.ch/record/2020886}{{CMS Collaboration}, ``Technical
  proposal for the phase-2 upgrade of the {Compact Muon Solenoid}'',} CMS
  Technical Proposal CERN-LHCC-2015-010, CMS-TDR-15-02, 2015.

\bibitem{CMS:2018rym}
\hrefCMSnoop {}{{CMS Collaboration}, ``Performance of the {CMS} muon detector
  and muon reconstruction with proton-proton collisions at
  $\sqrt{s}=13$\,{TeV}'',} \textit{ JINST} \textbf{ 13} (2018) P06015,
  \href{http://dx.doi.org/10.1088/1748-0221/13/06/P06015}{\doi{10.1088/1748-0221/13/06/P06015}},
  \href{http://www.arXiv.org/abs/1804.04528}{\texttt{arXiv:1804.04528}}.

\bibitem{CMS-DP-2020-021}
\href {https://cds.cern.ch/record/2717925}{{CMS Collaboration}, ``{ECAL} 2016
  refined calibration and {Run2} summary plots'',} CMS Detector Performance
  Summary CMS-DP-2020-021, 2020.

\bibitem{Cacciari:2008gp}
\hrefCMSnoop {}{M.~Cacciari, G.~P. Salam, and G.~Soyez, ``The anti-\kt jet
  clustering algorithm'',} \textit{ JHEP} \textbf{ 04} (2008) 063,
  \href{http://dx.doi.org/10.1088/1126-6708/2008/04/063}{\doi{10.1088/1126-6708/2008/04/063}},
  \href{http://www.arXiv.org/abs/0802.1189}{\texttt{arXiv:0802.1189}}.

\bibitem{Cacciari:2011ma}
\hrefCMSnoop {}{M.~Cacciari, G.~P. Salam, and G.~Soyez, ``{FastJet} user
  manual'',} \textit{ Eur. Phys. J. C} \textbf{ 72} (2012) 1896,
  \href{http://dx.doi.org/10.1140/epjc/s10052-012-1896-2}{\doi{10.1140/epjc/s10052-012-1896-2}},
  \href{http://www.arXiv.org/abs/1111.6097}{\texttt{arXiv:1111.6097}}.

\bibitem{Cacciari:2008gn}
\hrefCMSnoop {}{M.~Cacciari, G.~P. Salam, and G.~Soyez, ``The catchment area of
  jets'',} \textit{ JHEP} \textbf{ 04} (2008) 005,
  \href{http://dx.doi.org/10.1088/1126-6708/2008/04/005}{\doi{10.1088/1126-6708/2008/04/005}},
  \href{http://www.arXiv.org/abs/0802.1188}{\texttt{arXiv:0802.1188}}.

\bibitem{CMS:2016lmd}
\hrefCMSnoop {}{{CMS Collaboration}, ``Jet energy scale and resolution in the
  {CMS} experiment in {$\Pp\Pp$} collisions at 8\,{TeV}'',} \textit{ JINST}
  \textbf{ 12} (2017) P02014,
  \href{http://dx.doi.org/10.1088/1748-0221/12/02/P02014}{\doi{10.1088/1748-0221/12/02/P02014}},
  \href{http://www.arXiv.org/abs/1607.03663}{\texttt{arXiv:1607.03663}}.

\bibitem{CMS-PAS-JME-16-003}
\href {{http://cds.cern.ch/record/2256875}}{{CMS Collaboration}, ``Jet
  algorithms performance in 13{\TeV} data'',} {CMS Physics Analysis Summary}
  CMS-PAS-JME-16-003, 2017.

\bibitem{BTV-16-002}
\hrefCMSnoop {}{{CMS Collaboration}, ``Identification of heavy-flavour jets
  with the {CMS} detector in {$\Pp\Pp$} collisions at 13\,{TeV}'',} \textit{
  JINST} \textbf{ 13} (2018) P05011,
  \href{http://dx.doi.org/10.1088/1748-0221/13/05/P05011}{\doi{10.1088/1748-0221/13/05/P05011}},
  \href{http://www.arXiv.org/abs/1712.07158}{\texttt{arXiv:1712.07158}}.

\bibitem{Bols:2020bkb}
E.~Bols\hrefCMSnoop {}{ { et~al.}, ``Jet flavour classification using
  {DeepJet}'',} \textit{ JINST} \textbf{ 15} (2020) P12012,
  \href{http://dx.doi.org/10.1088/1748-0221/15/12/P12012}{\doi{10.1088/1748-0221/15/12/P12012}},
  \href{http://www.arXiv.org/abs/2008.10519}{\texttt{arXiv:2008.10519}}.

\bibitem{CMS-DP-2021-004}
\href {https://cds.cern.ch/record/2759970}{{CMS Collaboration}, ``B-tagging
  performance of the {CMS} legacy dataset 2018'',} CMS Detector Performance
  Summary CMS-DP-2021-004, 2021.

\bibitem{Sirunyan:2019kia}
\hrefCMSnoop {}{{CMS Collaboration}, ``Performance of missing transverse
  momentum reconstruction in proton-proton collisions at $\sqrt{s} = 13$\,{TeV}
  using the {CMS} detector'',} \textit{ JINST} \textbf{ 14} (2019) P07004,
  \href{http://dx.doi.org/10.1088/1748-0221/14/07/P07004}{\doi{10.1088/1748-0221/14/07/P07004}},
  \href{http://www.arXiv.org/abs/1903.06078}{\texttt{arXiv:1903.06078}}.

\bibitem{Kondo:1988yd}
\hrefCMSnoop {}{K.~Kondo, ``Dynamical likelihood method for reconstruction of
  events with missing momentum. 1: Method and toy models'',} \textit{ J. Phys.
  Soc. Jap.} \textbf{ 57} (1988) 4126,
  \href{http://dx.doi.org/10.1143/JPSJ.57.4126}{\doi{10.1143/JPSJ.57.4126}}.

\bibitem{CMS:2015enw}
\hrefCMSnoop {}{{CMS Collaboration}, ``Search for a standard model {Higgs}
  boson produced in association with a top-quark pair and decaying to bottom
  quarks using a matrix element method'',} \textit{ Eur. Phys. J. C} \textbf{
  75} (2015) 251,
  \href{http://dx.doi.org/10.1140/epjc/s10052-015-3454-1}{\doi{10.1140/epjc/s10052-015-3454-1}},
  \href{http://www.arXiv.org/abs/1502.02485}{\texttt{arXiv:1502.02485}}.

\bibitem{CMS:2023xjh}
\hrefCMSnoop {}{{CMS Collaboration}, ``Inclusive and differential cross section
  measurements of {$\PQt\PAQt\PQb\PAQb$} production in the lepton+jets channel
  at $\sqrt{s}=13$\,{TeV}'',} \textit{ JHEP} \textbf{ 05} (2024) 042,
  \href{http://dx.doi.org/10.1007/JHEP05(2024)042}{\doi{10.1007/JHEP05(2024)042}},
  \href{http://www.arXiv.org/abs/2309.14442}{\texttt{arXiv:2309.14442}}.

\bibitem{Sirunyan:2020kga}
\hrefCMSnoop {}{{CMS Collaboration}, ``Measurement of the cross section for
  {$\PQt\PAQt$} production with additional jets and {\PQb} jets in {$\Pp\Pp$}
  collisions at $\sqrt{s}=13$\,{TeV}'',} \textit{ JHEP} \textbf{ 07} (2020)
  125,
  \href{http://dx.doi.org/10.1007/JHEP07(2020)125}{\doi{10.1007/JHEP07(2020)125}},
  \href{http://www.arXiv.org/abs/2003.06467}{\texttt{arXiv:2003.06467}}.

\bibitem{Sirunyan:2019jud}
\hrefCMSnoop {}{{CMS Collaboration}, ``Measurement of the
  {$\PQt\PAQt\PQb\PAQb$} production cross section in the all-jet final state in
  {$\Pp\Pp$} collisions at $\sqrt{s}=13$\,{TeV}'',} \textit{ Phys. Lett. B}
  \textbf{ 803} (2020) 135285,
  \href{http://dx.doi.org/10.1016/j.physletb.2020.135285}{\doi{10.1016/j.physletb.2020.135285}},
  \href{http://www.arXiv.org/abs/1909.05306}{\texttt{arXiv:1909.05306}}.

\bibitem{Collaboration:2267573}
\href {{https://cds.cern.ch/record/2267573}}{{CMS Collaboration}, ``Object
  definitions for top quark analyses at the particle level'',} {CMS Note}
  CMS-NOTE-2017-004, 2017.

\bibitem{CMS:2024onh}
\hrefCMSnoop {}{{CMS Collaboration}, ``The {CMS} statistical analysis and
  combination tool: {\textsc{Combine}}'',} \textit{ Comput. Softw. Big Sci.}
  \textbf{ 8} (2024) 19,
  \href{http://dx.doi.org/10.1007/s41781-024-00121-4}{\doi{10.1007/s41781-024-00121-4}},
  \href{http://www.arXiv.org/abs/2404.06614}{\texttt{arXiv:2404.06614}}.

\bibitem{Conway:2011in}
\hrefCMSnoop {}{J.~S. Conway, ``Incorporating nuisance parameters in
  likelihoods for multisource spectra'',} in \textit{ {PHYSTAT 2011}}.
\newblock 3, 2011.
\newblock \href{http://www.arXiv.org/abs/1103.0354}{\texttt{arXiv:1103.0354}}.
\newblock
  \href{http://dx.doi.org/10.5170/CERN-2011-006.115}{\doi{10.5170/CERN-2011-006.115}}.

\bibitem{chollet2015keras}
\hrefCMSnoop {}{F.~Chollet { et~al.}, ``\textsc{Keras}''.}
  \url{https://keras.io}, 2015.

\bibitem{ml:general}
M.~Erdmann, J.~Glombitza, G.~Kasieczka, and U.~Klemradt, ``Deep Learning for
  Physics Research''.
\newblock WORLD SCIENTIFIC, 2021.
\newblock \href{http://dx.doi.org/10.1142/12294}{\doi{10.1142/12294}}.

\bibitem{snoek2012practical}
\hrefCMSnoop {}{S.~Jasper, L.~Hugo, and P.~A. Ryan, ``Practical bayesian
  optimization of machine learning algorithms'',} 2012.
  \href{http://www.arXiv.org/abs/1206.2944}{\texttt{arXiv:1206.2944}}.

\bibitem{pmlr-v70-franceschi17a}
\href {http://proceedings.mlr.press/v70/franceschi17a.html}{F.~Luca,
  D.~Michele, F.~Paolo, and P.~Massimiliano, ``Forward and reverse
  gradient-based hyperparameter optimization'',} in \textit{ Proceedings of the
  34th International Conference on Machine Learning}, D.~Precup and Y.~W. Teh,
  eds., volume~70 of \textit{ Proceedings of Machine Learning Research},
  p.~1165.
\newblock PMLR, 2017.

\bibitem{hinton2012improving}
G.~E. Hinton\hrefCMSnoop {}{ { et~al.}, ``Improving neural networks by
  preventing co-adaptation of feature detectors'',} 2012.
  \href{http://www.arXiv.org/abs/1207.0580}{\texttt{arXiv:1207.0580}}.

\bibitem{Wunsch:2018oxb}
\hrefCMSnoop {}{S.~Wunsch, R.~Friese, R.~Wolf, and G.~Quast, ``Identifying the
  relevant dependencies of the neural network response on characteristics of
  the input space'',} \textit{ Comput. Softw. Big Sci.} \textbf{ 2} (2018) 5,
  \href{http://dx.doi.org/10.1007/s41781-018-0012-1}{\doi{10.1007/s41781-018-0012-1}},
  \href{http://www.arXiv.org/abs/1803.08782}{\texttt{arXiv:1803.08782}}.

\bibitem{ghorbani2019data}
\hrefCMSnoop {}{A.~Ghorbani and J.~Zou, ``Data {Shapley}: Equitable valuation
  of data for machine learning'',} 2019.
  \href{http://www.arXiv.org/abs/1904.02868}{\texttt{arXiv:1904.02868}}.

\bibitem{Bjorken:1969wi}
\hrefCMSnoop {}{J.~D. Bjorken and S.~J. Brodsky, ``Statistical model for
  electron-positron annihilation into hadrons'',} \textit{ Phys. Rev. D}
  \textbf{ 1} (1970) 1416,
  \href{http://dx.doi.org/10.1103/PhysRevD.1.1416}{\doi{10.1103/PhysRevD.1.1416}}.

\bibitem{tagkey1979543}
\hrefCMSnoop {}{G.~C. Fox and S.~Wolfram, ``Event shapes in {$\Pep\Pem$}
  annihilation'',} \textit{ Nucl. Phys. B} \textbf{ 157} (1979) 543,
  \href{http://dx.doi.org/10.1016/0550-3213(79)90120-2}{\doi{10.1016/0550-3213(79)90120-2}}.

\bibitem{Goodfellow-et-al-2016}
I.~Goodfellow, Y.~Bengio, and A.~Courville, ``Deep Learning''.
\newblock MIT Press, 2016.
\newblock \url{http://www.deeplearningbook.org}.

\bibitem{Lindsey1996-nh}
J.~K. Lindsey, ``Parametric statistical inference''.
\newblock Clarendon Press, Oxford, England, 1996.

\bibitem{Skands:2014pea}
\hrefCMSnoop {}{P.~Skands, S.~Carrazza, and J.~Rojo, ``Tuning {\PYTHIA} 8.1:
  the {Monash} 2013 {Tune}'',} \textit{ Eur. Phys. J. C} \textbf{ 74} (2014)
  3024,
  \href{http://dx.doi.org/10.1140/epjc/s10052-014-3024-y}{\doi{10.1140/epjc/s10052-014-3024-y}},
  \href{http://www.arXiv.org/abs/1404.5630}{\texttt{arXiv:1404.5630}}.

\bibitem{Sirunyan:2018wem}
\hrefCMSnoop {}{{CMS Collaboration}, ``Measurement of differential cross
  sections for the production of top quark pairs and of additional jets in
  lepton+jets events from {$\Pp\Pp$} collisions at $\sqrt{s}=13$\,{TeV}'',}
  \textit{ Phys. Rev. D} \textbf{ 97} (2018) 112003,
  \href{http://dx.doi.org/10.1103/PhysRevD.97.112003}{\doi{10.1103/PhysRevD.97.112003}},
  \href{http://www.arXiv.org/abs/1803.08856}{\texttt{arXiv:1803.08856}}.

\bibitem{Stewart:2011cf}
\hrefCMSnoop {}{I.~W. Stewart and F.~J. Tackmann, ``Theory uncertainties for
  {Higgs} and other searches using jet bins'',} \textit{ Phys. Rev. D} \textbf{
  85} (2012) 034011,
  \href{http://dx.doi.org/10.1103/PhysRevD.85.034011}{\doi{10.1103/PhysRevD.85.034011}},
  \href{http://www.arXiv.org/abs/1107.2117}{\texttt{arXiv:1107.2117}}.

\bibitem{stxs_ttHbb_migration_uncertainties}
\href {{https://cds.cern.ch/record/2878797}}{{ATLAS Collaboration},
  ``Evaluation of {QCD} uncertainties for {Higgs} boson production through
  gluon fusion and in association with two top quarks for simplified template
  cross-section measurements'',} {ATLAS PUB Note} ATL-PHYS-PUB-2023-031, 2023.

\bibitem{Barlow:1993dm}
\hrefCMSnoop {}{R.~Barlow and C.~Beeston, ``Fitting using finite {Monte}
  {Carlo} samples'',} \textit{ Comp. Phys. Commun.} \textbf{ 77} (1993) 219,
  \href{http://dx.doi.org/10.1016/0010-4655(93)90005-W}{\doi{10.1016/0010-4655(93)90005-W}}.

\bibitem{CMS:2020utv}
\hrefCMSnoop {}{{CMS Collaboration}, ``First measurement of the cross section
  for top quark pair production with additional charm jets using dileptonic
  final states in {$\Pp\Pp$} collisions at $\sqrt{s}=13$\,{TeV}'',} \textit{
  Phys. Lett. B} \textbf{ 820} (2021) 136565,
  \href{http://dx.doi.org/10.1016/j.physletb.2021.136565}{\doi{10.1016/j.physletb.2021.136565}},
  \href{http://www.arXiv.org/abs/2012.09225}{\texttt{arXiv:2012.09225}}.

\bibitem{Heinemeyer:2013tqa}
\hrefCMSnoop {}{{LHC Higgs Cross Section Working Group}, ``Handbook of {LHC}
  {Higgs} cross sections: 3. {Higgs} properties'',} 2013.
  \href{http://www.arXiv.org/abs/1307.1347}{\texttt{arXiv:1307.1347}}.

\bibitem{Gritsan:2016hjl}
\hrefCMSnoop {}{A.~V. Gritsan, R.~R{\"o}ntsch, M.~Schulze, and M.~Xiao,
  ``Constraining anomalous {Higgs} boson couplings to the heavy flavor fermions
  using matrix element techniques'',} \textit{ Phys. Rev. D} \textbf{ 94}
  (2016) 055023,
  \href{http://dx.doi.org/10.1103/PhysRevD.94.055023}{\doi{10.1103/PhysRevD.94.055023}},
  \href{http://www.arXiv.org/abs/1606.03107}{\texttt{arXiv:1606.03107}}.

\bibitem{Demartin:2014fia}
F.~Demartin\hrefCMSnoop {}{ { et~al.}, ``{Higgs} characterisation at {NLO} in
  {QCD}: {CP} properties of the top-quark {Yukawa} interaction'',} \textit{
  Eur. Phys. J. C} \textbf{ 74} (2014) 3065,
  \href{http://dx.doi.org/10.1140/epjc/s10052-014-3065-2}{\doi{10.1140/epjc/s10052-014-3065-2}},
  \href{http://www.arXiv.org/abs/1407.5089}{\texttt{arXiv:1407.5089}}.

\end{thebibliography}\endgroup
\cleardoublepage \appendix\section{The CMS Collaboration \label{app:collab}}\begin{sloppypar}\hyphenpenalty=5000\widowpenalty=500\clubpenalty=5000
\cmsinstitute{Yerevan Physics Institute, Yerevan, Armenia}
{\tolerance=6000
A.~Hayrapetyan, A.~Tumasyan\cmsAuthorMark{1}\cmsorcid{0009-0000-0684-6742}
\par}
\cmsinstitute{Institut f\"{u}r Hochenergiephysik, Vienna, Austria}
{\tolerance=6000
W.~Adam\cmsorcid{0000-0001-9099-4341}, J.W.~Andrejkovic, T.~Bergauer\cmsorcid{0000-0002-5786-0293}, S.~Chatterjee\cmsorcid{0000-0003-2660-0349}, K.~Damanakis\cmsorcid{0000-0001-5389-2872}, M.~Dragicevic\cmsorcid{0000-0003-1967-6783}, P.S.~Hussain\cmsorcid{0000-0002-4825-5278}, M.~Jeitler\cmsAuthorMark{2}\cmsorcid{0000-0002-5141-9560}, N.~Krammer\cmsorcid{0000-0002-0548-0985}, A.~Li\cmsorcid{0000-0002-4547-116X}, D.~Liko\cmsorcid{0000-0002-3380-473X}, I.~Mikulec\cmsorcid{0000-0003-0385-2746}, J.~Schieck\cmsAuthorMark{2}\cmsorcid{0000-0002-1058-8093}, R.~Sch\"{o}fbeck\cmsorcid{0000-0002-2332-8784}, D.~Schwarz\cmsorcid{0000-0002-3821-7331}, M.~Sonawane\cmsorcid{0000-0003-0510-7010}, S.~Templ\cmsorcid{0000-0003-3137-5692}, W.~Waltenberger\cmsorcid{0000-0002-6215-7228}, C.-E.~Wulz\cmsAuthorMark{2}\cmsorcid{0000-0001-9226-5812}
% ...
\par}
\vskip\cmsinstskip
\dag:~Deceased\\
$^{1}$Also at Yerevan State University, Yerevan, Armenia\\
$^{2}$Also at TU Wien, Vienna, Austria\\
% ...
\end{sloppypar}
\end{document}